\documentclass[aps,rmp,reprint,amsmath,amssymb]{revtex4-2}
\usepackage[utf8]{inputenc}
\usepackage{graphicx}			
\usepackage{color}				
\usepackage[hidelinks]{hyperref}	

\usepackage[font=footnotesize]{caption}
\newcommand{\DHe}{D$^{3}$He}

\begin{document}

\title{Proton Imaging of High-Energy-Density Laboratory Plasmas}

\author{Derek B. Schaeffer}
    \email{derek.schaeffer@ucla.edu}
    \affiliation{Department of Physics and Astronomy, University of California - Los Angeles, Los Angeles, CA 90095, USA}
\author{Archie F.A. Bott}
    \affiliation{Department of Physics, University of Oxford, Parks Road, Oxford OX1 3PU, UK}
\author{Marco Borghesi}
    \affiliation{School of Mathematics and Physics, The Queen's University Belfast, Belfast BT7 1NN, United Kingdom}
\author{Kirk A. Flippo}
    \affiliation{Applied and Fundamental Physics (P-2), Los Alamos National Laboratory, Los Alamos, New Mexico 87544, USA}
\author{William Fox}
    \affiliation{Princeton Plasma Physics Laboratory, Princeton, NJ 08543, USA}
    \affiliation{Department of Astrophysical Science, Princeton University, Princeton, NJ 08540, USA}
\author{Julien Fuchs}
   \affiliation{LULI-CNRS, CEA, UPMC Univ Paris 06:Sorbonne Universit\'{e},\'{E}cole Polytechnique, Institut Polytechnique de Paris, Palaiseau, France}
\author{Chikang Li}
    \affiliation{Plasma Science and Fusion Center, Massachusetts Institute of Technology, Cambridge, Massachusetts 02139, USA}
\author{Hye-Sook Park}
    \affiliation{Lawrence Livermore National Laboratory, Livermore, California 94550, USA}
\author{Fredrick H. S\'{e}guin}
    \affiliation{Plasma Science and Fusion Center, Massachusetts Institute of Technology, Cambridge, Massachusetts 02139, USA}
\author{Petros Tzeferacos}
    \affiliation{Department of Physics and Astronomy, University of Rochester, Rochester, New York 14627, USA}
\author{Louise Willingale}
    \affiliation{G\'{e}rard Mourou Center for Ultrafast Optical Science, Department of Electrical Engineering and Computer Science, University of Michigan, Ann Arbor, Michigan 48109, USA}

\date{\today}

\begin{abstract}

Proton imaging has become a key diagnostic for measuring electromagnetic fields in high-energy-density (HED) laboratory plasmas.  Compared to other techniques for diagnosing fields, proton imaging is a non-perturbative measurement that can simultaneously offer high spatial and temporal resolution and the ability to distinguish between electric and magnetic fields.  Consequently, proton imaging has been used in a wide range of HED experiments, from inertial confinement fusion to laboratory astrophysics.  An overview is provided on the state of the art of proton imaging, including detailed discussion of experimental considerations like proton sources and detectors, the theory of proton-imaging analysis, and a survey of experimental results demonstrating the breadth of applications. Topics at the frontiers of proton imaging development are also described, along with an outlook on the future of the field.

\end{abstract}

\maketitle

\tableofcontents

\newpage

\section{Introduction} \label{sec:intro}

\subsection{Context and Principles} \label{sec:intro:overview}

\begin{figure}
    \centering
    \includegraphics[width=8cm]{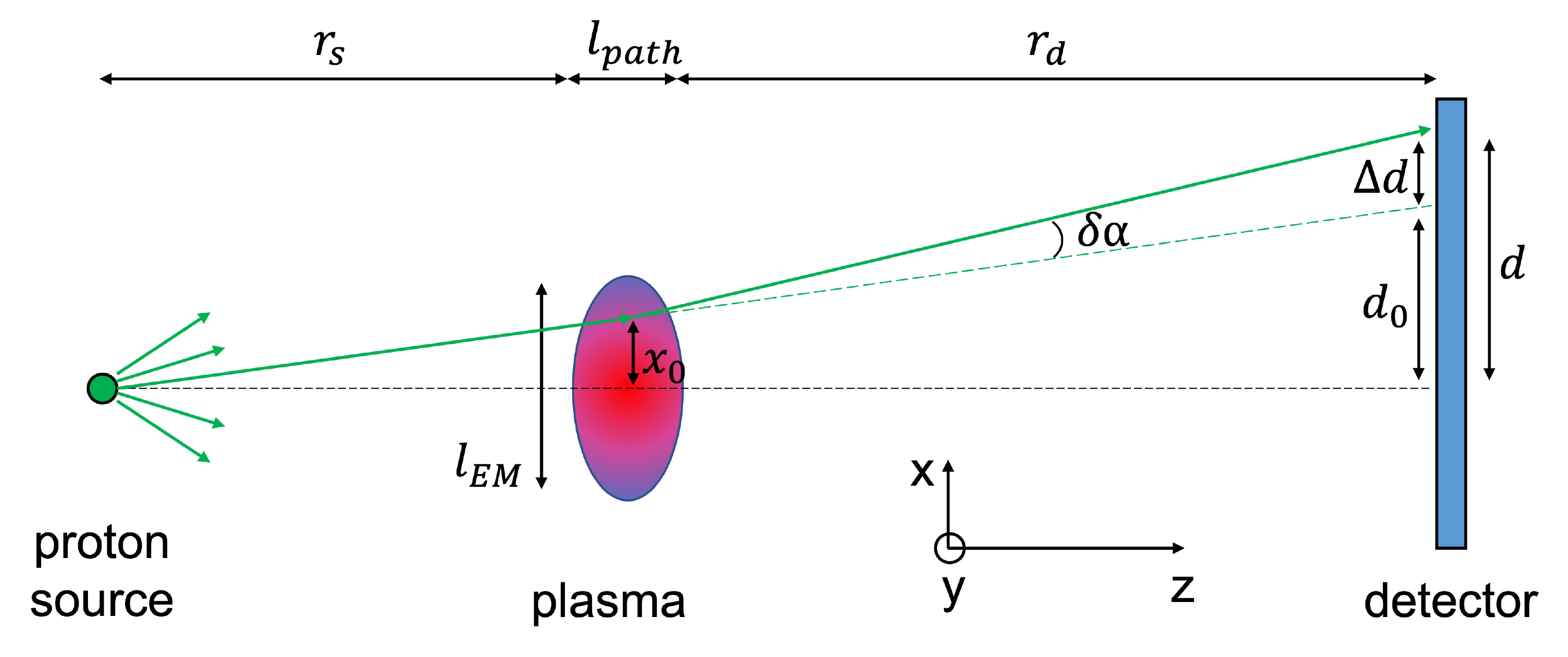}
    \caption{Schematic proton imaging setup.}
    \label{fig:simple_prad}
\end{figure}

Plasmas with energy densities exceeding $10^{11}$~J\,m$^{-3}$ -- or, equivalently, pressures above 1~Mbar -- are found in a wide range of contemporary laboratory experiments~\cite{Colvin_2014}. These ``high-energy-density'' (HED) plasmas are often created using high-powered laser beams or energetic pulsed power devices and, given the high pressures involved, the conditions typically exist over relatively short timescales ($\lesssim100$~ns) and small volumes ($\lesssim1$~cm$^{3}$)~\cite{Drake_2018}. Historically, HED plasmas were studied primarily in the pursuit of inertial confinement fusion (ICF), in which lasers are used to compress a small pellet of fusion material to extreme pressures, with the goal of initiating a self-sustaining burning plasma to harness as an energy source~\cite{Lindl_1995,Craxton_2015}. More generally, there have been a large number of experiments studying HED plasmas generated by the interaction of lasers with solid or gaseous targets, with applications to hydrodynamic instabilities, particle acceleration, and ultrafast field and particle dynamics. Because plasmas are also a key component of many astrophysical systems, more recently the field of laboratory astrophysics has utilized HED plasmas in scaled experiments to study a variety of astrophysical phenomena \cite{Gregori_2015,Lebedev_2019,Takabe_2021,Blackman_2022}.

An important component in many HED laboratory plasmas is the dynamics of the electromagnetic fields.  In ICF, the application of strong magnetic fields is sometimes used to help confine and heat the plasma \cite{Slutz_2010,Chang_2011}.  How these fields are compressed, diffuse, and seed instabilities are critical questions for controlling the fusion process.  Electromagnetic fields are also fundamental to many kinetic processes studied with HED experiments, including collisionless shocks, filamentary instabilities, jets, magnetic reconnection, and turbulence.  Measuring electromagnetic fields is thus vital for helping to answer many key open questions in HED plasma physics.  However, owing to the high plasma densities and temperatures, short timescales, and/or small volumes, measuring such fields with existing x-ray, optical and electronic diagnostics is extremely challenging.

Proton imaging is a diagnostic technique in which the deflection of a laser-driven proton probe by Lorentz forces in a plasma can be used to infer an image of the path-integrated strength of the electromagnetic fields.  For typical proton energies of several MeV, proton imaging is well-suited to studying HED plasmas with strong fields for several reasons: 1) the protons are ``stiff'' enough that they suffer only small deflections for typical field strengths, allowing the detected proton position to be related simply to the initial proton position in order to infer field strengths; 2) for many experiments, the protons traverse the experimental plasma on timescales short compared to dynamic timescales, providing a relatively static snapshot of the fields; 3) the proton images have high spatial resolution owing to (i) the small source size of laser-driven proton beams as well as (ii) their high laminarity; 4) the proton beam, being locally of much lower density than the  probed plasma, is non-perturbative; and 5) the dependence of the proton deflections on proton energy or geometry is different for electric and magnetic fields, enabling the contribution from each to be distinguished by using different proton energies or probing from different directions.  We note that proton imaging is also referred to as ``proton radiography'' or ``proton deflectometry'' in the literature, where the former can be used to describe the imaging of proton scattering and stopping from either density or electromagnetic fields, and the latter is often used when directly measuring proton deflections with, for example, a mesh or grid. In this review, we will primarily focus on proton deflections from electromagnetic fields rather than from collisions.

A schematic of a proton imaging setup is shown in Fig.~\ref{fig:simple_prad}. Protons from a point source pass through the plasma of interest, are deflected by electromagnetic fields, and then travel ballistically to a detector where they form an image of the field structures in the plasma.  Importantly, inhomogeneous electromagnetic fields in the plasma plane will give rise to an inhomogeneous proton fluence on the detector.  This in turn allows the path-integrated strengths of the electromagnetic fields to be estimated by relating the proton fluence variations on the detector to the displacement suffered by those protons as they pass through the fields in the plasma. 

The diagnostic is typically configured in the paraxial limit, in which the characteristic scale $\ell_{\rm EM}$ of electromagnetic fields in the plasma being probed is much smaller than the distance $r_{\rm s}$ between the source and the plasma ($\ell_{\rm EM}\ll r_{\rm s}$) and in a point-projection geometry, in which the distance $r_{\rm d}$ from the plasma to the detector greatly exceeds the path-length $l_{\rm path}$ of the protons through the plasma ($r_{\rm d} \gg l_{\rm path}$).  Consequently, for sufficiently large proton energies (with characteristic deflection velocities that are much smaller than the emitted velocities), the path-integrated electromagnetic field strengths can be related to the inhomogeneous distribution of proton fluence on the detector.  Under these approximations, and limiting to deflections along $\hat{\boldsymbol{x}}$ without loss of generality, the deflection angle $\delta \alpha$ of a proton is given by (see Fig.~\ref{fig:simple_prad})

\begin{equation}
\delta \alpha = \frac{e}{m_p v_p^2}\int_0^{l_{\rm path}} \mathrm{d}s \, \left[ E_x+(\boldsymbol{v}_p\times\boldsymbol{B})_{x} \right] ,
\end{equation}

\noindent where $e$ is the elementary charge, $m_p$ the mass of a proton, $\boldsymbol{v}_p$ ($v_p$) the protons' velocity (speed), and $\boldsymbol{E}$ and $\boldsymbol{B}$ the electric and magnetic fields in the plasma, respectively. The final position $d$ of the proton in the image plane at the detector will be

\begin{equation}
d =d_0 + \Delta d = \mathcal{M}x_0 + r_{\rm d}\delta \alpha ,
\end{equation}

\noindent where $x_0$ is the initial transverse position of the proton in the plasma, $d_0$ is the undeflected proton position in the detector plane accounting for magnification $\mathcal{M}\approx(r_{\rm s}+r_{\rm d})/r_{\rm s}$, and $\Delta d$ is the displacement due to the deflection of protons by electromagnetic fields in the plasma. Thus, the path-integrated fields can be inferred from

\begin{equation}
\int_0^{l_{\rm path}}\mathrm{d}s \, \left[ E_x+(\boldsymbol{v}_p\times\boldsymbol{B})_{x} \right] = \frac{m_p v_p^2}{e}\frac{r_{\rm s}+r_{\rm d}}{r_{\rm s} r_{\rm d}}(x - x_0) ,
\end{equation}

\noindent where $x =d/\mathcal{M}$ is the deflected position re-scaled to the plasma plane. 

A useful metric for classifying different types of proton-fluence inhomogeneities is the \textit{contrast parameter} 

\begin{equation}
\mu\equiv r_{\rm d}\delta\alpha/\mathcal{M} \ell_{\rm EM}\sim\delta\Psi/\Psi_0
\end{equation}

\noindent where $\Psi_0$ is the mean proton fluence, and $\delta\Psi$ is the magnitude of the inhomogeneities. For $\mu \ll 1$, the relation between the path-integrated fields and inhomogeneities is approximately linear, and the measured proton-fluence distribution is proportional to the path-integrated charge (for purely electrostatic fields) distribution, or current density (for purely magnetostatic fields) distribution, respectively.  As $\mu$ increases, the proton-fluence distribution becomes spatially distorted compared to the path-integrated charge-density and current-density distributions, with regions of focused and defocused fluence; however, qualitatively the image is still similar to these density distributions. When $\mu$ becomes larger than some critical value $\mu_c\sim1$, proton trajectories cross before reaching the detector, leading to the formation of caustics \cite{Kugland_2012} and making the interpretation of proton images more difficult.

Proton imaging has several advantages over other methods for measuring electromagnetic fields in HED plasmas, and is the only practical means for measuring electric fields.  Magnetic flux (``b-dot'') probes \cite{Everson_2009}, which consist of one or more loops of wire inserted into the plasma to measure magnetic flux through Faraday's law, are frequently used in plasma experiments.  However, they are perturbative in typical HED plasma experiments since their spatial extent is often a significant fraction of the size of such plasmas, which also make their spatial resolution poor.  Additionally, they do not measure electric fields and are sensitive to electromagnetic pulses (EMP) from high-intensity laser-target interactions \cite{Bradford_2018}. Faraday-rotation or Cotton-Mouton polarimetry~\cite{Segre_1999} are non-invasive laser-based optical probe diagnostics of magnetic fields that are insensitive to EMP, but since the former measures $\int n_e {B_{\|}} \mathrm{d}{z}$ (where $B_{\|}$ is the component of the magnetic field parallel to the probe beam), and the latter measures $\int n_e B_{\perp}^2 \mathrm{d}{z}$ (where $B_{\perp}$ is the perpendicular field's magnitude), both approaches require a simultaneous density measurement.  Similar to b-dot probes, polarimetry does not measure electric fields. Polarimetry measurements are also generally limited to underdense plasmas and can be difficult to implement due to refraction in plasmas with the large density gradients commonly found in HED experiments. The Zeeman effect can be used to measure magnetic fields \cite{Stamper_1991} by measuring the splitting of spectral lines, but field magnitudes in HED experiments are typically too small to be resolved with this technique, or the measurements are highly limited \cite{Rosenzweig_2020}.  Similarly, Thomson scattering, which measures scattered laser light from a plasma, can be used in principle to measure magnetic fields, but the required field strengths are much larger than achieved in most HED experiments~\cite{Froula_2011}. As a result, proton imaging has become a standard diagnostic of electromagnetic fields at many HED facilities.  

Additional historical context for the development of proton imaging is presented in the next section.  In Sec.~\ref{sec:exp} we discuss key components of experimental techniques and design for proton-imaging setups, including a comparison of proton sources and detectors.  In Sec.~\ref{sec:theory} we present a detailed overview of the theory of proton-imaging analysis, including both forward and inverse modeling.  In Sec~\ref{sec:apps} we briefly survey a wide variety of phenomena that have been investigated using proton imaging experiments.  In Sec.~\ref{sec:frontiers} we discuss the frontiers of proton imaging, including advanced proton sources, detectors, analysis techniques, and setup schemes.  Finally, in Sec.~\ref{sec:sum} we summarize our review and discuss the outlook for the field of proton imaging.

\subsection{Historical Development} \label{sec:intro:hist}

The first charged-particle-imaging experiments measuring electromagnetic fields in plasmas date back to the 1970s \cite{Mendel_1975} and utilized accelerators as a source of ions.  However, the long pulse length of ions from conventional accelerators and the difficulty of combining externally-produced ion beams with experiments limited the application of this technique to HED plasmas.  Not until the discovery of laser-driven, MeV proton sources was proton imaging regularly employed on HED facilities.

The development of multi-MeV, point-like proton sources useful for proton imaging was first demonstrated two decades ago \cite{Borghesi_2001}. The proton sources were generated by focusing high-intensity lasers onto thin foils; this generated MeV protons via a process called Target Normal Sheath Acceleration (TNSA) -- first described by \onlinecite{Wilks_2001}.  Radiochromic film stacks \cite{Borghesi_2001} and CR-39 nuclear track detectors \cite{Clark_2000, Maksimchuk_2000} were both initially used to image the protons, but the low fluence saturation limit of CR-39 and issues with data interpretation \cite{Gaillard_2006, Clark_2006} led to its disuse for TNSA protons.  Soon after these initial experiments, the first uses of TNSA-generated protons for measuring electromagnetic fields in HED plasmas were reported, with electric fields being characterized in ICF and laser-produced plasmas \cite{Borghesi_2001,Borghesi_2002}.  Meshes were first added a few years later to directly measure the proton deflections \cite{Mackinnon_2004}.

Around the same time that TNSA proton sources were being developed, a second type of laser-driven proton source based on capsules filled with \DHe{} gas was being developed in connection with direct-drive ICF experiments \cite{Li_2002,Smalyuk_2003}.  When imploded, these capsules emit $\sim$3.0 and $\sim$15.0~MeV protons as fusion byproducts. A distinctive feature of \DHe{}-capsule proton sources is their narrow energy spectra, which contrasts with the broadband proton energy-spectra generated by TNSA.  Compared to TNSA proton sources, the proton fluence from \DHe{} sources is significantly lower, requiring the use of low fluence CR-39 detectors \cite{Seguin_2003}.  In 2006, the use of a \DHe{} proton source to image electromagnetic fields in laser-produced plasmas was first reported \cite{Li_2006b}.

A key challenge of proton imaging is recovering the path-integrated electromagnetic fields based on the measured proton fluence.  The first approach chronologically, taken shortly after the initial deployment of high-intensity laser sources, was the development of numerical forward models that take a known electromagnetic field configuration and generate a synthetic proton fluence image that can be compared to the measured image. Quantitative analysis of such a comparison allowed for the optimal choice of characteristic parameters of the proposed electromagnetic field. These initial modeling efforts were employed to measure electric fields using data from TNSA proton sources \cite{Borghesi_2003,Romagnani_2005}, and with subsequent application, to determine electric and magnetic fields probed with \DHe{} sources \cite{Li_2006b}. Analytic models relating electromagnetic fields to their proton images were also developed around the same time \cite{Borghesi_2002,Romagnani_2005}, but the first detailed discussion of the analytic theory of proton imaging was not published until \onlinecite{Kugland_2012}. Obtaining direct measurements of the fields required the development of techniques to extract proton deflections from the proton fluence profiles.  This was first done through proton deflectometry \cite{Romagnani_2005,Li_2007,Petrasso_2009}, in which a mesh placed between the proton source and detector provided a direct reference for how the protons were deflected.  In many experiments, though, adding a mesh is not practical.  For these cases, a variety of numerical inversion schemes were developed and first reported in 2017 \cite{Bott_2017,Kasim_2017,Graziani_2017}. 

\section{Experimental Techniques} \label{sec:exp}

Proton imaging has been developed significantly over the past two decades, and is now commonly used at many HED experimental facilities.  In this section we describe each component needed to perform the measurement.  First, we discuss different proton sources and the methods for producing protons, as well as the properties of the protons generated.  Second, we describe the standard detectors used to measure the protons and the trade-offs associated with each.  Lastly, we discuss how the geometry of the experiment affects proton measurements and additional considerations when designing proton-imaging setups.

\subsection{Proton Sources} \label{sec:exp:sources}

There are two main types of proton sources that have been developed for proton imaging experiments: 1) proton beams accelerated by a high-intensity laser through the so-called Target Normal Sheath Acceleration (TNSA) mechanism, and 2) protons produced from nuclear fusion reactions resulting from laser-driven implosions of \DHe-filled targets. In the following we will briefly review the general characteristics of these two different sources, which differ significantly in terms of properties and capabilities. Table~\ref{tab:sources} summarizes comparatively the main properties of these sources.

\begin{table*}
\centering
\begin{tabular}{p{6cm} p{5cm} p{5cm}  l l l}
\hline
                                            & TNSA                                          & \DHe \\
\hline
Typical laser driver (energy, pulse width)  & $>50$~J, $\sim$~ps \newline $\sim1$~J, $\sim30$~fs                           & $\sim 10$~kJ, $\sim$~ns \\
Facility required                           & high-energy CPA laser                         & ICF facility (e.g., OMEGA, NIF, LMJ, Gekko and Shenguan)\\
\hline
Typical target                              & flat, metallic foil                           & \DHe-filled capsule (18~atm)\\
                                            & $\sim 10$--$25$~$\mu$m thick                  & capsule wall thickness $\sim 2.0$~$\mu$m\\
                                            &                                               & capsule diameter $\sim 420$~$\mu$m\\
\hline
Source size                                 & $\sim 10$~$\mu$m                              & $\sim 40$~$\mu$m (burn FWHM) \\
Source time cf. laser driver               & instantaneous                                 & $\sim 450$~ps (capsule bang time)\\
Proton temporal spread at source            & $\sim$~ps                                     & $100$~ps  \\
Spectral characteristics                    & Maxwellian-like \newline up to $\sim 60$~MeV  & DD, $\sim 3.3$~MeV \newline \DHe, $\sim 14.7$~MeV\\
Typical proton yield                        & $10^{11}$--$10^{13}$ (total in the beam)      & DD, $\sim 1 \times 10^9$\\
                                            &                                               & D$^3$He, $\sim 2 \times 10^{9}$\\
Proton directionality                       & Beam with $\sim 30^{\circ}$ divergence        & $4 \pi$ emission \\
Typical detector                            & RCF stack                                     & CR-39 \\
\hline
\end{tabular}
\caption{Comparison of the typical proton imaging source properties and characteristics.}
\label{tab:sources}
\end{table*}

\subsubsection{TNSA} \label{sec:exp:sources:tnsa}

Since the first reports of multi-MeV proton beams produced from laser-irradiated foils in 2000 \cite{Snavely_2000, Clark_2000,Maksimchuk_2000}, proton acceleration has been one of the most active fields of research employing high power, short-pulse lasers \cite{Macchi_2013}.
TNSA is the mechanism that has been most studied and has been widely employed for applications. TNSA was proposed as an interpretative framework \cite{Hatchett_2000, Wilks_2001} of the multi-MeV proton observations reported by \onlinecite{Snavely_2000}, obtained on the NOVA Petawatt laser at LLNL. The scheme typically employs mid-infrared (0.8--1~$\mu$m wavelength), multi-hundred-TW short-pulse (30~fs -- 10~ps pulse duration) laser systems that generate on-target intensities in the range of 10$^{19}$--10$^{21}$~W\,cm$^{-2}$.

\begin{figure*}
    \centering
    \includegraphics[width=15cm]{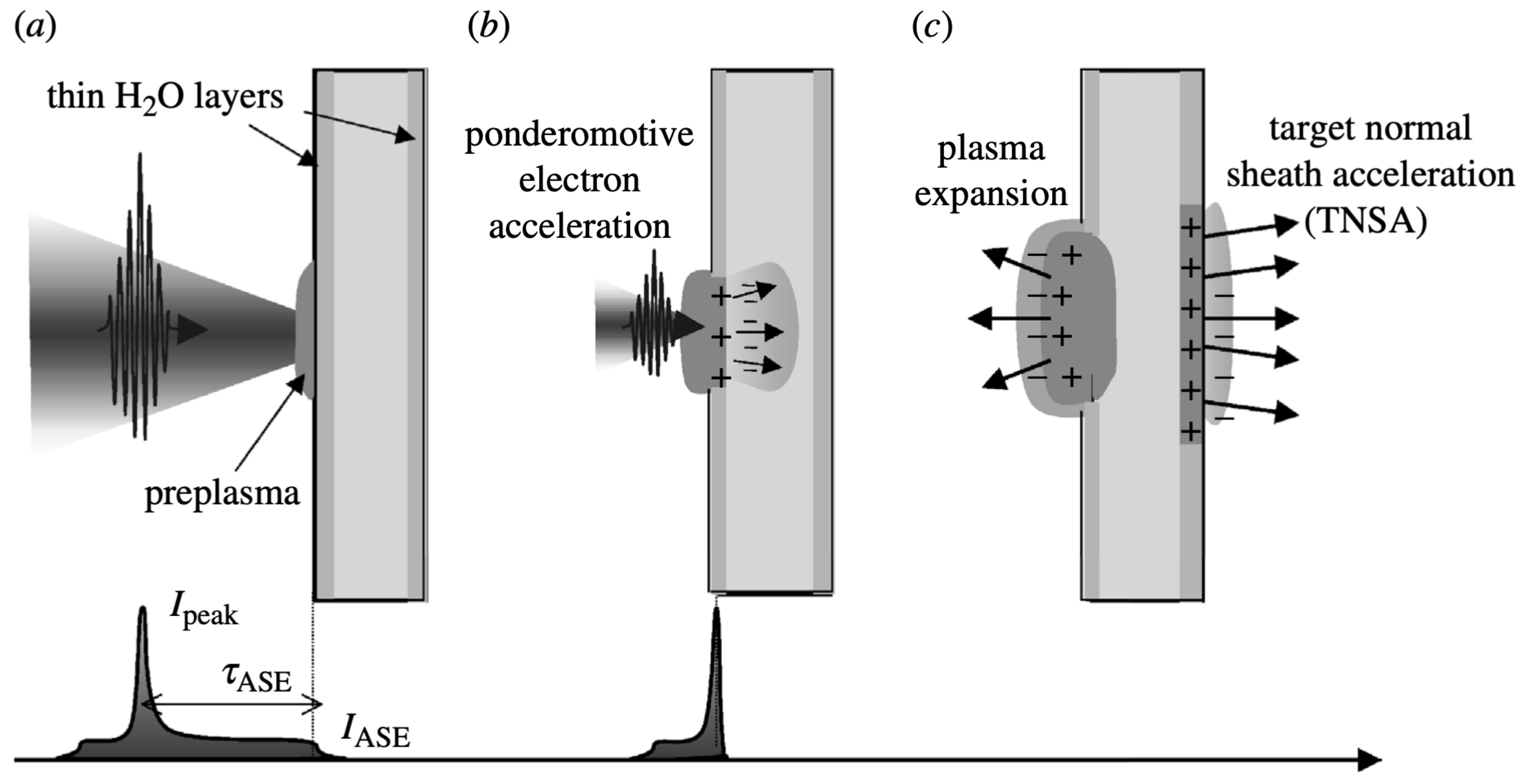}
    \caption{A schematic of the main processes involved in the TNSA mechanism. (a) First, a laser pre-pulse impinges upon and heats a thin target to form a pre-plasma.  The target contains layers of proton-rich hydrocarbons as common contaminants. (b) Second, the peak of the pulse arrives, efficiently heating electrons to relativistic temperatures. These electrons expand and propagate through the target. (c) Third, the hot electrons emerge into the vacuum and form an electron sheath of strength $\sim$~TV/m. This field ionizes the rear surface such that ions are accelerated to multi-MeV energies. Adapted from \onlinecite{McKenna_2006}.}
    \label{fig:TNSA}
\end{figure*}

A schematic of the TNSA process is shown in Fig.~\ref{fig:TNSA}.  A high-intensity laser pulse interacts with a solid foil target of thickness around a few microns. At these intensities, the laser pulse, focused on the foil surface, can efficiently couple energy into relativistic electrons, mainly through ponderomotive processes (e.g.\ $J \times B$ mechanism \cite{Kruer_1985}). The average energy of the electrons is typically of the order of MeV, e.g., their collisional range is much larger than the foil thickness, so that they can propagate to the rear of the target.
As the electrons expand into the vacuum they establish a space-charge field that ionizes the rear surface and drives the acceleration of ions from surface layers.
While a limited number of energetic electrons will effectively leave the target \cite{Link_2011}, most of the hot electrons are confined to within the target volume by the space charge and form a sheath extending by approximately a Debye length $\lambda_{D}=\sqrt{\varepsilon_{0}kT_{e,\mathrm{hot}} /(n_{e,\mathrm{hot}}e^{2})}$ from the initially unperturbed rear surface, where $n_{e,\mathrm{hot}}$ and $T_{e,\mathrm{hot}}$ are the density and temperature of the superthermal (\textit{hot}) electrons.
The electric field in the sheath is proportional to $(n_{e,\mathrm{hot}}T_{e,\mathrm{hot}})^{1/2}$ \cite{Mora_2003, Schreiber_2006}.  For a typical interaction, the sheath field reaches amplitudes in the TV/m range.  Under standard experimental conditions, contaminant layers (e.g., hydrocarbons, water) exist on the surface of any target \cite{Allen_2004}.  Therefore, protons are most efficiently accelerated by TNSA due to their favourable charge-to-mass ratio, and shield other ion species from experiencing the strongest accelerating fields.  This makes TNSA a very robust, efficient, and easily implementable mechanism for accelerating protons.

\begin{figure*}
    \centering
    \includegraphics[width=17cm]{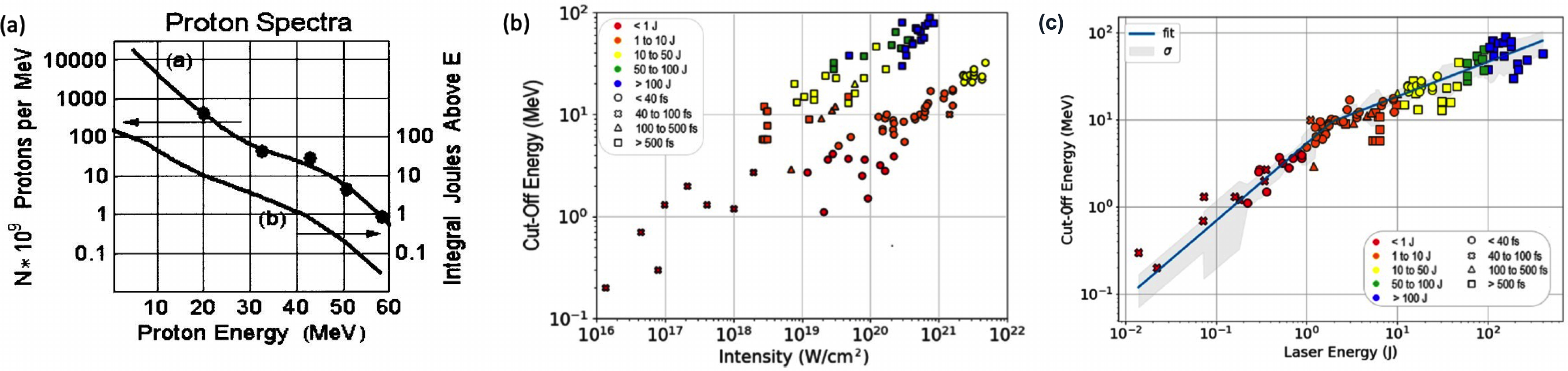}
    \caption{(a) TNSA spectrum obtained on the NOVA Petawatt laser at LLNL, expressed in number of protons per MeV (left scale).  Adapted from \onlinecite{Snavely_2000}. (b,c) TNSA cut-off energies plotted against (b) laser intensity on target and (c) laser energy. The data are taken from a selected number of experiments where a scan in laser energy was performed. Adapted from \onlinecite{Zimmer_2021}. See \onlinecite{Zimmer_2021} for references to the experiments used.}
    \label{fig:TNSA_scaling}
\end{figure*}

The energy spectra of TNSA proton beams are broadband, typically with an exponential profile up to a high energy cut-off (see Fig.~\ref{fig:TNSA_scaling}a).  The highest TNSA energies reported are of the order of 85~MeV \cite{Wagner_2016}, obtained with large PW-class laser systems, and available data generally shows that, at equal intensities, longer pulses ($\sim$~ps duration) containing more energy generally accelerate ions more efficiently than pulses with widths of tens of fs \cite{Macchi_2013}.  However, using state of the art fs systems and stringent control of the laser properties has recently allowed the energies of accelerated protons to be increased up to 70~MeV \cite{Ziegler_2021}.

Reported scaling laws for the proton energies as a function of laser intensity vary from a ponderomotive $I^{0.5}$ dependence for sub-ps pulses \cite{Macchi_2013} to a near-linear dependence observed for ultrashort laser pulses over restricted intensity ranges \cite{Zeil_2010} (see Fig.~\ref{fig:TNSA_scaling}b,c). Super-ponderomotive scaling for multi-kJ, multi-ps lasers has also been reported \cite{Flippo_2007, Mariscal_2019}. Nevertheless, secondary factors such as target thickness, target material, target size, laser contrast \cite{Kaluza_2004,Yogo_2008,Fuchs_2007,Schollmeier_2015}, etc. also play a very important role in TNSA accelerating energy performance.  Having a sharp density interface at the rear target surface is key to efficient TNSA acceleration.  For pulses with duration longer than a ps, the rear target surface evolves before the electrons associated with the peak intensity arrive, limiting the maximum acceleration \cite{Campbell_2019,Schollmeier_2015}.

If the laser pulse has a significant ``pre-pulse'', or energy arriving before the peak of the pulse, ionization of the material can begin before the main peak of the pulse arrives (see Fig.~\ref{fig:TNSA}).  The effect of the pre-pulse can be two-fold; it can create a plasma at the front of the target that alters the electron heating (usually enhancing the efficiency), and it can send a shock through the target that breaks out to form a pre-plasma on the rear surface.  Additionally, the interaction that is being probed may also cause pre-plasma at the rear of the target.  In either case, this pre-plasma at the rear surface can inhibit proton acceleration  \cite{Kaluza_2004,Fuchs_2007,Higginson_2021}. For this reason, a shield to protect the proton source foil is often used used to prevent these effects \cite{Zylstra_2012,Mackinnon_2006}.

The characteristics of the beams accelerated via TNSA are quite different from those of conventional radio frequency (RF) beams, with some superior properties that are particularly advantageous for use as a backlighter in proton imaging applications. These result from the short duration of the acceleration process \cite{Fuchs_2006,Schreiber_2006,Dromey_2016}, and from the fact that, unlike other ion sources, the protons are cold when accelerated with minimal transverse energy spread. 
The beams are therefore highly laminar \cite{Borghesi_2004} and characterized by ultralow transverse emittance (as low as 0.004~mm~mrad~\cite{Cowan_2004}) and by ultrashort ($\sim$~ps) duration at the source~\cite{Dromey_2016}. As a consequence of this, the emission properties of a TNSA beam can be described in terms of a \textit{virtual} source, much smaller than the region from which the protons are emitted, and typically located at a small distance in front of the target \cite{Borghesi_2004}.  The proton beam properties for imaging have been demonstrated to be optimum for $\sim$~ps duration laser pulses \cite{Campbell_2019} to limit emittance growth.
If the driving laser pulse duration is longer than $\sim 1$~ps, magnetic field instability growth on the rear surface deflects protons as they are accelerated  \cite{Nakatsutsumi_2018}.
Another key characteristic of TNSA proton beams is that they are bright, with $10^{11}$-–$10^{13}$ protons per shot with energies $>$ MeV, distributed across a broadband spectrum with a Boltzmann-like distribution.
The proton beam divergence is typically $\lesssim 30^{\circ}$, with the divergence decreasing with increasing energy \cite{Nurnberg_2009}.

The homogeneity of the transverse profile within a beam has been shown to be affected by the laser intensity profile at the target front \cite{Fuchs_2003}, as well as by instabilities occurring within the target, particularly within insulators, which tend to degrade the uniformity of the profile \cite{Fuchs_2003, Ruyer_2020}. Metallic targets typically induce smoother beams than insulators \cite{Quinn_2011}, and are therefore normally preferred for imaging applications.

\subsubsection{\DHe} \label{sec:exp:sources:dhe3}

\begin{figure*}
\centering
\includegraphics[width=14 cm]{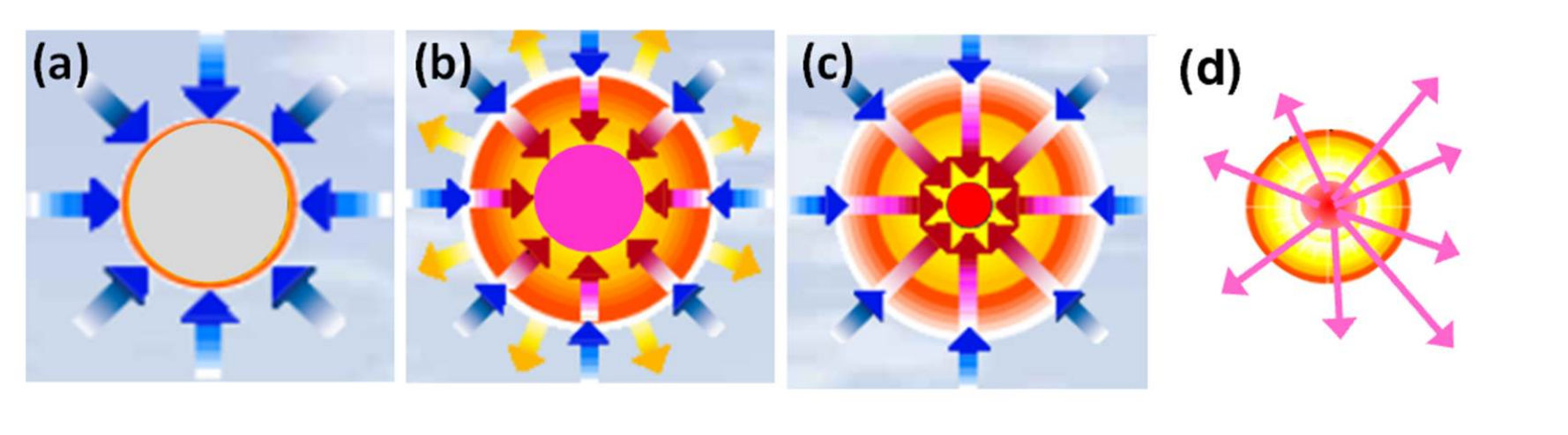}
\caption{Schematic of an exploding-pusher mode of capsule implosion and fusion in direct-drive inertial confinement fusion.  (a) Multiple laser beams directly and symmetrically illuminate the thin glass shell capsule surface. (b) The explosion of the shell caused by laser energy absorption drives a strong spherical shock propagating radially towards the capsule center. (c) The converging shock collapses in the center and bounces back, resulting in an increase of ion temperature and fuel density, and (d) the facilitation of nuclear fusion reactions and burn. }
\label{fig:dhe3_implosion}
\end{figure*}

\begin{figure*}
\includegraphics[width=16 cm]{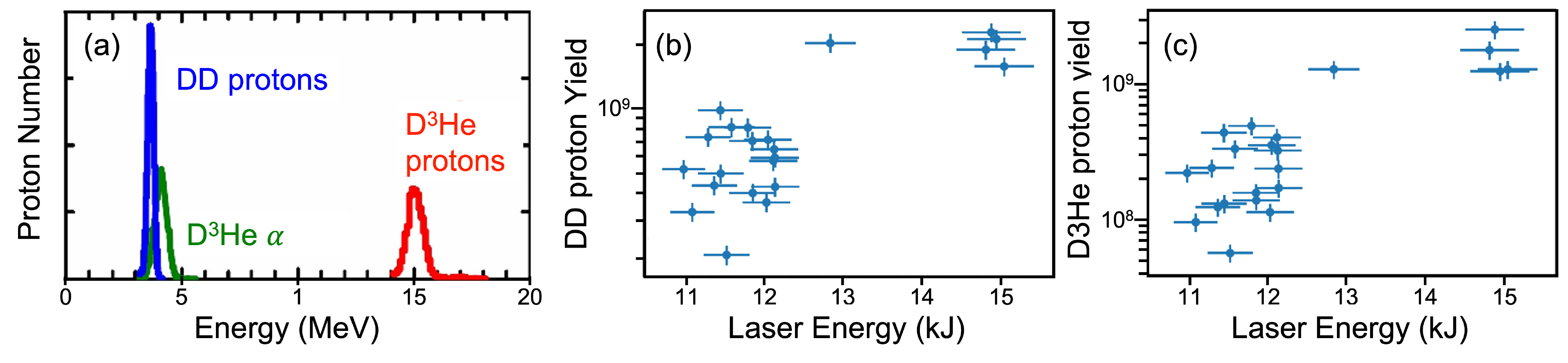}
\caption{(a) Typical spectra of fusion products generated in a \DHe-filled, thin-glass-shell, laser-driven exploding pusher as implemented for backlighters on the Omega laser facility. (b) DD and (c) \DHe{} proton yield as a function of laser energy on the capsule. Adapted from \onlinecite{Johnson_2021}.}
\label{fig:dhe3_spectrum}
\end{figure*}

A different approach to generating protons is to use the fusion reaction products from an inertial implosion. These sources were first developed in the context of proton backlighters for inertial confinement fusion (ICF) experiments on the Omega laser facility \cite{Li_2006a,Li_2006b} and have since been ported to the National Ignition Facility \cite{Zylstra_2020}. Contrary to the TNSA method, such a backlighter is formed by direct laser irradiation of a capsule filled with deuterium helium-3 (\DHe) gas.    

The \DHe{} backlighter platform uses a shock-driven implosion mode called ``exploding-pusher''. As schematically illustrated in Fig.~\ref{fig:dhe3_implosion}, the physical process involved in this scheme comprises three steps. First, multiple laser beams directly and symmetrically illuminate a thin glass shell capsule surface. Second, the strong laser absorption results in the explosion of capsule shell material, which drives a strong spherical shock wave propagating radially inwards towards the capsule center.  Finally, the converging shock collapses in the center and bounces back, resulting in an increase of the ion temperature and fuel density, which leads to nuclear fusion reactions and burn.  The nuclear ``bang time'' is usually defined as the time of peak fusion yield, and the nuclear ``burn time'' is defined by the full width at half maximum (FWHM) of the fusion product spectrum. 

The nuclear reaction results in the generation of mono-energetic 3~MeV DD protons [D + D$\rightarrow$T + p (3.0~MeV)] and 14.7~MeV \DHe~protons [D + $^3$He$\rightarrow$ $\alpha$ + p (14.7~MeV)], with typical yields of $\sim1\times10^9$. These fusion products and relative proton numbers are shown in Fig.~\ref{fig:dhe3_spectrum}a. More recently, a tri-particle backlighter platform utilizing a DT$^3$He capsule implosion has been developed, which provides 9.5~MeV deuterons from T + $^3$He$\rightarrow$ $\alpha$ + d (9.5~MeV) in addition to the 3~MeV DD and 14.7~MeV \DHe{} protons \cite{Sutcliffe_2021}.  Note that the interaction of the drive lasers with plasmas ablated from the capsule surface can generate hot electrons that escape from the capsule surface, which can lead to electric charging of the imploding capsule that can ``upshift'' the proton energies. For a typical implosion driven by a laser intensity of $10^{15}$~W\,cm$^{-2}$, $\sim$~MV electric potentials resulting in a $\sim0.5$ - $1.0$~MeV acceleration of fusion protons have been measured \cite{Hicks_2000, Rygg_2008}.

The typical implosion lasers consist of 0.6 -- 1~ns square pulses without phase plates and cumulative energies of $\sim$~10~kJ. The capsules have diameters of approximately 420 $\mu$m with a wall thickness of $\sim$~2~$\mu$m.  The capsule bang time is approximately 450~ps followed by a $\sim$~100~ps burn during which the protons are generated.  During the implosion the capsules reach a minimum burn size of $\sim$~40~$\mu$m (FWHM), which sets the spatial resolution of the resulting proton beams.

Recent studies have started to explore how proton yield from \DHe{} sources varies with laser and capsule parameters (see Fig.~\ref{fig:dhe3_spectrum}b,c).  By statistically sampling several hundred backlighter shots, it was found that total laser energy on the capsule and the asymmetry of the laser drive were the most important predictors of backlighter performance \cite{Johnson_2021}.  As a result, the best proton yields (both DD and \DHe{}) can be attained by using as many drive beams as possible (as least 9~kJ is recommended) while keeping the capsule illumination as symmetric as possible (see \onlinecite{Johnson_2021} for details).  In general, the combination of high asymmetry and a small number of beams should be avoided whenever possible.

\DHe{} protons have several unique features compared to TNSA protons. First, the fusion-generated protons are mono-energetic, with a typical energy uncertainty of about 3\%~\cite{Li_2006a} due to the finite nuclear burn region and energy straggling on the backlighter. Second, the different characteristic energies of the DD and \DHe{} protons naturally results in distinct times of flight for each proton energy, which can provide temporal resolutions of $\sim$~100~ps.  Third, a uniform and symmetric emission of fusion products provides a $4\pi$ solid angle isotropic proton fluence, though electric charging of the capsule may distort this~\cite{Manuel_2012} .

\subsection{Detectors} \label{sec:exp:detectors}

Each proton source is associated with a corresponding detector, namely radiochromic film (RCF) for TNSA protons and CR-39 for \DHe{} protons.  In the following sections we discuss the properties and characteristics of these detectors, which play a key role, along with the beam properties, in determining the features of the proton images. Briefly mentioned are other detectors which have been used, albeit less frequently.

\subsubsection{Film} \label{sec:exp:detectors:film}

\begin{figure}
\includegraphics[width=8 cm]{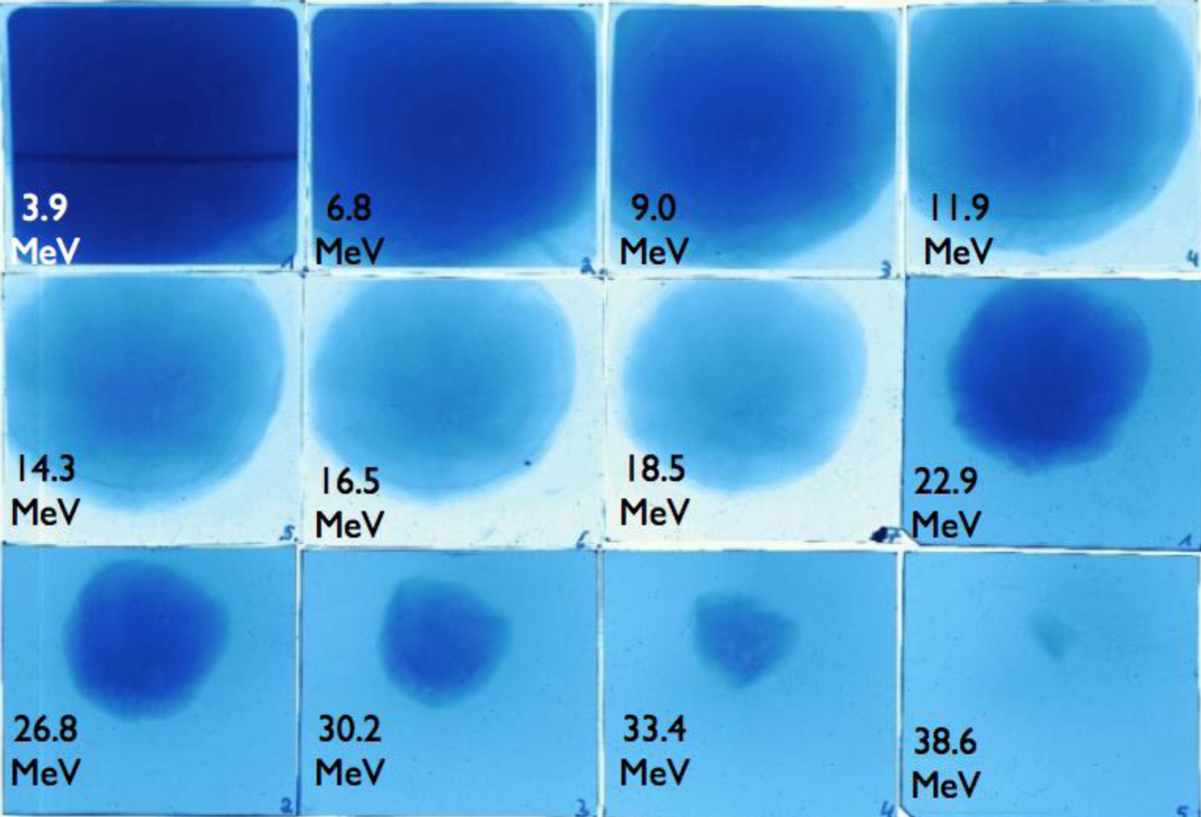}
\caption{RCF stack obtained on PHELIX, consisting of 7 films of type HD-810 and 5 films of type MD-55. The proton Bragg peak energy is given for each film layer. Adapted from \onlinecite{Bolton_2014}.}
\label{fig:RCFstack}
\end{figure}

Radiochromic films (RCF) are commonly used in dosimetry for a wide range of radiation sources (electrons, protons, photons) for medical, industrial, and scientific applications. This is a high-dose, high-dynamic range film, widely used in the clinical context for x-ray dosimetry \cite{Niroomand-Rad_1998}. RCF has become a popular choice for spectral and angular characterization of laser-driven proton beams \cite{Nurnberg_2009, Schollmeier_2014}, and the main detector of choice for TNSA-based proton imaging, thanks to its ease of use and effective performance at the particle fluences of typical experimental arrangements. The films consist of one or more active layers containing a microcrystalline monomeric dispersion buried in a clear plastic substrate. Different types are available, under the commercial GafChromic\textsuperscript{\texttrademark} name, that have varying active layer thickness and composition and consequently different sensitivity to ionizing radiation. Currently popular varieties are HD-V2 and EBT3.

There are a number of features that make RCF particularly attractive. RCF is a passive detector, the color and optical density of which is immediately, permanently, and visibly changed upon irradiation as a consequence of polymerization processes in the active layer, without the need for processing. The subsequent change in optical density can be calibrated against the radiation dose absorbed in the active layer of the film. Therefore, it is possible to extract information on particle fluence within the layer.

RCF can be digitized using inexpensive commercial photoscanners (photo-type flatbed scanners), which are fast and offer high spatial resolution (1600 dpi, or 63~dots/mm, resulting in a resolution of 16~$\mu$m, in most cases) and 16-bits per channel.
The intrinsic spatial resolution of RCF is higher (typically of micron scale) than the resolution of the scanners. RGB scanning provides separate color channels and produces images with different contrast/sensitivity, and provides options for extending further the dynamic range of the film. Conversion of the scanned images into dose requires a prior calibration of the film, which is typically obtained by exposing the films to known doses delivered by well-characterized fluxes of protons in conventional accelerators (e.g.,\ see \onlinecite{Chen_2016,Bin_2019,Xu_2019}).

In standard experimental configurations, RCFs are used in a stack arrangement, so that each layer acts as a filter for the following ones in the stack. Sometimes additional filter layers, typically aluminum foils, are used as spacers. The signal in a given film within the stack will only be due to protons having energy E $\geq{E_{B}}$, where $E_{B}$ is the energy reaching the Bragg peak within the active layer of the film. In first approximation, for a Boltzmann-like spectrum such as those typically produced by TNSA, the dose deposited in a layer can be taken as mostly deposited by protons with $E\sim{E_{B}}$. As we will see in Sec.~\ref{sec:exp:diag:temporal}, this property is at the basis of the unique temporal characterization capabilities of TNSA proton imaging. 
An example RCF stack is shown in Fig.~\ref{fig:RCFstack} and illustrates the color change of the film and the reduction in the beam divergence at higher proton energies.

\subsubsection{CR39} \label{sec:exp:detectors:cr39}

The \DHe~backlighter is ideally complemented by imaging detectors made of CR-39 \cite{Seguin_2003}. Although the process of reading out the data recorded on CR-39 is complicated (discussed below), the great advantage is that it records the exact position of every individual incident charged particle in the detector plane to an accuracy of $\sim2$~$\mu$m, as long as the maximum incident particle fluence is smaller than about $10^6$ per cm$^2$.  This fluence limit and saturation effects at higher flux \cite{Gaillard_2007} are the reason why CR-39 is not typically used for TNSA proton beam detection.

CR-39 (allyl diglycol carbonate, Columbia Resin \#39) polymer is part of a class of solid state nuclear track detectors (SSNTDs) that have been used for decades in many high energy particle counting applications, from radioactive dating to cosmic rays and neutrons (see \onlinecite{Fleisher_1965} and references therein). It has the useful property of being relatively insensitive to other forms of ionizing radiation, like gamma rays, x rays, or electrons, and is nearly 100\% efficient at detecting ions in a given energy range. Consequently, CR-39 has become the work horse for \DHe{} capsule backlighter experiments. It has also been used to calibrate other detectors due to its high efficiency and known response \cite{Harres_2008,Mancic_2008}.

CR-39 is a transparent plastic with chemical composition C$_{12}$H$_{18}$O$_7$ \cite{Seguin_2003,Seguin_2016,Fews_1982}.  A charged particle of appropriate energy passing through it leaves a trail of damage along its path in the form of broken molecular chains and free radicals. The amount of local damage along the path is related to the local rate at which energy is lost by the particle ($dE/dx$, where $x$ is distance along the path). The length of the path is the range of the particle in the plastic.  Particle paths can be made visible by etching the CR-39 in NaOH (e.g., see \onlinecite{Fews_1982, Gaillard_2007}); the etch time is typically between 0.5 and 5 hours (based on characteristics of the experiment such as the expected backlighter yield). The surface of the plastic is etched away at a ``bulk etch rate,'' while damaged material along a particle path etches at a faster ``track etch rate.'' If a particle path is normal to the plastic surface, the result of etching is a conical pit, or ``track,'' with a sharply defined, round entrance hole. 

\begin{figure}
\includegraphics[width=8cm]{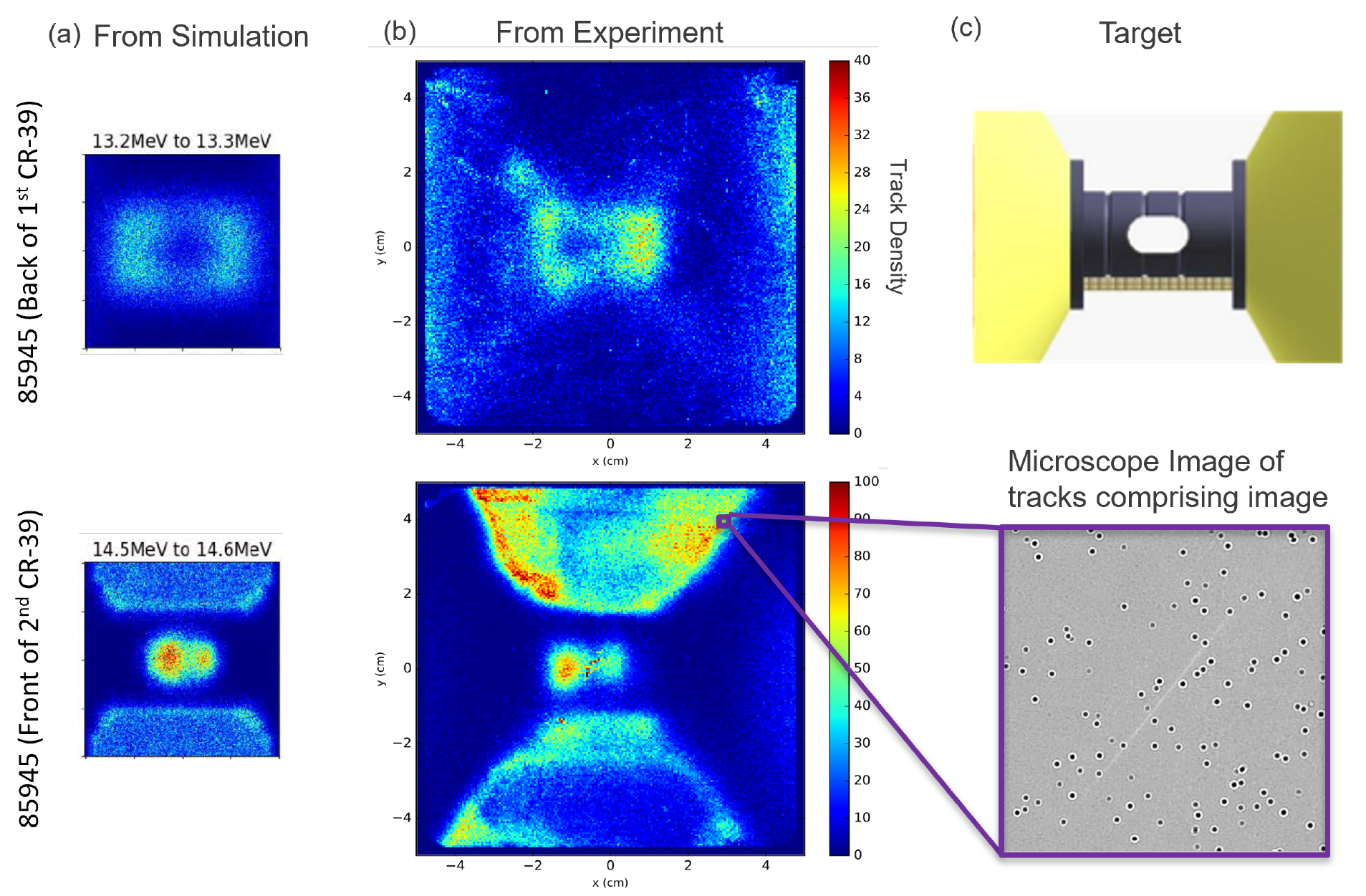}
\caption{Example proton images using a \DHe{} source. Column (a) shows synthetic proton images of the target shown at the top of column (c), several ns after lasers have driven shocks into the grey tube from either end. The top image is of 13~MeV protons, and the bottom image is of 14.7~MeV protons. These are the detected energies from protons born at 14.7~MeV and down scattered in energy by the target, allowing different aspects of the target to be imaged with a mono-energetic source. Column (b) shows the same images from the experiment, taken from a two-piece stack of CR-39.  The top image of downs-scattered 13~MeV protons is from the rear-side of the first piece of CR-39, while the bottom image of 14.7~MeV protons is from the front-side of the second piece of CR-39. The bottom image of column (c) shows an enlargement of the CR-39. Each dark circle is a particle track, and the faint diagonal line is due to laser light from a microscope’s autofocus mechanism. This image corresponds to about $1.6\times 10^{-4}$~cm$^2$, equivalent to 15\% of the area of one pixel in the experimental images. Adapted from \onlinecite{Lu_2020}.}
\label{fig:CR39_microscope}
\end{figure}

\begin{figure}
\includegraphics[width=8 cm]{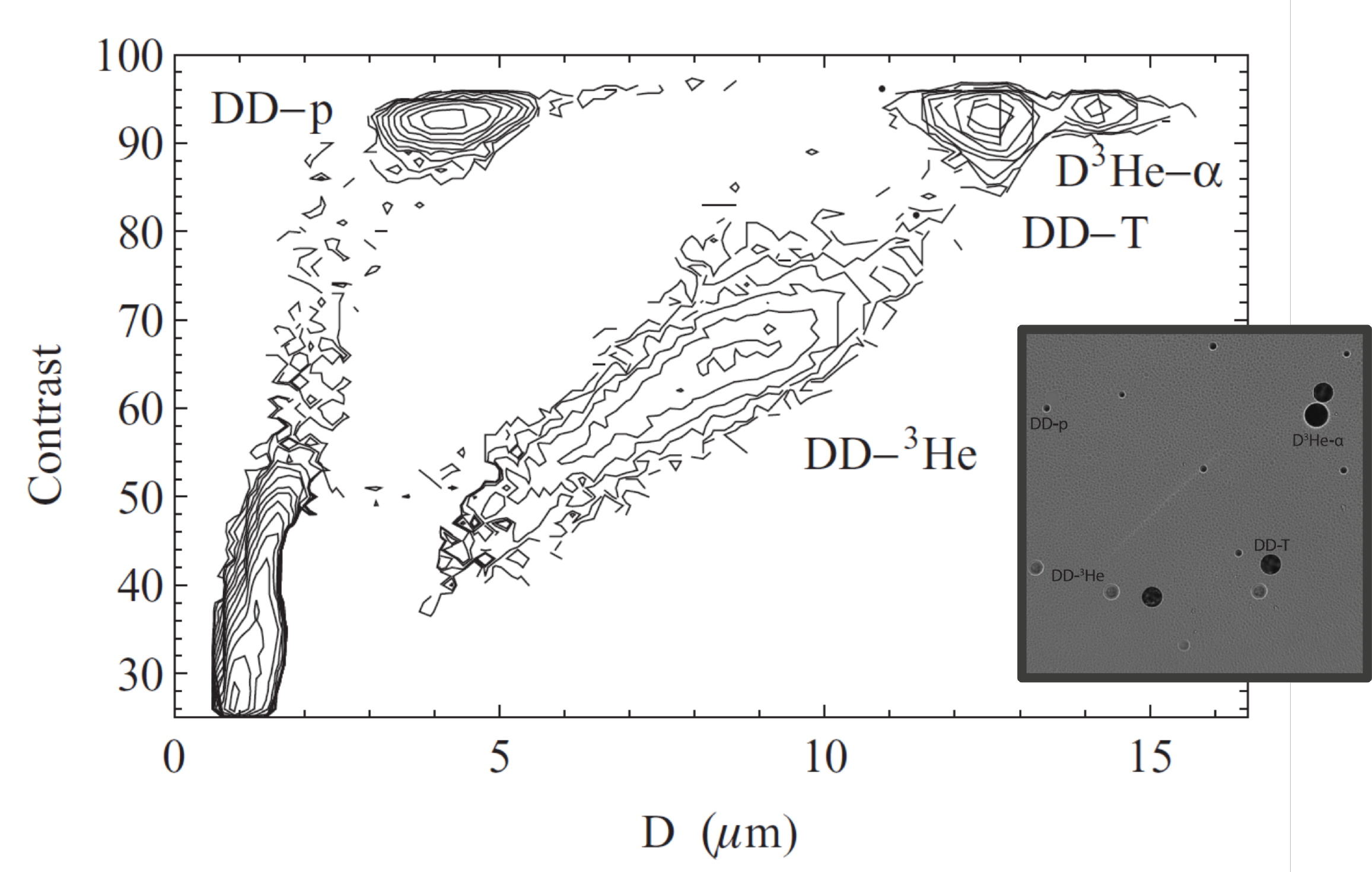}
\caption{Contour plot of the number of tracks versus track contrast and diameter for the piece of CR-39 shown in the inset.  The four particle species visible are labeled on the plot (compare to the inset image). Intrinsic CR-39 noise appears in the low-contrast low-diameter regime. Contours represent a constant number of tracks per unit contrast and diameter; the values of this quantity corresponding to plotted contours form a geometric series with a ratio of 2. As defined in this work a high contrast number is a dark track, while a low contrast number is a light track. Adapted from \onlinecite{Zylstra_2011}.}
\label{fig:CR39_tracks}
\end{figure}

Retrieving information about all individual particle tracks in an exposed piece of CR-39 involves scanning the entire CR-39 surface with an automated microscope system. Figure~\ref{fig:CR39_microscope}c bottom shows a sample microscope image of \DHe{}-proton tracks, each of which appears as a dark circle on a light background.  The location of each pit shows where a proton entered, and its diameter provides a measure of $dE/dx$ for the proton.  Since $dE/dx$ is different for particles of a given type but different energies, the diameter can provide a measure of particle energy (after passing through any filters in the detector pack).  $dE/dx$ is also different for different particle types, so diameters can often be used to identify the particle type if the energy is known (see Fig.~\ref{fig:CR39_tracks}), or to estimate the energy if the particle type is known \cite{Zylstra_2011,Sienian_2011}.
Not only can different particles or source energies be used to form images at different times (due to time of flight, see for example \onlinecite{Li_2009}), but also one can use the known down-scattered energies of one of the monoenergetic particles to produce separate images of the same target at the same time (see Fig.~\ref{fig:CR39_microscope}).

The optical magnification used in the scanning microscope system is usually (but not necessarily) chosen so that one camera frame covers the area that will be used for one pixel in the final desired proton image of particle fluence versus position.  That area is often chosen to be about 300~$\mu$m $\times$ 300~$\mu$m.  Each such camera image is evaluated with special algorithms that identify every individual track and determine its position coordinates, its diameter, its optical contrast, and its eccentricity \cite{Seguin_2003}. All of these measured parameters are recorded, and the microscope moves on to the next frame, continuing until the entire surface is covered. The resultant ``scan data'' file is saved for later processing, in which the final proton image is made by going through all of the recorded track information after deciding what display resolution is desired (frequently one microscope frame for each pixel) and counting the number of tracks in each ``pixel'' area that satisfies carefully chosen limits on diameter, contrast, and eccentricity \cite{Seguin_2003}.  Examples can be seen in Sec.~\ref{Sec:apps:ICF}.

\subsubsection{Others} \label{sec:exp:detectors:other}

While passive, single-use detectors such as RCF and CR-39 have been used in the vast majority of proton imaging experiments so far, the use of Micro Channel Plates (MCP) is also reported, albeit rarely, in the literature. MCPs, which are high-gain, spatially resolved electron multipliers \cite{Bolton_2014}, have been used often in proton acceleration experiments, mostly in the dispersion plane of a magnetic spectrometer or Thomson parabola \cite{Harres_2008}. An arrangement reported by \onlinecite{Sokollik_2008} extends this use to a streaked deflectometry approach, in which a TNSA beam is analyzed, after backlighting a target, in a magnetic spectrometer coupled to an MCP.  Use of MCPs as a proton imaging detector in a standard  arrangement is more challenging, e.g., requiring fast gating to obtain a single temporal snapshot out of a TNSA beam, and, to our knowledge, has not been pursued so far. 

Initial tests with scintillator plates \cite{Tang_2020} have indicated that, by selecting appropriate detector parameters, these may be used as an alternative to RCF, with the advantage of being suited to repeated use. The main disadvantages of scintillator detectors for proton imaging are 1) the energy resolution is reduced compared to RCF due to the thickness of the detector material and 2) it is difficult to extract the signal from different detector layers. A novel setup by \onlinecite{Huault_2019} using a concertina design of scintillators has been used to observe the proton energy spectra and proton beam divergence simultaneously. See Sec.~\ref{sec:frontiers:detectors} for more discussion.

\subsection{Diagnostic Geometry and Other Considerations} \label{sec:exp:diag}

\begin{figure*}
\includegraphics[width=18cm]{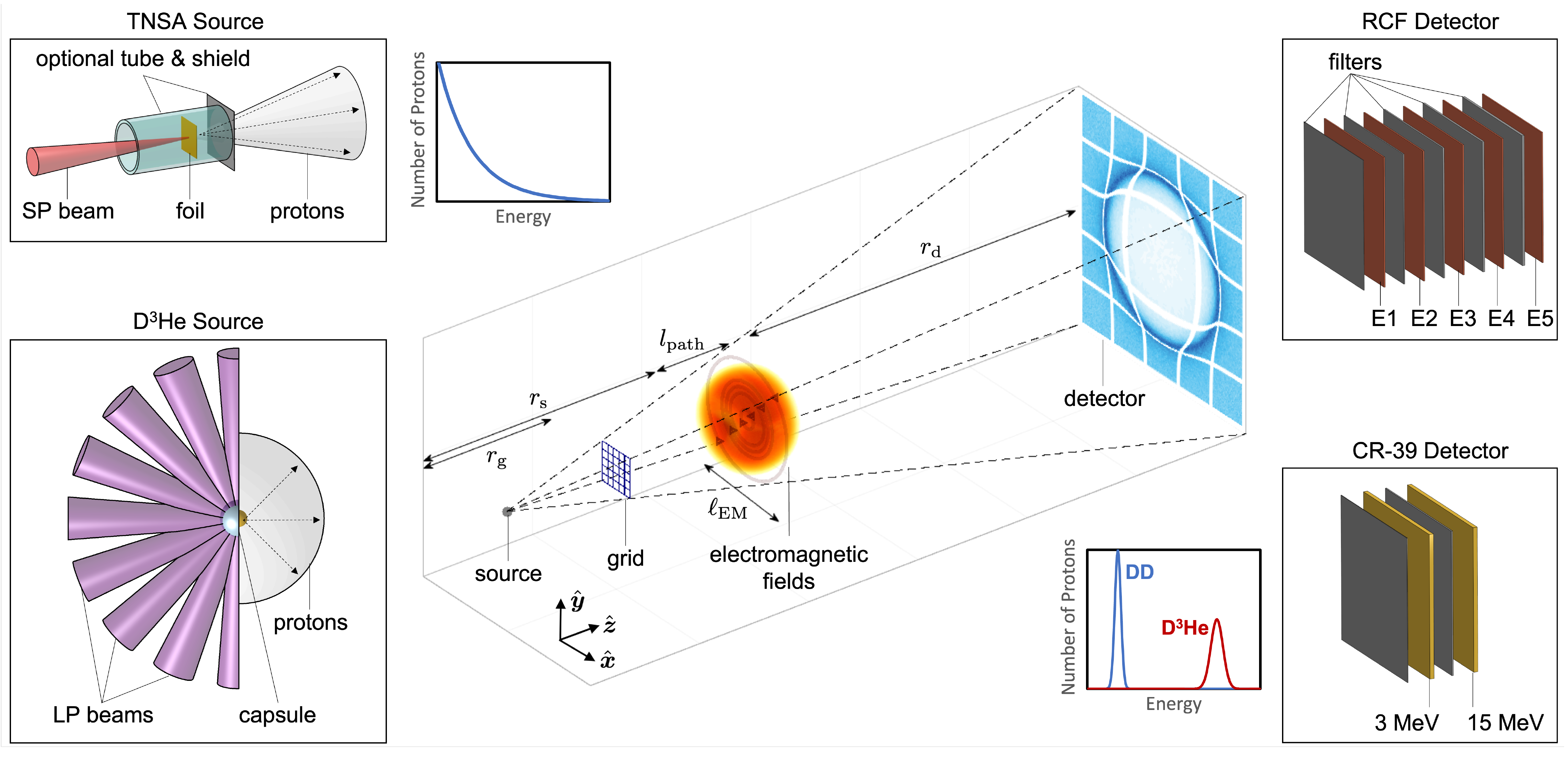}
\caption{Diagram illustrating the main components of a proton imaging diagnostic. \textit{Left:} Typical proton sources.  TNSA protons are generated by using a short pulse (SP) laser to irritate a thin foil, which emits protons in a beam with a broadband energy profile.  \DHe{} protons are generated by using long pulse (LP) lasers to drive the implosion on a thin-shell capsule, which isotropically emits mono-energetic DD and \DHe{} protons as fusion byproducts.  \textit{Right:} Typical proton detectors.  TNSA protons are collected on a stack of RCF that provides energy resolution.  \DHe{} protons are collected on CR-39, one for each proton energy. \textit{Center:} Typical proton imaging setup in a magnified point-source configuration, including source, optional mesh, the plasma under study, and detector (not to scale).} 
\label{fig:PRAD_diagram}
\end{figure*}

A diagram of a typical proton imaging setup as deployed in an experiment is shown in Fig.~\ref{fig:PRAD_diagram}.  The source can be either TNSA- or \DHe{}-generated protons, with corresponding detectors of either RCF or CR-39, respectively.  During an experiment, the protons are emitted by the source, propagate a distance $r_{\rm s}$ to the interaction region where they receive small deflections due to the electromagnetic fields, and then travel ballistically a distance $r_{\rm d}$ to the detector.

Typical implementations of TNSA and \DHe{} sources, and example detector stacks, are also shown in Fig.~\ref{fig:PRAD_diagram}.  A standardized TNSA source target has been developed on OMEGA EP \cite{Zylstra_2012}, in which a thin foil is mounted within a plastic tube, with a thin protective foil mounted over the end.  This shields the TNSA foil from radiation and plasma emerging from the object under study.  The tubes are transparent, which allows alignment of the laser focus to the foil via target chamber cameras.  The TNSA foil is driven by a short pulse laser, which can be moderately off-axis to allow some setup flexibility.  The resulting protons are emitted in a cone normal to the TNSA foil with a broadband energy distribution.

Likewise, a standard \DHe{} source capsule has been developed for both OMEGA \cite{Li_2006a,Li_2006b} and NIF \cite{Zylstra_2020}. The capsules are mounted on stalks and driven by a relatively symmetric set (typically $> 20$) of long pulse beams, resulting in protons emitted into $4\pi$ with mono-energetic energy distributions.  A comparison of TNSA and \DHe{} proton sources and detectors is summarized in Table~\ref{tab:sources}.

\subsubsection{Magnification} \label{sec:exp:diag:mag}

Typical setups take advantage of the small source size of the protons and obtain a magnified image onto a larger detector, with magnification

\begin{equation}
    \mathcal{M} = (r_{\rm d} + r_{\rm s}) / r_{\rm s}.
\end{equation}
 
\noindent Such a setup is often used to magnify the image from the plasma size (mm to 1  cm) to the detector size (typically several cm).  The magnification also improves by a factor $\mathcal{M}$ the spatial resolution at the plasma plane compared to the detector's spatial resolution.  Note that $r_{\rm d}$ is in principle different for each layer in the detector stack.  This can be especially important for the analysis of TNSA detector stacks, which can have a large number of layers.  Additionally, in experiments where the interaction length $l_{\rm path}$ is large, there can also be significant variation in $\mathcal{M}$.  An example of this is discussed in Sec.~\ref{Sec:apps:Dynamics}.

Experimental design should consider the size of the interaction such that the proton beam has expanded to overfill the region of interest. A small angle approximation is often used to assume that the proton beam along the probing axis travels the same distance as the protons at the edge of the detector (or the beam if smaller).  TNSA proton beams typically have a divergence of less than $30^{\circ}$, meaning the small angle approximation is reasonable in most cases, whereas the \DHe{} implosion is an isotropic source, so a limited solid angle should be used.  Similarly, when calculating the energy of the protons for a particular RCF stack layer, the extra distance within material traveled by the protons at the edge of the beam is usually ignored for calculations.

\subsubsection{Meshes/grids} \label{sec:exp:diag:mesh}

An optional mesh can be used to break the initial proton beam into beamlets, which, in a \textit{proton deflectometry} approach \cite{Mackinnon_2004}, facilitates measuring the fields via directly tracking beamlet deflections. The meshes used are typically commercial transmission electron microscopy grids that are available in a variety of pitches, hole widths, and bar widths, and are manufactured from relatively high-Z metals such as copper, nickel, or gold. The thickness of the meshes is typically such that a shadow is imprinted on the proton beam via multiple scattering in the mesh bars \cite{Borghesi_2004}.  By geometric arguments the mesh magnification to the detector is (see Fig.~\ref{fig:PRAD_diagram})

\begin{equation}
    \mathcal{M}_{\rm mesh,d} = \frac{r_{\rm d} + r_{\rm s}}{r_{\rm g}}.
\end{equation}

\noindent The spatial resolution, in turn, is set by the projection of the mesh period \textit{p} onto the plasma plane, i.e.$ (r_{\rm s}/r_{\rm g})\times p$.

The period of the mesh should be ideally chosen so that a sufficient number of mesh elements is projected across the probed region of interest. The period of the mesh should also be larger than the source size so that the mesh is not overly smeared out when projected.

A variation on the beamlet technique is to use an object (such as a mesh, mask, or Pepperpot) to sub-aperture the proton beam into many beamlets \cite{Sokollik_2008,Johnson_2022}, or down to a few ``pencil'' beamlets \cite{Lu_2020}, or even just a single beamlet \cite{Chen_2020}.  This allows one to probe areas of specific interest in a limited fashion which is more easily detectable (in terms of deflection) or to streak the beamlet in time.

\subsubsection{Spatial Resolution} \label{sec:exp:diag:spatial}

As is typical of all projection backlighting schemes, the intrinsic and ultimate spatial resolution of proton images is determined by the size of the proton source.  For \DHe{} capsules this is set by the burn volume of the implosion, which has been measured to be typically 40~$\mu$m FWHM \cite{Manuel_2012} (see Sec.~\ref{sec:exp:sources:dhe3}). For TNSA targets the relevant size is instead the ``virtual'' source size resulting from the beam's laminarity and emittance \cite{Borghesi_2004} (see Sec.~\ref{sec:exp:sources:tnsa}). This is typically of order 10~$\mu$m FWHM \cite{Borghesi_2004,Wang_2015,Li_2021}, set by the size of the laser focal spot, but can vary from experiment to experiment.

Scattering of the protons, which can occur in the plasma probed, in any protective foil (TNSA sources) and, in principle, in the detector, can degrade the spatial resolution from the values given above.  Scattering in the plasma will depend on its density and dimensions, as well as on the proton energy, and typically leads to a Gaussian distribution of angles with some $1/e$ radius $\theta_{\rm SC}$, which can be evaluated by Monte-Carlo calculations \cite{Ziegler_2010} or through empirical formulae \cite{Highland_1975}. This causes a resolution degradation characterized by a $1/e$ spatial width of order $r_{\rm d}\theta_{\rm SC}/\mathcal{M}$ in units of distance in the plasma plane \cite{Li_2006a}.  A similar effect will be caused by scattering in any protective foil, although the foil thickness is typically chosen in order to minimize this effect. Scattering in the detector, occurring for example when protons cross a stack on the way to the layer where they are detected, normally leads to negligible resolution loss once the magnification is taken into account.

Similarly, the intrinsic spatial resolution of the detector is typically very high, of order $\mu$m, and therefore does not contribute to the spatial resolution of the diagnostic when registered back to the plasma plane.

\subsubsection{Temporal Resolution} \label{sec:exp:diag:temporal}

There are three primary factors contributing to the temporal resolution of proton images \cite{Sarri_2010}:

1) The temporal duration of the source $\delta{t_{\rm p}}$. Ss discussed in the previous sections, this is of order 1~ps for TNSA beams for ps drivers (shorter for fs drivers \cite{Fuchs_2006}) and 100~ps for the \DHe{} capsules. This is the factor that determines the ultimate temporal resolution possible for a proton image, and the dominant factor for probing with \DHe{} protons.

2) The transit time $\delta{t_{\rm t}}$ of the protons through the region where the transient fields are located. This is related to the spatial scale over which the fields under investigation extend and is therefore intrinsic to the phenomenon under investigation. If the fields change on the timescale of the proton transit, the information will be temporally averaged over a time
$\delta t_{\rm t} \sim l_{\rm path}/v_{p} \sim l_{\rm path}(m_{p}/ 2 \epsilon_{p})^{0.5}$, where $\epsilon_{p}$ and $v_{p}$ are the energy and velocity of the protons, respectively. For example, for 10~MeV protons crossing a 100~$\mu$m region, one has $\delta{t_{\rm t}}\sim{2}$~ps.

3) The time-of-flight uncertainty (from the source to the plasma being probed) $\delta{t_{\rm d}}$ resulting from the energy resolution $\delta \epsilon_{p}$ of the detector. This is given by $\delta t_{\rm d} \sim r_{\rm s}({m_{p}/{2 \epsilon_{p}^{3})\delta{\epsilon_{p}}}}$.
For example, for a layer in a RCF stack one typically has $\delta{\epsilon_{p}}\sim 0.5$~MeV.
Taking, for example, 10~MeV protons and $r_{\rm s}= 3$~mm, one has $\delta{t_{\rm d}}\sim{1.5}$~ps.

While $\delta t_{\rm d}$ and $\delta t_{\rm t}$ are typically not relevant to determine the resolution for \DHe{} proton images (where the source duration is the dominant factor), they all can contribute significantly to the temporal resolution for experiments employing TNSA protons. Under typical experimental conditions, and depending on the specific experimental arrangement, this is typically in the range of 1--5~ps.

\subsubsection{Multi-Frame Capability } \label{sec:exp:diag:frame}

\begin{figure*}
\includegraphics[width=17 cm]{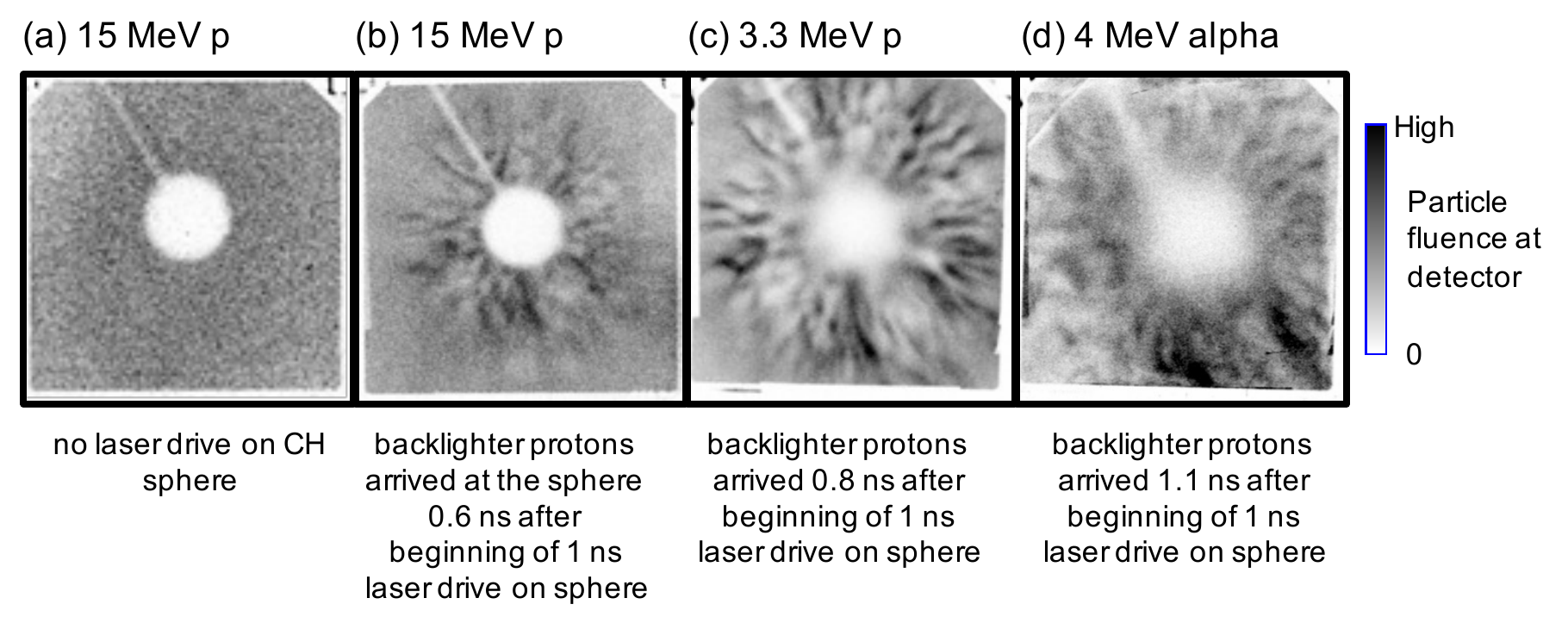}
\caption{Proton images of a laser-driven, solid 840 $\mu$m diameter CH sphere, made using a setup similar to Fig.~\ref{fig:PRAD_diagram}. Image (a) was recorded with no laser drive on the CH sphere, while images (b), (c) and (d) were recorded with laser drive for three different particle types and energies.  Adapted from \onlinecite{Seguin_2012a}.}
\label{fig:dhe3_radiographs}
\end{figure*}

Both TNSA and \DHe{} sources emit protons in a burst, which is typically shorter (or much shorter in the case of TNSA) than the time-of-flight to the plasma $r_{\rm s} / v_{\rm p}$. For $r_{\rm s}=1$~cm, for example, this would be $\sim 180$~ps for 15~MeV protons and $\sim 400$~ps for 3~MeV protons.  Consequently, a multi-frame capability can be achieved by using energy-resolving detectors (as RCF or CR-39 stacks), where stacking up images from different proton energies provides information on the temporal dynamics of the system over time intervals of order hundreds of ps.  

For \DHe{} sources, different frames can be obtained by employing the different fusion products produced during the implosion (see Fig.~\ref{fig:PRAD_diagram}). An example of the application of this capability is provided in Fig.~\ref{fig:dhe3_radiographs}.  The structure of the detector pack involves two metal filters and two separate layers of CR-39. The first CR-39 layer is proceeded by one of the metal filters, which helps protect the CR-39 from debris while still allowing the detection of $\sim$~3~MeV DD protons.  A second filter is placed before the second CR-39 layer and acts to help slow down the $\sim$~15~MeV \DHe{} protons to energies of 1--6~MeV, which is the best energy range for detecting protons on the CR-39.

\begin{figure}
\includegraphics[width=7 cm]{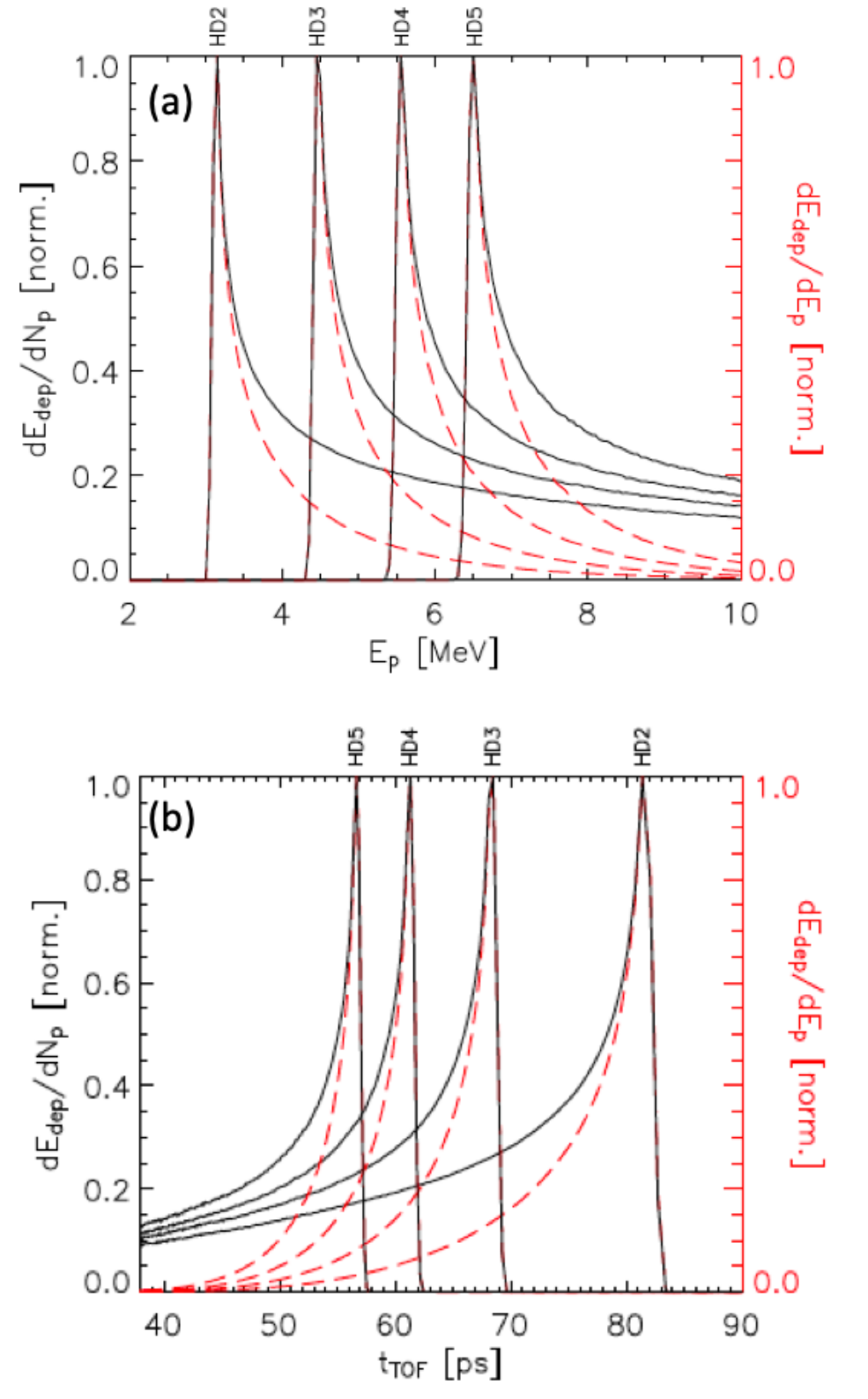}
\caption{(a) Normalized energy response curves for a RCF stack made of several layers of HD810 (black solid curve). (b) Normalized temporal response curve for the stack configuration in (a) and a source-plasma distance of 3~mm (black curve). The red dashed curves in both graphs are response curves multiplied by a typical TNSA exponential spectrum with temperature of 2~MeV. Adapted from \onlinecite{Romagnani_2005b}}.
\label{fig:multiframe}
\end{figure}

The broadband spectrum of TNSA sources allows sequential temporal frames to be recorded in consecutive layers of a RCF stack. When using high energy TNSA protons from a PW-class laser system, one can obtain up to several tens of temporally separated proton images of the interaction.  In the multi-frame approach, every layer is labelled temporally with the time-of-flight (calculated from the source to the center of the film pack) of the energy at which the relevant response curve is maximized (essentially the energy reaching the Bragg peak in the active layer of the RCF).  An example of energy and temporal response for four consecutive layers (2nd to 5th) in an RCF stack is shown in Fig.~\ref{fig:multiframe} \cite{Romagnani_2005b}, based on SRIM (Stopping and Range of Ions in Matter) \cite{Ziegler_2010} calculations. The temporal multi-frame capability is highlighted in Fig.~\ref{fig:multiframe}(b), and the energy and temporal resolution of the different layers can be obtained as FWHM values. 

\begin{figure*}
\includegraphics[width=15 cm]{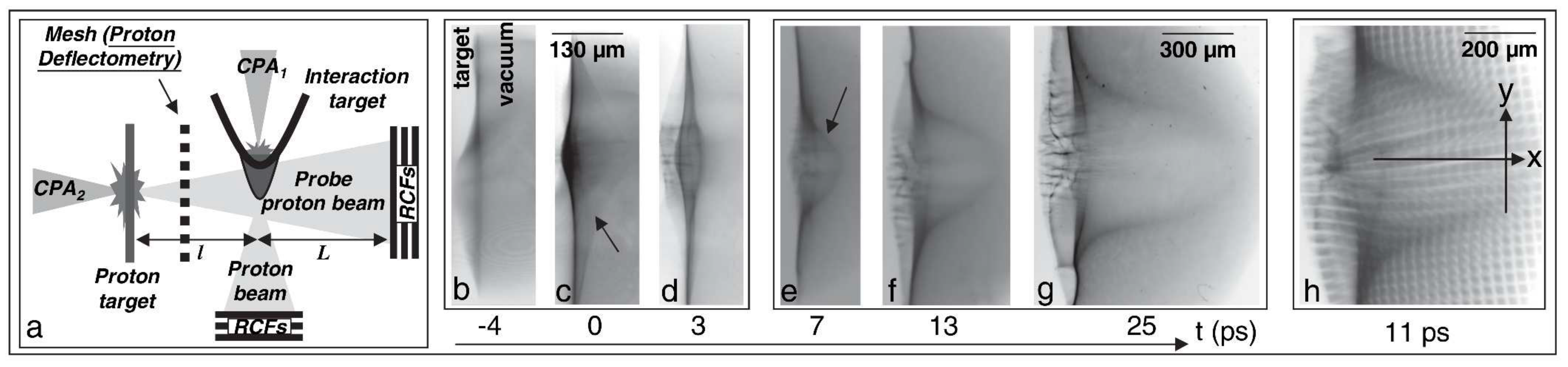}
\caption{Proton probing of the expanding sheath at the rear surface of a laser-irradiated target. (a) Setup of the experiment. A proton beam is used as a transverse probe of the sheath. (b)-(g) Temporal series of proton images in a time-of-flight arrangement. The probing times are relative to the peak of the interaction. (h) A deflectometry image where a mesh is placed between the probe and the sheath plasma for a quantitative measure of proton deflections. Adapted from \onlinecite{Romagnani_2005}.}
\label{fig:TNSA probing}
\end{figure*}

A suitable choice of parameters allows inter-frame time-steps of order ps to be obtained, as for example achieved in the data of Fig.~\ref{fig:TNSA probing} \cite{Romagnani_2005}.  Obtaining multiple snapshots also enables the temporal dynamics of the same event to be followed, which is particularly useful under conditions where there is a pronounced shot-to-shot variability. 

When probing ultrafast phenomena, it is often necessary to consider time-of-flight variations across a single RCF layer. These arise from the longer path of protons propagating obliquely and intercepting the RCF layer at an angle, compared to the protons propagating on axis, which can lead to temporal differences of order $\sim$~ps across the RCF layer.  This is important, for example, when imaging field structures moving at speeds close to $c$  across the field of view of the proton images \cite{Kar_2007, Ahmed_2016}.

\section{Theory of Proton-Imaging Analysis} \label{sec:theory}

\subsection{Basics} \label{sec:theory:basics}

As explained in the Introduction, the physics that underpins the proton-imaging diagnostic is quite simple. With the exception of interactions with dense HED plasma/matter, the characteristic speeds of imaging-beam protons are sufficiently large that collisional interactions between the beam protons and the plasma being probed are generally negligible~\cite{Kugland_2012,Bott_2017}. 
In addition, the characteristic density of proton-imaging beams is sufficiently low that the beam does not perturb the plasma via either collisionless plasma interactions or space-charge effects~\cite{Kugland_2012}. As a result, the protons that constitute typical imaging beams behave like test particles, being deflected by electromagnetic forces associated with fields already present in the plasma prior to the arrival of the proton beam. Thus, the proton beam's profile post interaction encodes information about the inherent electric and magnetic fields of the plasma.  
 
The trajectory of charged particles through electric and magnetic fields (and the final velocity of those particles post interaction) can be rather complicated in the general case of arbitrary proton speeds and characteristic field strengths; proton imaging setups typically overcome this issue by their use of fast multi-MeV protons (see Sec.~\ref{sec:exp:sources}) and careful geometric design to restrict the set of possible proton trajectories. For most laser-plasma experiments currently performed, the magnitude of deflection angles due to plasma-generated electromagnetic fields is small for multi-MeV protons, and thus the electromagnetic fields in the plasma are approximately sampled along the unperturbed, linear trajectories of the beam protons\footnote{Several subtle caveats exist to this statement; we discuss these subsequently.}. Therefore, for an incident proton with velocity $\tilde{\boldsymbol{v}}$ (and whose unperturbed trajectory has position vector $\tilde{\boldsymbol{x}}$), it can be shown by time-integrating the proton's equation of motion that the velocity perturbation $\Delta \boldsymbol{v}_{\perp}$ acquired in the directions perpendicular to $\tilde{\boldsymbol{v}}$ as the proton passes through a plasma containing an electric field $\boldsymbol{E}$ and a magnetic field $\boldsymbol{B}$ is  

\begin{equation}
\Delta \boldsymbol{v}_{\perp} \approx \frac{e}{m_p v_{0}} \int_0^{l_{\rm path}} \mathrm{d}s \, \left\{\boldsymbol{E}_{\perp}\!\left[\tilde{\boldsymbol{x}}(s)\right] + \frac{\tilde{\boldsymbol{v}} \times \boldsymbol{B}_{\perp}\!\left[\tilde{\boldsymbol{x}}(s)\right]}{c}\right\} 
\, , \label{deflection_vel_eqn}
\end{equation} 

\noindent where $e$ is the elementary charge, $m_p$ is the proton mass, $c$ is the speed of light, $l_{\rm path}$ is the distance covered by the proton as it traverses the plasma, $v_0 \equiv |\tilde{\boldsymbol{v}}|$ is the proton's 
initial speed, and $s$ is the path length. The deflection angle $\delta \alpha$ of each proton is $\delta \alpha \approx |\Delta \boldsymbol{v}_{\perp}|/v_0$. Because the unperturbed trajectories of beam protons are linear, angular deflections of the proton beam are thus directly relatable to line-integrated electromagnetic fields in the plasma (or, more specifically, to the components of the fields that are perpendicular to the proton beam's incident direction of motion).   
As a given proton is interacting with electromagnetic fields, it will also acquire a perpendicular displacement $\Delta \boldsymbol{x}_{\perp}$ in addition to a velocity displacement $\Delta \boldsymbol{v}_{\perp}$, which in principle complicates the interpretation of a non-uniform proton beam profile. However, by ensuring that the distance $r_{\rm d}$ from the plasma to the detector is much larger than $l_{\rm path}$ (a geometric setup of this form is known as \emph{point-projection geometry}), it follows that the measured displacement ${\Delta \boldsymbol{d}}_{\perp}$ 
of protons from their projected position $\boldsymbol{d}_{\perp 0}$ in the absence of any electromagnetic fields is dominated by the displacement acquired as protons free-stream at their 
(slightly) perturbed velocity: ${\Delta \boldsymbol{d}}_{\perp} \approx r_{\rm d} \Delta \boldsymbol{v}_{\perp}/v_0$, with $|{\Delta \boldsymbol{d}}_{\perp}| \approx r_{\rm d} \delta \alpha \gg |\Delta \boldsymbol{x}_{\perp}|$. 

Historically, this conclusion has been leveraged to discern properties of the electromagnetic fields in the plasma using a proton beam in two ways. Simplest of these is to introduce a well defined spatial modulation to the profile of the proton beam prior to its interaction with any electromagnetic fields using a grid (see Sec.~\ref{sec:exp:diag:mesh}): only protons that do not intersect the grid are subsequently detected.  
Any distortions ${\Delta \boldsymbol{d}}_{\rm g}$ to the grid-induced profile detected post interaction (which provide a direct measure of ${\Delta \boldsymbol{d}}_{\perp}$) can then be attributed to angular deflections caused by electromagnetic fields in the plasma, and the line-integrated values of two components of those fields estimated via

\begin{equation}
\int_0^{l_{\rm path}} \mathrm{d}s \, \left\{\boldsymbol{E}_{\perp}\!\left[\tilde{\boldsymbol{x}}(s)\right] + \frac{\tilde{\boldsymbol{v}} \times \boldsymbol{B}_{\perp}\!\left[\tilde{\boldsymbol{x}}(s)\right]}{c}\right\} 
\approx \frac{m_p v_0^2}{e r_{\rm d}} {\Delta \boldsymbol{d}}_{\rm g} \, 
. \label{path_int_EMField_eqn}
\end{equation} 

This technique is typically known as \emph{proton deflectometry}, and has been successfully used in a number of different laser-plasma experiments to provide measurements of electromagnetic fields (e.g., see \onlinecite{Romagnani_2005,Li_2007,Petrasso_2009,Willingale_2011a,Tubman_2021}). The main advantage of this approach is its conceptual simplicity. However, it does also have a few issues. Determining the exact projection of the initial profile in the absence of any deflections is not always a trivial matter, because confounding factors such as imperfect target fabrication can mean that a deflectometry grid's position is not always consistent from shot to shot. 
Blurring of the mesh due to the ablation of actual physical grids by strong X-ray radiation that inevitably arises during the course of laser-plasma experiments can also inhibit successful tracking of the grid's distortion~\cite{Johnson_2022,Malko_2022}. In some circumstances, the grid itself can become charged, resulting in apparently distorted grids when there is in fact no interaction of the proton beam with plasma electromagnetic fields~\cite{Palmer_2019}. The resolution of electromagnetic field measurements is also limited to that of the grid; this constraint is inevitably much larger than the theoretical resolution that can be achieved given typical proton source sizes (see Sec.~\ref{sec:exp:sources}). Finally, in cases of highly non-uniform deflections, successfully tracking the grid's distortion is not always possible~\cite{Willingale_2010a}. 

A second approach which attempts to overcome these issues is to assume approximate transverse uniformity of such beams prior to their interaction with a plasma being imaged -- a property of proton-beam sources which typical experimental setups aim to realize (see Sec.~\ref{sec:exp:diag}) -- and thereby quantitatively relate inhomogeneities in the beam profile detected post-interaction on a proton image to electromagnetic fields in the plasma~\cite{Kugland_2012,Graziani_2017,Bott_2017,Kasim_2017}. 
The successful interpretation of detected non-uniformities in proton images in terms of the electromagnetic fields associated with them using either of these approaches requires a theoretically grounded analysis methodology. Historically, there have been two methodologies that have been used for this interpretation: particle-tracing simulations and analytic modeling. We discuss both approaches in Secs.~\ref{sec:theory:forward} and \ref{sec:theory:inverse}, respectively.

\subsection{Particle-tracing simulations} \label{sec:theory:forward}

\subsubsection{Overview} \label{sec:theory:forward:overview}

Analyzing proton images using particle-tracing simulations is typically done as follows. A candidate model for an electromagnetic field structure in a particular laser-plasma experiment is proposed; the interaction of the proton beam (whose parameters are chosen to be the same as those used experimentally) with that field structure is simulated using a (test) particle-tracing code; a simulated proton image associated with that proton beam is then generated; finally, the simulated image is compared with the experimental one, with the candidate model deemed to be reasonable if there is qualitative -- or, ideally, quantitative -- agreement. Particle-tracing simulations provide a powerful approach for analyzing proton-imaging data, because they make relatively few assumptions about the nature of the interaction between the proton beam and the plasma being imaged. 

Arguably, the most important question which must be considered when using particle-tracing simulations to analyze proton images is how to construct an appropriate candidate model for the electromagnetic field. There are two approaches to addressing this question that have been used for analyzing data from previous laser-plasma experiments. The first is to use the electromagnetic fields generated by a high-energy-density-physics (HEDP) code of the relevant laser-plasma experiment. The second involves introducing a physically motivated parameterized model, and optimizing the model's parameters using an algorithmic best-fit procedure. Often, these approaches are used complementarily, with the output of a HEDP code serving as an inspiration for a simpler, parameterized model. The two approaches are discussed in Secs.~\ref{sec:theory:forward:combined} and \ref{sec:theory:forward:models}, respectively. 
Irrespective of the approach used to construct the candidate electromagnetic field model, the successful use of particle-tracing simulations relies upon efficient particle-tracing algorithms; we therefore discuss these algorithms first. 

\subsubsection{Particle-tracing algorithms} \label{sec:theory:forward:tracing}

\begin{figure*}
\includegraphics[width=6.5in]{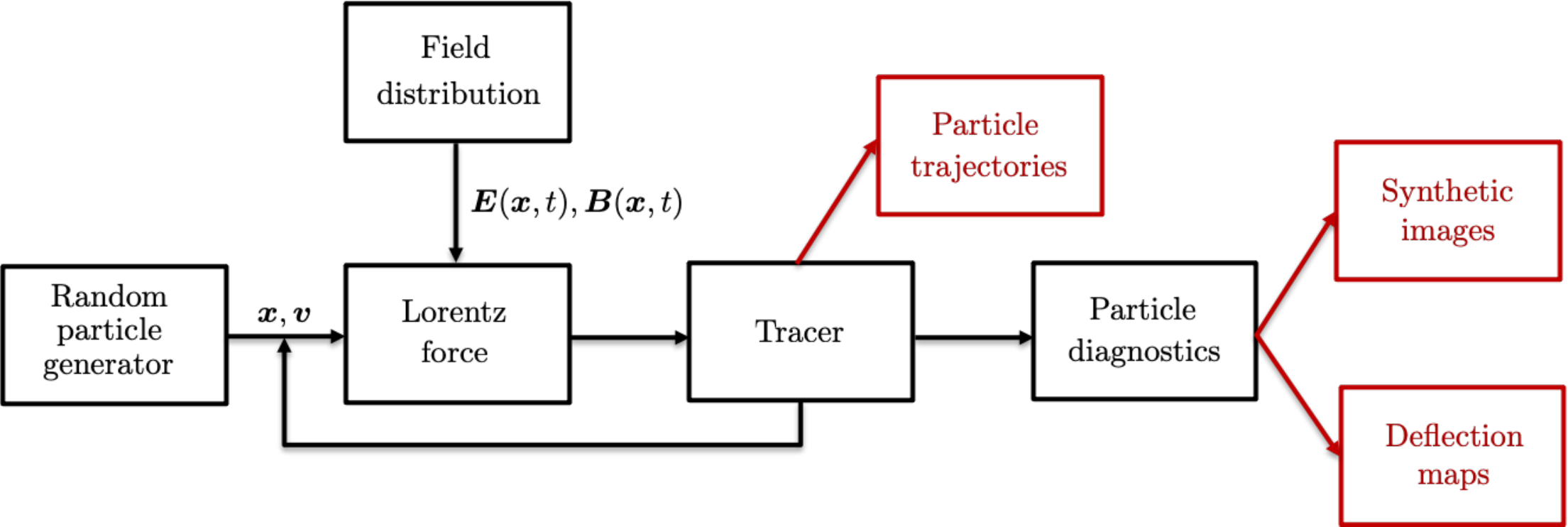}
\caption{Workflow for a typical particle-tracing algorithm. Adapted from \onlinecite{Romagnani_2005}.}
\label{fig:particlesim_codeworkflow}
\end{figure*}

The process underpinning a typical particle-tracing algorithm is illustrated in Fig.~\ref{fig:particlesim_codeworkflow}. 

Particle-tracing simulations typically employ a Monte Carlo method. First, synthetic protons are generated at the location of the proton source, and assigned a velocity which, aside from being constrained to have a pre-specified magnitude and an orientation with a cone of a certain solid angle, is random. These particles are then traced to the compact domain in which the (possibly time-dependent) electromagnetic fields are defined. In this domain, the non-relativistic equation of motion for protons under the action of the Lorentz force associated with the electromagnetic fields is numerically integrated along particle trajectories. This integration is implemented using efficient numerical schemes in typical particle-tracing simulations, in order that the simulations can be run quickly for millions of synthetic protons~\cite{Birdsall_1985,Welch_2004,Vay_2008}.

Once a given synthetic proton has completed its interaction with the electromagnetic field, the output can then be included in various particle diagnostics: most immediately, synthetic proton images, but also other outputs such as deflection maps. While the most basic particle-tracing codes typically assume a point source of monoenergetic protons created instantaneously with a smooth spatial profile, it is a simple matter to relax these assumptions, and to include a finite source size or emission time, use a pre-defined spectrum of proton energies, or incorporate realistic random departures from laminarity. It is also not too challenging to include some additional physics beyond the simple action of Lorentz forces: for example, multiple small-angle scattering or energy loss due to Coulomb collisions in dense plasmas~\cite{Lu_2020}. Particle-tracing simulations of proton beams have also been performed using full particle-in-cell (PIC) codes (e.g., see \onlinecite{Huntington_2015}) that are capable of including the beam's feedback on the electromagnetic fields being imaged via collisionless interaction mechanisms (although this is effect is usually not important).

\subsubsection{Combined modeling with HEDP codes} \label{sec:theory:forward:combined}

\begin{figure*}
\includegraphics[width=16cm]{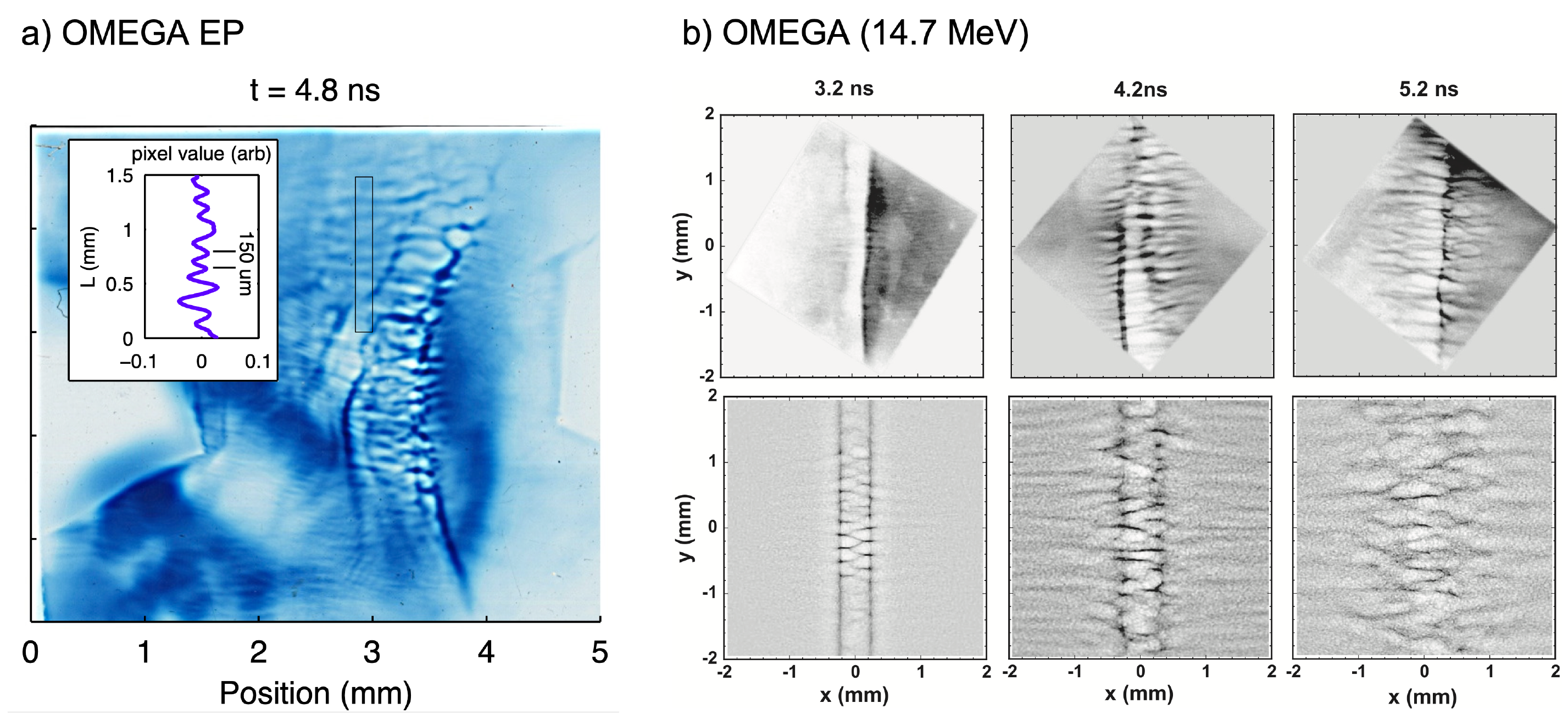}
\caption{Observations of filamentary magnetic field generation by ion-Weibel instability in counterstreaming plasmas.  The filamentation scale is at the order of the ion-skin depth.  (a) Observations at OMEGA EP using TNSA protons at approximately 5~MeV. Adapted from \onlinecite{Fox_2013}. (b) [Top Row] Observations at OMEGA using \DHe{} protons at 14.7~MeV. [Bottom Row] Synthetic proton images from 3D PIC simulations modeling the experiments from the top row. Adapted from \onlinecite{Park_2015}.}
\label{fig:Weibel_prad_fig}
\end{figure*}

Because of the complexity of the physics inherent in most laser-plasma experiments, as well as the difficulties involved in diagnosing such experiments, HEDP simulation codes are typically used to help design, implement, and interpret their results. Depending on the experiment, the state-of-the-art codes that are run at the present time are either magnetized fluid codes (e.g., LASNEX~\cite{Zimmerman_1977}, FLASH~\cite{Fryxell_2000, Tzeferacos_2015}, GORGON~\cite{Chittenden_2004}, RAGE~\cite{Gittings_2008}, HYDRA~\cite{Langer_2015}), particle-in-cell codes (e.g., OSIRIS~\cite{Fonseca_2002}, EPOCH~\cite{Arber_2015}, PSC~\cite{Germaschewski_2016}, Smilei~\cite{Derouillat_2018}, VPIC~\cite{Bird_2022}), or hybrid codes (e.g., ZEPHIROS~\cite{Kar_2009,Ramakrishna_2010}, Chicago~\cite{Thoma_2017}, dHybrid~\cite{Gargate_2007}), all of which output electromagnetic fields. Thus, choosing to use the outputs from such codes as inputs for candidate electromagnetic fields in particle-tracing simulations of proton images is a natural approach. For the outputs of such particle-tracing simulations to provide a plausible comparison with experimental data, the HEDP simulation should either be three dimensional or two dimensional with symmetry, with good spatial/temporal resolution over sufficiently large spatial/temporal scales. Aside from ease of implementation if HEDP simulations have already been completed, this approach can be particularly advantageous if complex electromagnetic field geometries arise (see Fig.~\ref{fig:Weibel_prad_fig} in Sec.~\ref{Sec:apps:Weibel} for an example); constructing parameterized electromagnetic field models from scratch in such situations is laborious. That being said, relying solely on synthetic images derived from HEDP simulations can become problematic if those images turn out to be qualitatively and/or quantitatively distinct from the experimental data they are meant to model; if this situation arises, it is often challenging to determine how to ``correct'' the outputs from HEDP simulations systematically.

\subsubsection{Parameterized field models} \label{sec:theory:forward:models}

Provided the morphology of experimentally observed proton-fluence inhomogeneities are not too complex, it is often the case that a simple parameterized analytical model for a candidate electromagnetic field -- motivated by considerations of the physical mechanism(s) responsible for generating that field -- can be constructed. The optimum choice of the parameters can then be found iteratively using particle-tracing simulations: given a first guess of parameters, a synthetic image is generated and then compared with the experimental image, with the quantitative differences between the outputs then used to determine a revised set of parameters, and so on~\cite{Romagnani_2005,Romagnani_2008}. This approach can prove to be helpful if 3D HEDP simulations of a given experiment have not been performed, or are producing outputs that are discrepant with experimental data. By construction, the technique will recover a good fit to the experimental data for simple proton-fluence inhomogeneities; however, for inhomomegeneities lacking symmetry, successfully devising an appropriate analytic model with only a few parameters becomes very difficult. Examples of this approach being applied to proton-imaging data are presented in Figs.~\ref{fig:Cecchetti_fig3} and \ref{fig:schaeffer_2019} in Sec.~\ref{sec:apps}. 

\subsection{Analytic modeling} \label{sec:theory:inverse}

\subsubsection{Overview} \label{sec:theory:inverse:overview}

The second methodology for interpreting proton images that has been utilized historically is analytic modeling: that is, relating the line-integrated values of electromagnetic fields to inhomogeneous distributions of the detected proton fluence analytically under a set of simplifying assumptions~\cite{Romagnani_2005, 
Kugland_2012}. While analytic relations of this type can be used to test particular candidate electromagnetic field models (analytic forward-modeling), they have proven to be particularly helpful in two key regards. First, they provide a direct interpretation of proton-fluence inhomogeneities in terms of either physical properties of the plasma (specifically, path-integrated charge and current structures) or features inherent in point-projection imaging (specifically, caustics); for both cases, see Sec.~\ref{sec:theory:inverse:physicalinterp}. Secondly, analytic theory has been used to show the conditions under which the determination of line-integrated electromagnetic field structures from proton-fluence inhomogeneities (which we refer to as field reconstruction) is a mathematically well-posed inversion problem, and if those conditions are met, how such field reconstruction can be carried out systematically.

An analytical theory of proton imaging is not really tractable unless simplifying assumptions about the imaging setup are made; these assumptions are outlined in Sec.~\ref{sec:theory:inverse:theory}, as is the theory that follows directly from them. Once the analytical theory has been established, we then explain in Sec.~\ref{sec:theory:inverse:physicalinterp} how that theory can be used for the direct interpretation of proton-fluence inhomogeneities. Finally, the possibility and implementation of field-reconstruction analysis is discussed in Sec.~\ref{sec:theory:inverse:alg}.

\subsubsection{Analytic theory of proton imaging} \label{sec:theory:inverse:theory}

In addition to the (justified) assumption that the imaging protons behave as test particles, most analytic theories of proton imaging make seven key assumptions: 

\begin{itemize}
\item \emph{Small-angle deflections}: $\delta \alpha \ll 1$. As discussed in Sec.~\ref{sec:theory:basics},
this assumption (generally) allows for the trajectories of beam protons to be treated as linear, and thus
for deflection angles to be linearly related to line-integrated electromagnetic fields (viz., Eqn.~(\ref{deflection_vel_eqn})). Using Eqn.~(\ref{path_int_EMField_eqn}), it can be shown that this condition is equivalent to assuming that the (transverse) path-integrated electric and/or magnetic field is much smaller than some critical value; specifically, $|\int_0^{l_{\rm path}} \mathrm{d}s \, \boldsymbol{E}_{\perp}|  \ll m_p v_0^2/e$, or $|\int_0^{l_{\rm path}} \mathrm{d}s \, \boldsymbol{B}_{\perp}|  \ll m_p c 
v_0/e$. Relative to, for example, 3.3~MeV protons (one of the two main types of fusion protons produced by D$^3$He capsules), these bounds are

\begin{eqnarray}
  \left|\int_0^{l_{\rm path}} \mathrm{d}s \, \boldsymbol{E}_{\perp}\right|  & \ll & 6.6  
  \left[\frac{W_0(\mathrm{MeV})}{3.3 \, \mathrm{MeV}}\right] \, \mathrm{MV}  , \\ 
  \left|\int_0^{l_{\rm path}} \mathrm{d}s \, \boldsymbol{B}_{\perp}\right|  
 & \ll & 0.26 \left[\frac{W_0(\mathrm{MeV})}{3.3 \, \mathrm{MeV}}\right]^{1/2}  \, \mathrm{MG \, cm} \, 
  ,
\end{eqnarray}

where $W_0$ is the initial energy the imaging protons. This implies that electric fields with strengths of $\sim$~MV/cm or magnetic fields of $\sim$~MG strengths permeating the full extent of a millimeter-scale plasma (a typical size for plasmas created during HED experiments) are required for the small-angle deflection assumption to become invalid. Though such large electric and magnetic fields are routinely realized, for example, during the interaction of medium-energy, high-intensity lasers with solid targets, generating them across such a volume has only been realized on the very highest-energy laser facilities such as the National Ignition Facility (e.g., see \onlinecite{Meinecke_2022}). 

\item \emph{Point-projection}: $l_{\rm path} \ll r_{\rm d}$. The importance of this assumption 
was also outlined in Sec.~\ref{sec:theory:basics}: it allows for proton displacements observed at the detector to be treated as being due to velocity perturbations (as opposed to spatial perturbations) acquired whilst interacting with the plasma's electromagnetic fields.  

\item \emph{Small source size}: $a \ll \ell_{\rm EM}$, where $\ell_{\rm EM}$ is the characteristic length scale of the electromagnetic field in the direction transverse to the trajectory of the proton beam. This assumption allows 
the proton beam source to be treated as a point source. 

\item \emph{Monoenergetic beam}: $\Delta v_0 \ll v_0$, where $\Delta v_0$ is the characteristic spread of proton speeds in the detected imaging beam. This assumption means that the deflection angles of any constituent protons of the imaging beam that pass along the same trajectory can be treated as being the same. 

\item \emph{Instantaneous transit \& short pulse}: $\delta{t_{\rm p}} \ll \tau_{\rm EM}$ and $\delta{t_{\rm p}} \sim l_{\rm path}/v_0 \ll \tau_{\rm EM}$, where $\delta{t_{\rm p}}$ is the characteristic duration of the proton beam, $\tau_{\rm EM}$ is the characteristic time scale over which the electromagnetic field evolves in the plasma, and $\delta{t_{\rm t}}$ is the transit time of the protons through the plasma. If both the transit time and pulse duration of the proton beam are short compared to $\tau_{\rm EM}$, then the electromagnetic field can be treated as electrostatic and/or magnetostatic. 

\item \emph{Paraxial approximation}: $\ell_{\rm EM} \ll 2 r_{\rm s}$. This approximation allows for the proton beam to be treated as a expanding planar `sheet' as it passes through the plasma. 
\end{itemize}

We note that particle-tracing simulations do not necessarily have to make any of these assumptions when generating artificial proton images; however, if these assumptions are not valid, the correct interpretation of proton images is much more challenging. More detailed discussions of these assumptions can be found elsewhere~\cite{Kugland_2012,Bott_2017}. 

Under these seven approximations, the effect of the electromagnetic fields on the proton beam can be modeled as a ``re-mapping'' of the beam's (two-dimensional) transverse profile prior to it reaching the detector: any proton with an initial perpendicular position ${\boldsymbol{x}}_{\rm \perp 0} \equiv \tilde{\boldsymbol{x}}_{\rm \perp}(0)$
that in the absence of any electromagnetic fields would arrive at the detector plane at the position $\boldsymbol{d}_{\perp 0} = \mathcal{M} {\boldsymbol{x}}_{\rm \perp 0}$ 
(where $\mathcal{M} \equiv (r_{\rm d}+r_{\rm s}+l_{\rm path})/r_{\rm s} \approx (r_{\rm s}+r_{\rm d})/r_{\rm s}$ is the magnification) instead arrives at the (re-mapped) position 

\begin{equation}
{\boldsymbol{d}}_{\perp}({\boldsymbol{x}}_{\rm \perp 0}) = \mathcal{M} {\boldsymbol{x}}_{\rm \perp 0} + {\Delta \boldsymbol{d}}_{\perp}({\boldsymbol{x}}_{\rm \perp 0}) \, , \label{plasma_detector_map}
\end{equation}
where
\begin{eqnarray}
{\Delta \boldsymbol{d}}_{\perp}({\boldsymbol{x}}_{\rm \perp 0}) & =  & \frac{e r_{\rm d}}{m_p v_0^2} \int_0^{l_{\rm path}} \mathrm{d}z \, \bigg\{\boldsymbol{E}_{\perp}\!\left[\boldsymbol{x}_{\perp 0} \left(1+\frac{z}{r_{\rm s}}\right)+z \hat{\boldsymbol{z}}\right] \nonumber \\
&+& \frac{\tilde{\boldsymbol{v}}}{c}\times \boldsymbol{B}_{\perp}\!\left[\boldsymbol{x}_{\perp 0} \left(1+\frac{z}{r_{\rm s}}\right)+z \hat{\boldsymbol{z}}\right]\bigg\} 
\, . \label{plasma_det_displacement} 
\end{eqnarray}

Here, $\hat{\boldsymbol{z}}$ is the unit vector normal to the plane of the detector, and $z$ the coordinate along that axis. Conservation of proton number within any (infinitesimal) surface element of the beam's transverse profile then implies that the distribution $\Psi(\boldsymbol{d}_{\perp})$ of protons measured by the detector at position $\boldsymbol{d}_{\perp}$ is related to the initial distribution $\tilde{\Psi}_0(\boldsymbol{x}_{\rm \perp 0})$ via 

\begin{equation}
\Psi[\boldsymbol{d}_{\perp}(\boldsymbol{x}_{\rm \perp 0})] = \sum_{\boldsymbol{x}_{\rm \perp 0}:\boldsymbol{d}_{\perp}=\boldsymbol{d}_{\perp}(\boldsymbol{x}_{\rm \perp 0})}\frac{\tilde{\Psi}_0(\boldsymbol{x}_{\rm \perp 0})}{\left|\det{\nabla_{\perp 0}\left[\boldsymbol{d}_{\perp}(\boldsymbol{x}_{\rm \perp 0})\right]}\right|} 
\, . \label{RKimagefluxrel}
\end{equation}

We note that the denominator of Eqn.~(\ref{RKimagefluxrel}) is the absolute value of the Jacobian determinant of the 
mapping defined by Eqn.~(\ref{plasma_detector_map}), and that the sum accounts for the fact that protons from multiple different initial positions can in principle contribute to the proton-fluence distribution at the same position on the detector if Eqn.~(\ref{plasma_detector_map}) describes a non-injective mapping. Equation (\ref{RKimagefluxrel}) is the key analytic relationship between inhomogeneities in the detected proton fluence and path-integrated electromagnetic fields. For both TNSA and D$^{3}$He proton sources, the initial fluence distribution $\tilde{\Psi}_0(\boldsymbol{x}_{\rm \perp 0})$ is to a good approximation uniform over small solid angles (see Sec.~\ref{sec:exp}); $\tilde{\Psi}_0(\boldsymbol{x}_{\rm \perp 0})$ is therefore often assumed to be uniform: $\tilde{\Psi}_0(\boldsymbol{x}_{\rm \perp 0}) \approx \mathcal{M}^2 \Psi_0$, where $\Psi_0$ is the mean detected proton fluence. 

Naively, the mapping Eqn.~(\ref{plasma_detector_map}) seems to depend on four path-integrated components of the electromagnetic field being imaged via the displacement term Eqn.~(\ref{plasma_det_displacement}). However, 
Eqn.~(\ref{plasma_det_displacement}) has a convenient mathematical property: it can be expressed as the gradient of a (two-dimensional) scalar potential that is a linear combination of path-integrated electromagnetic potentials. More specifically, it can be shown that~\cite{Kugland_2012,Bott_2017}

\begin{eqnarray}
&& {\Delta \boldsymbol{d}}_{\perp}({\boldsymbol{x}}_{\rm \perp 0}) \approx -\frac{e r_{\rm d}}{m_p v_0^2}  \nonumber \\
&& \qquad \times \Bigg( \nabla_{\perp 0} \int_0^{l_{\rm path}} \mathrm{d}z \, \bigg\{\phi\!\left[\boldsymbol{x}_{\perp 0} \left(1+\frac{z}{r_{\rm s}}\right)+z \hat{\boldsymbol{z}}\right] \nonumber \\
&&\qquad - \frac{v_0}{c} A_{\|}\!\left[\boldsymbol{x}_{\perp 0} \left(1+\frac{z}{r_{\rm s}}\right)+z \hat{\boldsymbol{z}}\right]\bigg\}  ,\quad \label{displacement_pot} 
\end{eqnarray}

\noindent where $\phi$ is the electromagnetic scalar potential, $\boldsymbol{A}$ is the electromagnetic vector potential, and $A_{\|}$ the component parallel to $\tilde{\boldsymbol{v}}$. 
We deduce that Eqn.~(\ref{plasma_detector_map}) can be written as
\begin{equation}
{\boldsymbol{d}_{\perp}({\boldsymbol{x}}_{\rm \perp 0}) \approx \nabla_{\perp 0} \psi(\boldsymbol{x}}_{\rm \perp 0}) \, , \label{plasma_detector_map_pot}  
\end{equation}
where
\begin{eqnarray}
 \psi(\boldsymbol{x}_{\rm \perp 0}) & \equiv & \frac{1}{2} \mathcal{M} \boldsymbol{x}_{\rm \perp 0}^2 + \varphi(\boldsymbol{x}_{\rm \perp 0}) \, ,  \label{pot_defs_a} \\
 \varphi(\boldsymbol{x}_{\rm \perp 0}) & \equiv &\frac{e r_{\rm d}}{m_p v_0^2} \int_0^{l_{\rm path}} \mathrm{d}z \, \bigg\{-\phi\!\left[\boldsymbol{x}_{\perp 0} \left(1+\frac{z}{r_{\rm s}}\right)+z \hat{\boldsymbol{z}}\right] \nonumber \\
& + &  \frac{v_0}{c} A_{\|}\!\left[\boldsymbol{x}_{\perp 0} \left(1+\frac{z}{r_{\rm s}}\right)+z \hat{\boldsymbol{z}}\right]\bigg\} . \quad \label{pot_defs_b}  
 \end{eqnarray}
 
Thus, provided the assumptions underpinning standard analytical theories of proton imaging are valid, detected proton-fluence inhomogeneities are a function of just two path-integrated scalar functions pertaining to the electromagnetic field: a property of vital importance for successfully realizing field reconstruction (see Sec.~\ref{sec:theory:inverse:alg}). 

\subsubsection{Analytic interpretations of proton-fluence inhomogeneities} \label{sec:theory:inverse:physicalinterp}

Using the relation Eqn.~(\ref{RKimagefluxrel}) between path-integrated electromagnetic fields and the distribution of proton fluence -- a relation which is in turn a function of the two-dimensional mapping Eqn.~(\ref{plasma_detector_map}) -- it becomes possible to construct a framework that systematically characterizes into a few different regimes all classes of proton-fluence inhomogeneities that can arise in images of arbitrary electromagnetic fields. The key dimensionless parameter that underpins this framework is the \emph{contrast parameter} $\mu$, which is defined by~\cite{Kugland_2012,Bott_2017}

\begin{equation}
    \mu \equiv \frac{r_{\rm d} \delta \alpha}{\mathcal{M} \ell_{\rm EM}} \, .
\end{equation}

Physically, this parameter quantifies the relative magnitude $\ell_{\rm EM}^{(\mathrm{d})} \equiv \mathcal{M} \ell_{\rm EM}$ of the electromagnetic structures being imaged (including magnification) and the displacements $\Delta d_{\perp} \equiv r_{\rm d} \delta \alpha$ of protons at the detector acquired due to their interaction with those electromagnetic structures [mathematically, $\mu$ quantifies the relative magnitude of the two terms in the mapping Eqn.~(\ref{plasma_detector_map}) when their gradient is taken in the denominator of the fraction present on the right-hand side of Eqn.~(\ref{RKimagefluxrel})]. Depending on the size of $\mu$, there are three regimes of qualitatively distinct nature for electromagnetic fields with a single characteristic scale\footnote{The characterization of multi-scale electromagnetic fields, or fields with sharp gradients, is more subtle; see~\onlinecite{Kugland_2012,Bott_2017}}:

\begin{enumerate}
\item \emph{Linear regime} ($\mu \ll 1$): in this regime, $\Delta d_{\perp} \ll \ell_{\rm EM}^{(\mathrm{d})}$, and so the characteristic scale of proton-fluence inhomogeneities is similar to that of the electromagnetic fields being imaged. As a result, the relationship between proton-fluence inhomogeneities and path-integrated electromagnetic fields becomes to a good approximation linear (hence the regime's name), with the characteristic size $\delta \Psi$ of those inhomogeneities being small compared with the mean proton fluence $\Psi_0$: $\delta \Psi/\Psi_0 \sim \mu \ll 1$. Indeed, in the linear regime, proton-fluence inhomogeneities have a simple physical interpretation in terms of path-integrated \emph{charge} and \emph{current} densities; for purely electrostatic fields, $\delta \Psi/\Psi_0 \propto -\int_0^{l_{\rm path}} \mathrm{d}s \, \rho$, where $\rho$ is the charge density in the plasma~\cite{Romagnani_2005}, while for purely magnetic fields, $\delta \Psi/\Psi_0 \propto -\int_0^{l_{\rm path}} \mathrm{d}s \, j_{\|}$, where $\boldsymbol{j}$ is the magnetohydrodynamic (MHD) current density~\cite{Graziani_2017}. 

\item \emph{Nonlinear injective regime} ($\mu \lesssim \mu_{\rm c} \sim 1$): in this regime, $\Delta d_{\perp} \lesssim \ell_{\rm EM}^{(\mathrm{d})}$, but with the additional constraint that $\mu$ is not larger than some critical value $\mu_c$ at which the proton beam self-intersects prior to reaching the detector on account of spatially inhomogeneous deflections [\emph{viz.}, the mapping Eqn.~(\ref{plasma_detector_map}) remains injective]. As a result of the comparatively large magnitude of $\Delta d_{\perp}$ compared with $\ell_{\rm EM}^{(\mathrm{d})}$, the characteristic scales of proton-fluence inhomogeneities are distorted away from those of the path-integrated electromagnetic fields -- inhomogeneities with $\delta \Psi > \Psi_0$ are focused, while those with $\delta \Psi < \Psi_0$ are defocused -- and the magnitude of proton-fluence inhomogeneities in this regime is typically comparable to the mean proton fluence ($\delta \Psi \sim \Psi_0$). The simple physical interpretation of proton-fluence inhomogeneities in terms of path-integrated charge and current structures is no longer quantitative in the nonlinear injective regime, but such relationships still hold qualitatively. We note that the value of $\mu_{\rm c}$ depends on the particular electromagnetic field structure being imaged, but is typically of order unity. 

\item \emph{Caustic regime} ($\mu \geq \mu_{\rm c}$): in this regime, $\Delta d_{\perp} \gtrsim \ell_{\rm EM}^{(\mathrm{d})}$, with spatial gradients being sufficiently large that the proton beam self-intersects prior to being detected. This self-intersection leads to the emergence of proton-fluence inhomogeneities known as \emph{caustics}. Caustics have a specific structure that is unrelated to the electromagnetic fields responsible for them: they attain very large magnitudes ($\delta \Psi \gg \Psi_0$) in isolated regions, and typically occur in pairs (see \onlinecite{Kugland_2012} for a detailed discussion of caustics). It follows that the interpretation of proton-fluence inhomogeneities in terms of path-integrated electromagnetic fields is more challenging in the presence of caustics than in their absence, though successful measurements of simple field structures in this circumstance have been made~\cite{Kugland_2012b,Kugland_2013,Morita_2016, Levesque_2021}. 
\end{enumerate}

Because $\mu$ is directly proportional to the deflection angle $\delta \alpha$, it is linear in the characteristic strength of the electromagnetic field being imaged. By contrast, $\mu$ is inversely related to the initial proton energy: for magnetic fields, $\mu \propto W_0^{-1/2}$, while for electric fields, $\mu \propto W_0^{-1}$. Thus, a given electromagnetic field structure can be in any of the contrast regimes, depending on its strength and the energy of protons being used to performing imaging. Varying the dimensional parameters that describe the imaging diagnostic setup (e.g., $r_{\rm s}, r_{\rm d}$) also affects the contrast regime.

\begin{figure}
\includegraphics[width=\linewidth]{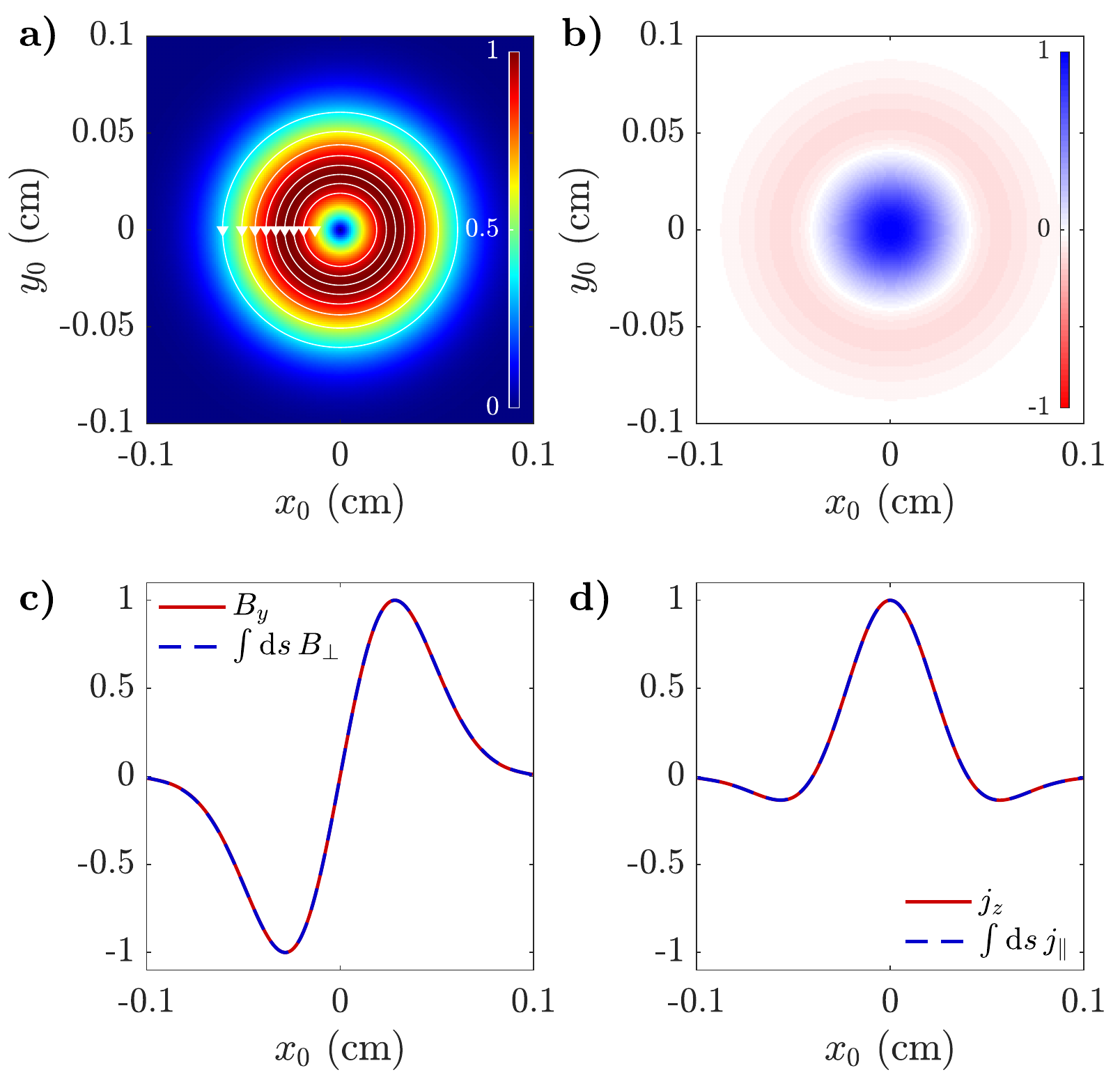}
\caption{Plots of an ``ellipsoidal blob'' magnetic field used to illustrate contrast regimes. $(x_0,y_0)$-slice plots of a) magnetic field strength $B = |\boldsymbol{B}|$ and b) MHD current density $j_z$ in the center of the ellipsoidal blob (at $z = z_0$), normalized by the maximum field strength $B_{\rm max}$ and maximum current density, respectively. On panel a), magnetic field lines and the field's orientation are shown in white. Here, $\ell_{\rm M\|} = \ell_{\rm M\perp} = 0.04 \, \mathrm{cm}$. On panels c) and d), normalized lineouts of $B_y$ and $j_z$ in $x_0$ (at $y_0 = 0$) are plotted with the values of $B_{\perp}$ and $j_{\|}$ line-integrated along the trajectories of protons that originate from a source at $(x,y,z) = (0,0,-r_{\rm s})$, and pass through positions $(x,y,z) = (x_0,y_0,0)$.}
\label{fig:anayticPRAD_cocoonfield}
\end{figure}

We illustrate the key features of the three contrast regimes with a simple numerical example. In this case, we compare the three regimes by choosing one particular field structure and then generating a sequence of (synthetic) proton images at increasing characteristic field strengths. We choose a ``ellipsoidal blob'' magnetic field~\cite{Kugland_2012} given by

\begin{equation}
   \boldsymbol{B} = \frac{B_{\rm max}}{\sqrt{2}} \frac{\boldsymbol{x}_{\perp 0} \times \hat{\boldsymbol{z}}}{\ell_{\rm M\perp}} \exp{\left[-\frac{|\boldsymbol{x}_{\perp 0}|^2}{\ell_{\rm M\perp}^2}-\frac{(z-z_{\rm c})^2}{\ell_{\rm M\|}^2}-\frac{1}{2}\right]} , 
\end{equation}

where $B_{\rm max}$ is the maximum strength of the field, $\ell_{\rm M\perp}$ its perpendicular scale length, $\ell_{\rm M\|}$ its parallel scale, and $z_{\rm c}$ the $z$-coordinate of the field's central point. The field is visualized in Fig.~\ref{fig:anayticPRAD_cocoonfield}a.

The spatial distribution of the $z$ component of the MHD current density, which is given by

\begin{eqnarray}
   j_z&  = & \frac{c B_{\rm max}}{8 \sqrt{2} \pi \ell_{\rm M\perp}} \left(1-\frac{|\boldsymbol{x}_{\perp 0}|^2}{\ell_{\rm M\perp}^2}\right) \nonumber \\
   && \quad \times \exp{\left[-\frac{|\boldsymbol{x}_{\perp 0}|^2}{\ell_{\rm M\perp}^2}-\frac{(z-z_{\rm c})^2}{\ell_{\rm M\|}^2}-\frac{1}{2}\right]} , 
\end{eqnarray}

is visualized in Fig.~\ref{fig:anayticPRAD_cocoonfield}b. Note that, for this particular choice, both the path-integrated magnetic field and the MHD current density have approximately the same perpendicular spatial structure as the three-dimensional field itself (see Figs.~\ref{fig:anayticPRAD_cocoonfield}c and \ref{fig:anayticPRAD_cocoonfield}d). 

\begin{figure}
\includegraphics[width=\linewidth]{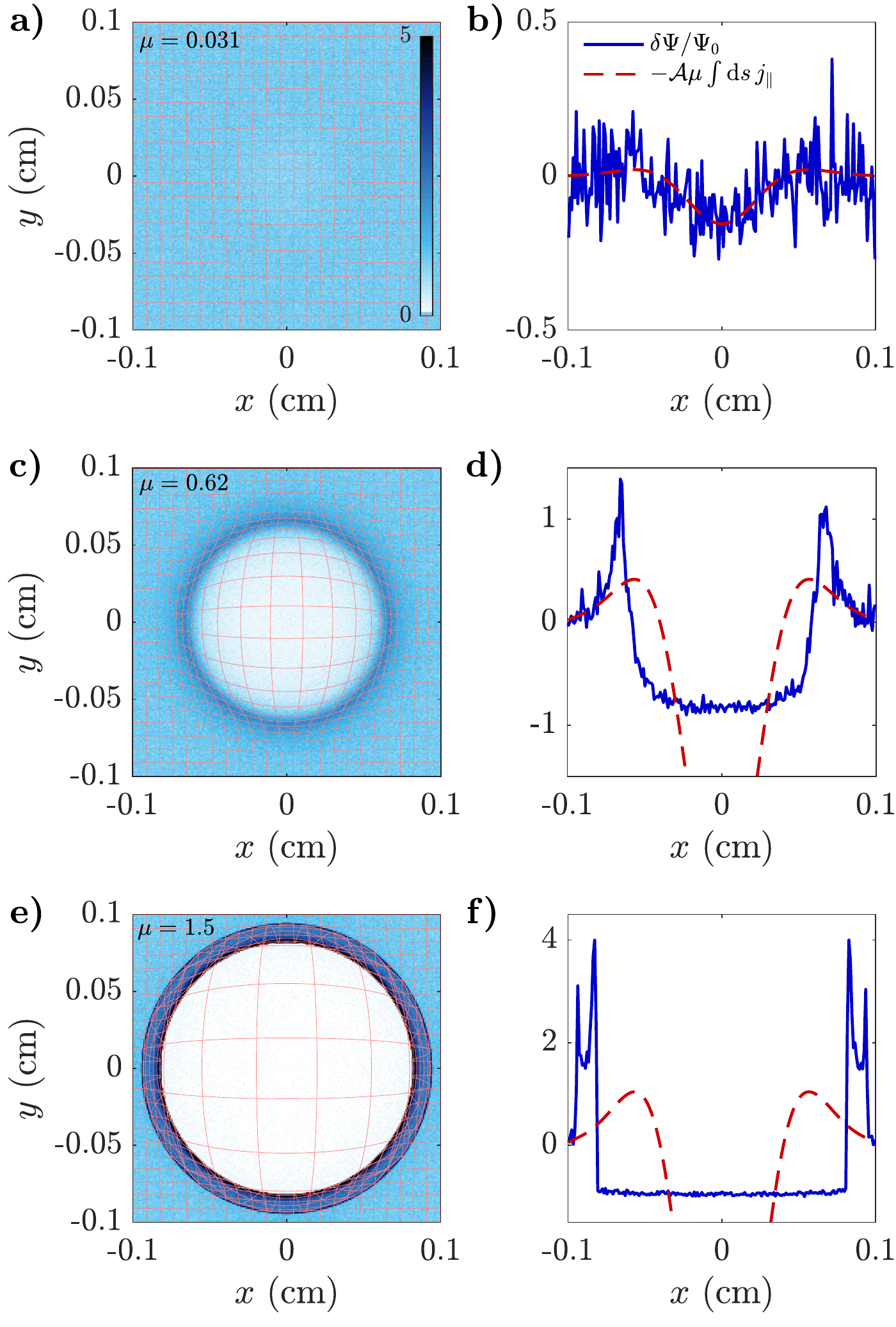}
\caption{Comparison of contrast-parameter ($\mu$) regimes for proton images of the ellipsoidal blob magnetic field described in Fig.~\ref{fig:anayticPRAD_cocoonfield}. Images in the a) linear, c) nonlinear injective, and e) caustic regimes are shown, as are proton-fluence lineouts along $x$ (at $y = 0$) in b), d), and f), respectively. Superimposed onto the images is the mapping $\boldsymbol{d}_{\perp} = \boldsymbol{d}_{\perp}(\boldsymbol{x}_{\perp 0})$ for each case (red lines). To generate the images, protons were simulated with $v_0 = 5.31 \times 10^{9} \, \mathrm{cm/s}$ (corresponding to 14.7~MeV protons), a setup with $r_{\rm s} = 1 \, \mathrm{cm}$, and $r_{\rm s} = 30 \, \mathrm{cm}$, and a field $B_{\rm max} = 10, 200, 500 \, \mathrm{kG}$ for the linear, nonlinear injective, and caustic regime images, respectively. The resolution of the images is $200 \times 200$~pixels, with a mean proton density per pixel of 100. Following previous conventions for the ellipsoidal blob~\cite{Kugland_2012}, $\mu$ is defined by $\mu = \sqrt{\pi} e r_{\rm s} B_{\rm max} \ell_{\rm M\|}/m_p c v_0 \mathcal{M} \ell_{\rm M\perp}$. Here, $\mathcal{A}$ is an order-unity constant of proportionality.}
\label{fig:anayticPRAD_contrastregime}
\end{figure}

Corresponding proton images of this magnetic field in the linear, nonlinear injective and caustic regimes are shown in Fig.~\ref{fig:anayticPRAD_contrastregime}.

In the linear regime, Figs.~\ref{fig:anayticPRAD_contrastregime}a and b demonstrate that the proton-fluence inhomogeneity $\delta \Psi$ is indeed small in magnitude compared to the mean fluence $\Psi_0$, with that inhomogeneity being approximately proportional to the MHD current. In the nonlinear injective regime, the central part of the ellipsoidal blob (which has $j_{z} > 0$) appears larger in the proton image than its true size (Fig.~\ref{fig:anayticPRAD_contrastregime}c), and the fluence and MHD current profile no longer agree quantitatively (Fig.~\ref{fig:anayticPRAD_contrastregime}d). Finally, in the caustic regime (see Figs.~\ref{fig:anayticPRAD_contrastregime}e,f), two high-amplitude caustic structures demarcate the edge of the ellipsoidal blob, whose structure does not resemble the true value of $j_z$.    

\subsubsection{Inverse analysis using EM field reconstruction algorithms} \label{sec:theory:inverse:alg}

In addition to providing a framework for the general interpretation of proton-fluence inhomogeneities in terms of the path-integrated fields creating them, further consideration of the relation Eqn.~(\ref{RKimagefluxrel}) reveals the conditions under which direct inversion of path-integrated electromagnetic fields from proton images is possible: that is, determining ${\Delta \boldsymbol{d}}_{\perp}({\boldsymbol{x}}_{\rm \perp 0})$ directly from  $\Psi(\boldsymbol{d}_{\perp})$. The key result of previous studies~\cite{Graziani_2017,Bott_2017} is that direct inversion of Eqn. (\ref{RKimagefluxrel}) from a single image is a well-posed mathematical problem provided that the mapping Eqn. (\ref{plasma_detector_map}) is injective or, equivalently, that there are no caustics present in the images. In terms of contrast regimes, inversion can be performed in either the linear or nonlinear injective regimes. This finding follows from the observation that relation Eqn. (\ref{RKimagefluxrel}) can be written as a Monge-Amp\'ere equation if the mapping Eqn.~(\ref{plasma_detector_map}) is injective: 

\begin{equation}
\Psi[\nabla_{\perp 0}\psi(\boldsymbol{x}_{\rm \perp 0})] = \frac{\tilde{\Psi}_0(\boldsymbol{x}_{\rm \perp 0})}{\det{\nabla_{\perp 0}\nabla_{\perp 0}\psi(\boldsymbol{x}_{\rm \perp 0})}} 
\, , \label{MongAmpEqn}
\end{equation}

\noindent where $\psi(\boldsymbol{x}_{\rm \perp 0})$ is the scalar function defined by Eqn.~(\ref{pot_defs_a}) in Sec.~\ref{sec:theory:inverse:theory}. In spite of their nonlinearity, Monge-Amp\'ere equations have unique solutions for $\nabla_{\perp 0}\psi(\boldsymbol{x}_{\rm \perp 0})$ [and thus $\boldsymbol{d}_{\perp}({\boldsymbol{x}}_{\rm \perp 0}) \approx \nabla_{\perp 0}\psi(\boldsymbol{x}_{\rm \perp 0})$] given appropriate (Neumann) boundary conditions. In the case of general electromagnetic fields, more information is needed to distinguish between path-integrated electrostatic and magnetic fields; but in the case where one dominates over the other, the path-integrated electrostatic or magnetic field in a plasma can be reconstructed. 

Various different ``field-reconstruction'' algorithms for recovering path-integrated electromagnetic fields directly from proton-fluence inhomogeneities have been proposed. In the linear regime, it can be shown that the inversion problem is just equivalent to solving a Poisson equation for the scalar function $\varphi(\boldsymbol{x}_{\rm \perp 0})$~\cite{Romagnani_2005,Kugland_2012}: $\nabla_{\perp 0}^2\psi(\boldsymbol{x}_{\rm \perp 0}) = -\mathcal{M} \delta \Psi/\Psi_0$. However, a later study by \onlinecite{Graziani_2017} found that inversion quickly fails for anything but very small values of $\mu$; the authors of that study proposed overcoming this by including first-order terms in the $\mu \ll 1$ expansion, solving the resulting equations with the PRALINE code~\cite{Graziani_2017}. In the $\mu \lesssim 1$ regime, descriptions of three different algorithms have been published: a Voronoi-diagram method to reconstruct path-integrated magnetic fields by \onlinecite{Kasim_2017}; the PROBLEM code by \onlinecite{Bott_2017}, which uses a nonlinear diffusion method proposed by \onlinecite{Sulman_2011} to solve the same problem; and finally, a trained neural network by \onlinecite{Chen_2017} (the authors of this study also trained their network to resolve 3D structure of ellipsoid blobs, though it is unlikely this approach is applicable to more general electromagnetic fields).
Others algorithms have been used to reconstruct electromagnetic fields in particular experiments (e.g., \onlinecite{Schaeffer_2019,Campbell_2020,Levesque_2022}), though full details of these codes have not yet been published. Comparing the outputs of these codes is currently an active research effort (e.g.,  see \onlinecite{Davies_2022}). 

By contrast, the possibility of performing direct inversion-analysis from a single proton image if caustics are present has been shown to not be a well-posed mathematical problem: multiple path-integrated electromagnetic-field ``solutions'' exist for a single proton-fluence distribution $\Psi$. In this situation, it can be proven that the solution to the Monge-Amp\`ere equation, Eqn.~(\ref{MongAmpEqn}), minimizes the functional

\begin{equation}
    \mathcal{C}\{{\Delta \boldsymbol{d}}_{\perp}\} = \int \mathrm{d}^2 \boldsymbol{x}_{\perp 0} \, \left|{\Delta \boldsymbol{d}}_{\perp}({\boldsymbol{x}}_{\rm \perp 0})\right|^2 \tilde{\Psi}_0(\boldsymbol{x}_{\rm \perp 0}) \, .
\end{equation}

In the case of a uniform initial proton-fluence distribution, the Monge-Amp\`ere solution therefore minimizes the root-mean-square of proton displacements over the space of all possible solutions. In practice, if $|\mu - \mu_{\rm c}| \ll 1$, the ``family'' of possible solutions associated with a particular distribution $\Psi$ is typically constrained to be quite similar to the Monge-Amp\`ere solution. So outputs of reconstruction algorithms are usually close to the ``true'' result for a point source of protons~\cite{Kasim_2017}, though this rule-of-thumb becomes much less robust if realistic proton-sources sizes are taken into account~\cite{Bott_2017}. If $|\mu - \mu_{\rm c}| \gtrsim 1$, then previous studies~\cite{Bott_2017} have shown that the Monge-Amp\`ere solution can return significant underestimates of characteristic path-integrated electromagnetic-field strengths compared to those of the ``true'' electromagnetic field. Systematically extracting information about path-integrated electromagnetic fields from proton images that contain caustics is therefore an outstanding research problem in proton-image analysis (although very recently, some progress has been made on this -- see Sec.~\ref{sec:frontiers:schemes}).     

Although field-reconstruction algorithms have been applied successfully to real experimental data~\cite{Tzeferacos_2018,Schaeffer_2019,Campbell_2020,Bott_2021,Bott_2021b,Levesque_2022}, these efforts have shown that several technical issues can arise in the process of performing such analysis. First of these is the finding that field-reconstruction algorithms are very sensitive to any large-scale variations in the initial proton-fluence profile, $\tilde{\Psi}_0(\boldsymbol{x}_{\perp 0})$. For example, it has been shown that two qualitatively different path-integrated magnetic fields can give the same proton image with only subtly different initial profiles~\cite{Bott_2017}. While both TNSA and \DHe3{} proton sources can produce beams whose transverse spatial inhomogeneities are small compared to the mean fluence on sub-mm plasma scales~\cite{Manuel_2012}, it has proven challenging to avoid significant inhomogeneities on larger scales. Studies aimed at overcoming this problem are ongoing, but possible remedies include high-pass filtering of either images or reconstructed path-integrated fields in order to isolate only those outputs for which uncertainties are not too large~\cite{Kasim_2017,Bott_2017}, applying constrained polynomial or Gaussian (as opposed to uniform) models for the initial profile~\cite{Palmer_2019,Fox_2020}, or using Bayesian inference conditioned on the (well characterized) properties of the initial proton beam inhomogeneities~\cite{Kasim_2019}. Another issue that is particularly important for fusion-capsule proton 
sources is the effect of a finite source size. It has been demonstrated~\cite{Bott_2017} that the source’s finite size reduces the characteristic value of $\mu_c$ below which field-reconstruction algorithms return accurate results compared with the case of a genuine point-source; in the referenced study, the Lucy-Richardson deconvolution algorithm was proposed as a way to mitigate this issue, but further study of more robust techniques is warranted.

\subsection{Comparing particle-tracing and analytic modeling techniques} \label{sec:theory:comparison}

In Secs.~\ref{sec:theory:forward} and \ref{sec:theory:inverse}, we have reviewed the use of particle-tracing simulations and analytic theory, respectively, for analyzing proton images; providing a comparative discussion of the two methodologies with respect to each other is therefore apt. The main advantage that particle-tracing simulations have over analytic modeling is the possibility of avoiding approximations which analytic modeling has to make in order to be tractable. These approximations involve the physics underpinning the interaction of the beam with the plasma, the precise properties of the proton beam's source, and the geometry of the imaging setup (see Sec.~\ref{sec:theory:inverse:theory}). Avoiding some of these approximations is vital for certain categories of laser-plasma experiments, such as those investigating ultrafast laser-plasma dynamics (see Sec.~\ref{PRADexp_ultrafast}). On the other hand, it is challenging for particle-tracing simulations to overcome one of the central challenges for all forward-modeling techniques -- the possibility of multiple qualitatively distinct solutions that are all consistent with the input data -- without recourse to analytically derived results (e.g., uniqueness). As well as this, field-reconstruction algorithms based on analytic modeling allow for images of complicated electromagnetic field structures to be analyzed in situations when HEDP simulations are either unavailable or are not able to reproduce the relevant physics correctly. All being said and done, we emphasize that either technique can be highly effective, but with the most robust analysis (usually) involving both.

\section{Proton Imaging Experiments} \label{sec:apps}


To demonstrate the variety of phenomena that can be investigated using proton imaging, we provide here a survey of the different types of experiments that have been performed using the diagnostic.  Examples include applications with spontaneously generated magnetic fields (Sec.~\ref{Sec:apps:Field_Generation}), magnetic reconnection (Sec.~\ref{Sec:apps:Reconnection}), Weibel instabilities (Sec.~\ref{Sec:apps:Weibel}), shocks (Sec.~\ref{Sec:apps:Shocks}), jets (Sec.~\ref{Sec:apps:Jets}), turbulence and dynamos (Sec.~\ref{Sec:apps:Dynamos}), ultrafast dynamics (Sec.~\ref{Sec:apps:Dynamics}), hydrodynamic instabilities (Sec.~\ref{Sec:apps:Hydro}), and ICF (Sec.~\ref{Sec:apps:ICF}).

\subsection{Magnetic Field Generation}
\label{Sec:apps:Field_Generation}

Magnetic fields can be spontaneously generated by several different mechanisms in initially unmagnetized plasmas, and proton imaging has been used to explore and characterize these processes in various laser-plasma experiments. Understanding these possible sources of magnetic fields is an important research area in HED plasma physics, because basic processes such as heat transport can be profoundly altered if magnetic fields become strong enough to magnetize the plasma's constituent particles (that is, reduce their Larmor radii below their respective Coulomb mean free paths). A detailed discussion of the many sources of magnetic fields in hot laser-produced plasma can be found in Haines~\cite{Haines_1986}; here, we focus on the most notable ones and their investigation using proton imaging.

One of the first mechanisms for generating magnetic fields in plasmas to be identified theoretically -- and also one of the first to be observed in experiments~\cite{Stamper_1971} -- is the Biermann battery, whereby magnetic fields are generated by misaligned electron density and pressure gradients~\cite{Biermann_1951}. Within the framework of extended MHD, the Biermann battery can be modeled as a source term in the induction equation: 

\begin{eqnarray}
    \frac{\partial \boldsymbol{B}}{\partial t}  & = & -\frac{\nabla n_e \times \nabla p_e}{e n_e^2} -\nabla \times \left(  \frac{\beta_{\parallel} \nabla T_e}{e} \right) 
    \nonumber \\
    &&  + \nabla \times (\boldsymbol{v_{B}} \times \boldsymbol{B})- \nabla\times \left( \eta \nabla \times \boldsymbol{B} \right). \quad  \label{XMHD_inductioneqn}
\end{eqnarray}
Here, the first source term on the right hand side is the Biermann battery (the second source term, which is often neglected in modeling, is associated with ionization). It can be shown that the Biermann battery term generates a field $\delta B$ in a time interval $\delta t$ of magnitude $\delta B \sim \delta t k_{\rm B} \nabla T_e \times \nabla n_e/e n_e$ (where $k_{\rm B}$ is Boltzmann's constant). Once generated, this Biermann field then evolves through advection at a characteristic bulk-flow velocity $\boldsymbol{v_{B}}$ (the third term of Eqn. (\ref{XMHD_inductioneqn})) and through  diffusion by the resistivity $\eta$ (the fourth term of Eqn. (\ref{XMHD_inductioneqn}))~\cite{Haines_1986}.  
Since non-parallel plasma temperature and density gradients are common in plasmas, magnetic field generation by the Biermann battery is ubiquitous in HED experiments and a frequent subject of proton imaging. For a laser pulse interacting with a solid target, the electron density gradient is primarily in the target normal direction, whereas the electron temperature gradient is primarily radial, meaning an azimuthal magnetic field is generated around the laser focal spot. The rate of field generation is therefore dependent on processes like the energy transfer from the laser to the plasma, and parameters such as the focal spot size and intensity profile. 

\begin{figure}
\includegraphics[width=7 cm]{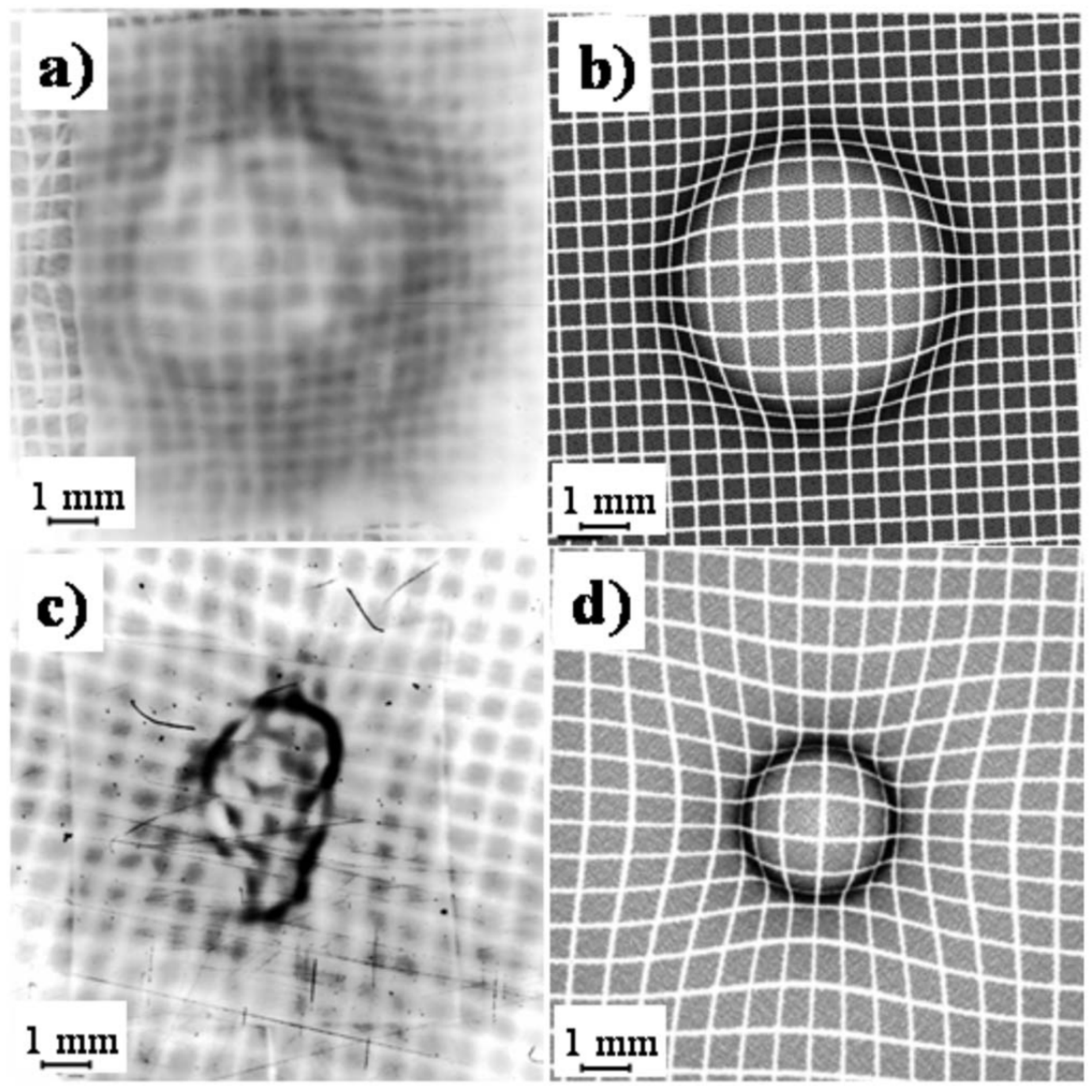}
\caption{Example experimental images of TNSA proton deflectometry are shown for (a) front-projection and (c) rear-projection geometries. Corresponding synthetic proton images were created with the particle-tracing code, PTRACE, using an idealized magnetic field torus in (b) front and (d) rear geometries, respectively. Adapted from \onlinecite{Cecchetti_2009}.}
\label{fig:Cecchetti_fig3}
\end{figure}

With the advent of proton imaging, a number of experiments have studied the generation of magnetic fields near the surface of laser-driven targets by the Biermann battery~\cite{Nilson_2006, Li_2006b, Petrasso_2009, Cecchetti_2009, Willingale_2010a, Willingale_2011a, Lancia_2014, Gao_2012, Gao_2013, Gao_2015, Campbell_2020}. The first measurements used grid deflectometry, measuring the deflections of a known periodicity mesh to infer the path integrated magnetic fields (see Sec.~\ref{sec:exp:diag:mesh}). The proton beam deflections are affected by the direction of the projection of the protons. For a ``front'' projection geometry, where the protons travel from the interaction side of the main target to the rear, the $\mathbf{v} \times \mathbf{B}$ Lorentz force primarily deflects the proton beam radially inwards. A ``rear'' projection geometry, where the protons first pass through the target before observing the front side magnetic fields, produces an outward deflection. This is illustrated in the work of Cecchetti \textit{et al.}; Fig.~\ref{fig:Cecchetti_fig3} presents experimental data using a TNSA source a) in front-projection and c) in rear-projection geometries. Figure \ref{fig:Cecchetti_fig3} b) and d) are the particle-tracking calculations for an idealized magnetic torus in front- and rear-projection geometries, respectively. Similar data using a \DHe~source is presented by \onlinecite{Petrasso_2009}. While the proton images naively make the extent and magnitude of the fields appear different for the two geometries, comparisons to analytical field maps show that the strength and scale of the fields are in fact similar.

\begin{figure}
\includegraphics[width=8 cm]{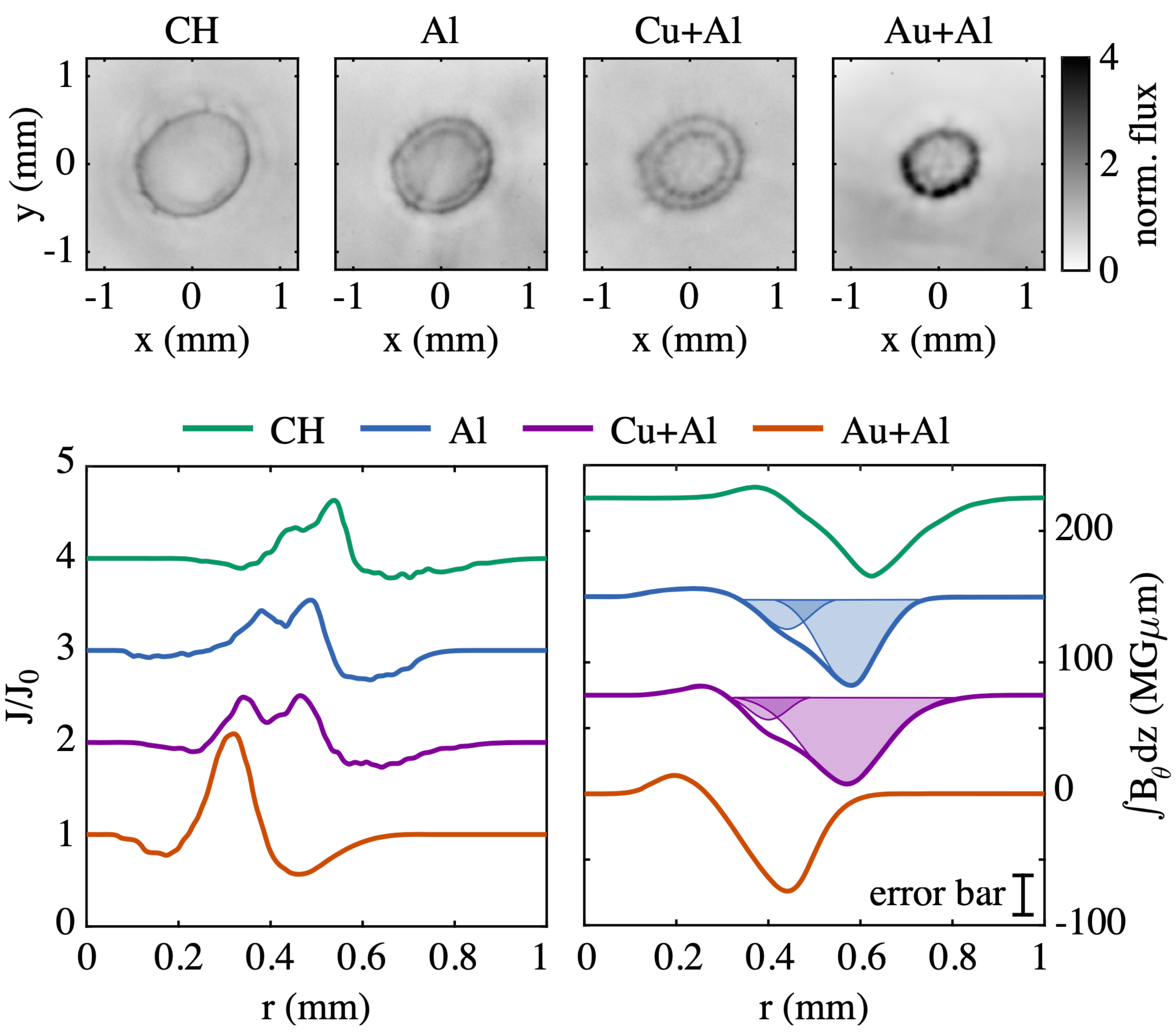}
\caption{Top row: proton images of the fields generated from different target materials at a time of 0.75~ns into a 1~ns interaction. The proton energy is 37.3~MeV for CH, Al, and Cu~$+$~Al, and 30.7~MeV for Au~$+$~Al.
Bottom left: radial lineouts of the proton fluence ($J$), normalized by the mean inferred reference profile ($J_0$). Bottom right: the resulting reconstructed field profiles. For Al and Cu~$+$~Al, the results of double-Gaussian fitting are shown with shaded regions. Adapted from \onlinecite{Campbell_2020}.}
\label{fig:Campbell_fig}
\end{figure}

Biermann-battery generated fields can be up to a MG or more and evolve on ns timescales. These measurements are to within order of magnitude, but not necessarily in exact agreement with, simulation predictions \cite{Li_2006b}. Numerical modeling typically consists of MHD simulations that include a Biermann battery source term, resistive magnetic diffusion and fluid advection [cf. Eqn.~(\ref{XMHD_inductioneqn})], and often Nernst advection, Righi-Leduc heat flow, and radiation. Measurements have confirmed that, once generated, magnetic fields can indeed be advected by the bulk plasma motion, i.e.\ at the ion fluid velocity, or the hot electron flux can transport the magnetic field at a faster speed through the Nernst effect \cite{Nishiguchi_1984, Willingale_2010a} and other effects \cite{Lancia_2014, Gao_2015}. Proton imaging experiments by Campbell \textit{et al.} have shown that varying the target material alters the field generation (see Fig.~\ref{fig:Campbell_fig}) and even the development of a double ablation front for mid-Z materials \cite{Campbell_2020}. Careful analysis of the field measurements to quantify total magnetic flux show that kinetic effects can suppress Biermann battery field generation in laser-plasma interactions \cite{Campbell_2022}.

In 1992, Wilks \textit{et al.} proposed that magnetic fields much stronger than those generated by the Biermann battery can be created by relativistic laser interactions ($>10^{18}$~W~cm$^{-2}$) due to currents produced by supra-thermal electrons accelerated in the evanescent region of the laser wave, which propagate deep into the interior of the plasma \cite{Wilks_1992}. This magnetic field is in the azimuthal direction about the laser axis of propagation, and the peak field extends for about an anomalous skin depth into the plasma (i.e., $d = [(c/\omega_{pe})(v_{te}/\omega_{0})]^{\frac{1}{2}}$ where $v_{te}$ is the electron thermal velocity). \onlinecite{Mason_1998} predicted the generation of fields up to 250~MG in the overdense plasma for moderately relativistic interactions.  Measurements of these short-pulse, relativistic intensity generated magnetic fields have been measured using TNSA protons by \onlinecite{Sarri_2012}. One significant difference compared to the lower intensity measurements is that it is expected that large fields will be present on both the front and rear sides of the target. Hot electrons rapidly move through the target to expand into the vacuum at the front and rear, creating time-varying sheath fields that generate opposing magnetic fields on the front and rear target surfaces. This means on one side of the target the proton beam is deflected radially inwards while on the other side it is deflected outward from the interaction region, significantly complicating the analysis and interpretation of the proton data.

A different method for creating magnetic fields with laser-plasma interactions is through laser-driven coils \cite{Gao_2016,Peebles_2022}.  In this approach, a laser is used to heat and eject electrons from a plate so that a current is drawn through a loop connected to the plate.  The interaction region within the loop, a volume of the order 1~mm$^3$, contains a strong axial magnetic field that can be used as an externally applied field for other experiments. Peebles \textit{et al.} measured axial fields of up to $65 \pm 15$~T \cite{Peebles_2020}.

\subsection{Magnetic Reconnection}
\label{Sec:apps:Reconnection}

Magnetic reconnection \cite{Yamada_2010} is a physical process whereby the magnetic field topology is rearranged, dissipating magnetic energy in a plasma into kinetic energy. It is a prevalent phenomena throughout the universe, occurring under many different conditions: for example, within the solar corona, where it leads to solar flares and coronal mass ejections \cite{Parker_1957}, between the solar wind and the Earth's magnetosphere, and during fusion plasma instabilities \cite{Taylor_1986}. Breaking and reconnecting magnetic field lines at observed rates require dissipation mechanisms to function at rates greater than allowed by classical resistivity \cite{Yamada_2010}. Consequently, there are many open questions to be investigated, including the temporal and spatial scales of the reconnection, the role of dynamical processes like plasmoid formation, and the final energy partition of the system. Furthermore, there are a wide range of reconnection regimes to explore due to how the magnetization, collisionality, and symmetry of the system affects the mediation of the reconnection process.

\begin{figure}
\includegraphics[width=8 cm]{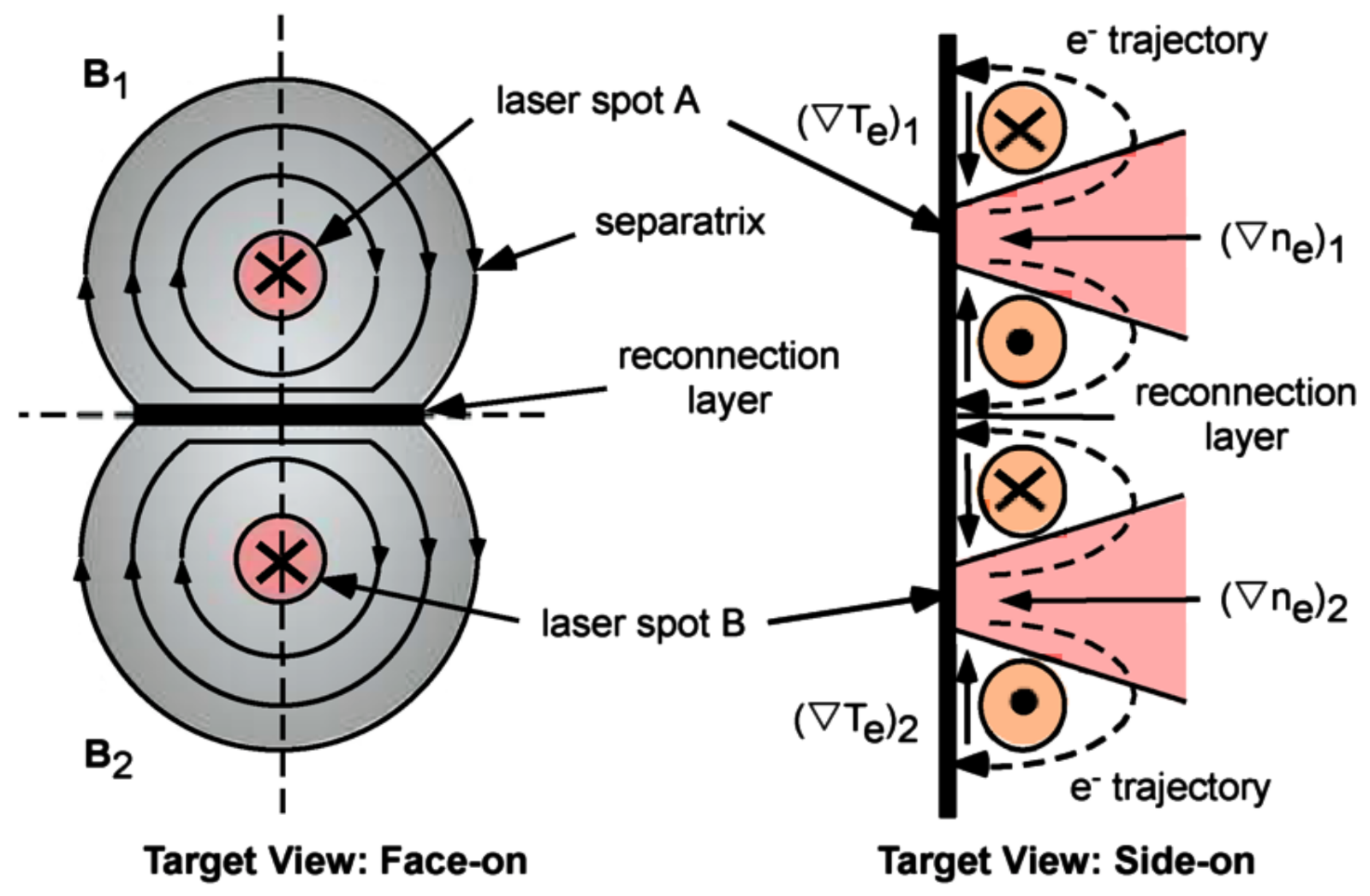}
\caption{Basic laser-driven magnetic reconnection geometry. The fields can be probed at different times to observe the dynamics. Adapted from \onlinecite{Nilson_2006}.}
\label{fig:recon_geometry}
\end{figure}

Laser-driven magnetic reconnection is a convenient way to study impulsive, strongly driven reconnection physics in the laboratory. Using proton imaging to diagnose the magnetic fields, the first experimental demonstration using lasers was performed by \onlinecite{Nilson_2006}, with the basic experimental configuration shown in Fig.~\ref{fig:recon_geometry}. These experiments used two neighboring high-energy, ns duration laser pulses to produce self-generated azimuthal magnetic fields through the Biermann-battery mechanism \cite{Nilson_2006, Nilson_2008, Willingale_2010b, Li_2007, Zhong_2010, Fox_2011, Fox_2012, Dong_2012, Rosenberg_2012} (see Sec.~\ref{Sec:apps:Field_Generation}). The magnetic fields were then advected out either by the frozen-in-flow or by heat transport via the Nernst effect, leading the opposing magnetic fields to be driven together in the mid-plane between the two focal spots. A key feature of such experiments is that the so-called plasma $\beta$ -- defined as the ratio of thermal to magnetic pressure -- is typically large. 

Numerous high-quality proton-imaging measurements have been made of magnetic fields in reconnection laser-plasma experiments of this type. For example, \onlinecite{Li_2007} and \onlinecite{Willingale_2011a} observed the rearrangement of the magnetic field's topology using proton imaging (as well as elevated plasma temperatures in the mid-plane region using Thomson scattering, and plasma jets emanating from the reconnection plane using optical probing). Experiments by Rosenberg \textit{et al.} used proton imaging to observe the slowing of the reconnection rate as the plasma inflows weaken \cite{Rosenberg_2015b}, and investigated the effect of asymmetric field structures \cite{Rosenberg_2015a}. Experimental measurements using proton deflectometry by Tubman \textit{et al.} showed anomalously fast reconnection in weakly collisional colliding laser plasmas \cite{Tubman_2021}.

These measurements (and also concurrent measurements from other complimentary diagnostics of the plasma conditions) have prompted new theoretical and numerical modeling studies of the high-$\beta$ reconnection regime, in turn helping to advance our understanding of reconnection processes more generally. For example, Fox \textit{et al.} performed numerical modeling of laser-driven experiments using particle-in-cell simulations (both with and without a collision operator) and noted the importance of the flux pile up to the reconnection process \cite{Fox_2011, Fox_2012}. Joglekar \textit{et al.} used a fully implicit 2D Vlasov-Fokker-Planck code to show that in high-$\beta$ laser generated plasmas, heat flows rather than Alfv\'{e}nic flows dictate the reconnection rate \cite{Joglekar_2014}. Supporting modeling identified the role of anisotropic electron pressure in explaining the enhanced reconnection rate seen in the experiments reported by \onlinecite{Tubman_2021}.

\begin{figure}
\includegraphics[width=8 cm]{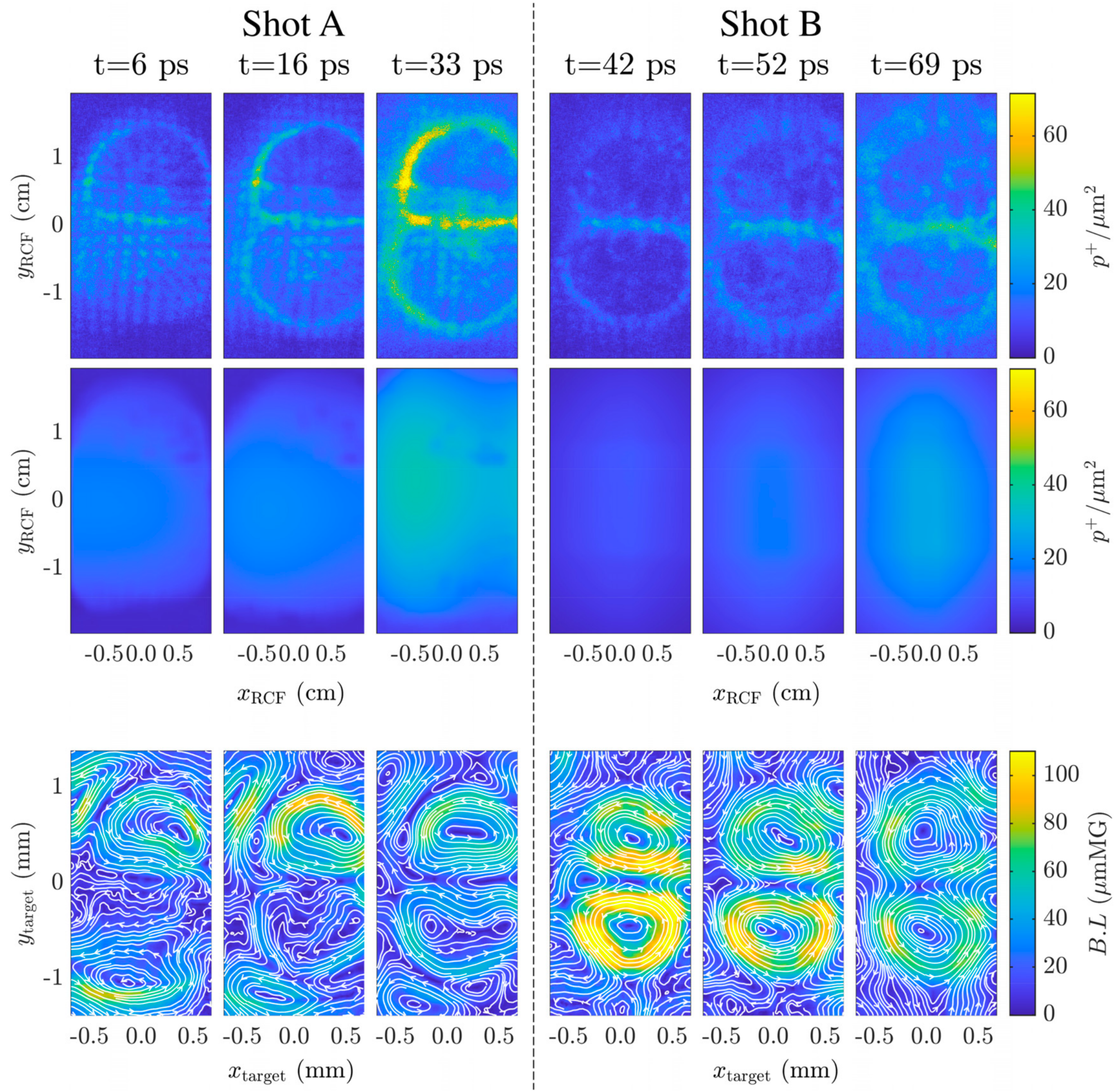}
\caption{Time series from a high-intensity laser-plasma-driven reconnection experiment. The top row shows the measured proton fluence at the detector plane, and the middle row shows the calculated undisturbed beam fluence at the detector plane. The bottom row presents the retrieved path-integrated magnetic fields at the interaction plane. The white contours with arrows show the topology of the calculated magnetic fields. Adapted from \onlinecite{Palmer_2019}.}
\label{fig:Palmer_recon}
\end{figure}

Proton imaging has also been used successfully to diagnose magnetic fields in other types of laser-driven reconnection experiments. For example, Palmer \textit{et al.} explored reconnection of fields generated by higher (${\sim}10^{18} \, \mathrm{W} \; \mathrm{cm}^{-2}$) laser intensities through proton deflectometry measurements \cite{Palmer_2019}. Figure~\ref{fig:Palmer_recon} illustrates a time history of data taken along with the 2D magnetic field maps reconstructed from proton images using a field-reconstruction algorithm (see Sec.~\ref{sec:theory:inverse}).
These maps showed faster dissipation of magnetic fields at the mid-plane compared to the outer plane, confirming that reconnection was taking place in the experiment on a timescale of tens of ps\footnote{We note that the large magnification used for this experiment meant that the interaction image extended close to the edge of the proton beam, necessitating detailed modeling of the assumed unperturbed proton fluence; by reducing the magnification so that the unperturbed region around the interaction is larger, it becomes easier to infer the unperturbed proton fluence and so reduces the potential error of the reconstruction method.}. 
 
Alternative laser-driven reconnection geometries have also been developed and studied using proton imaging. Fiksel \textit{et al.} employed externally applied opposing magnetic fields driven together by expanding laser plumes \cite{Fiksel_2014}, and  Chien \textit{et al.} used laser-interactions to drive currents through U-shaped coils configured in a reconnection geometry \cite{Chien_2019}.

\subsection{Weibel Instabilities}
\label{Sec:apps:Weibel}

Weibel-type filamentation instabilities \cite{Weibel_1959, Fried_1959, Davidson_1972} are ubiquitous in laboratory and astrophysical plasmas. They arise in plasmas whose particle distribution functions have significant velocity-space anisotropy.  The velocity space anisotropy includes cases where the temperature $T_{j} = \langle mv_j^2/2 \rangle$ differs among the three directions, where $j$ is one of $(x,y,z)$, and can be driven by counterstreaming particle beams which produce an effective anisotropy.  The counterstreaming between a ``hot'' forward particle population, balanced by a ``cold'' return current, which arises in situations of large heat flux, is another source of anisotropy important for electron-driven Weibel. The instability grows predominantly with wavenumber $\mathbf{k}$ aligned along the ``cold'' direction(s). The instability can play a broad role in plasmas, including magnetic field generation in the early universe and magnetic field generation and amplification in high-Mach number shocks.

The fundamental Weibel mechanism is that the large counterstreaming currents along the ``hot'' direction tend to pinch and coalesce into current-carrying filaments, and the forces driving coalescence are sufficient to overcome the transverse plasma pressure along the ``cold'' directions. Transverse magnetic fields associated with the 
current filaments then deflect the particle trajectories, reinforcing the filamentation and leading to a positive feedback. The non-linear regime includes rich physics such as the kinking and re-merging of magnetized flux tubes.

Ion-driven Weibel instabilities are important in astrophysical plasmas, as the large bulk-flow energy-density $M_i n_i V_i^2/2$ of ions can be greater than analogous energy-densities of the electron population and can therefore be a larger reservoir of free-energy for the Weibel process, producing stronger magnetic fields at larger scales. The ion-Weibel instability was identified in laboratory laser-driven experiments using proton imaging \cite{Fox_2013, Huntington_2015, Park_2015}. In the experiments, two plasma plumes were ablated from opposing targets and collided. The high temperature and low density of the ablation flows sets up counterstreaming ion populations in the interaction region. Proton imaging directly imaged the magnetized filaments produced in this interaction region by the ion-Weibel instability. Fox~\textit{et al.} observed the time-history of the development of the Weibel process and showed that the fast growth and typical filament wavelengths were consistent with the ion-Weibel dispersion relation. Huntington \textit{et al.} and Park \textit{et al.} measured statistics of the observed filamentary structures, which compared favorably to non-linear kinetic simulations.

The electron-Weibel instability is important in relativistic plasmas with strong beam currents, and is an important energy-coupling process that can lead to anomalous stopping of relativistic electron beams driven by short-pulse laser-plasma interactions. This type of instability was observed in face-on and side-on proton probing experiments \cite{Borghesi_2002b,Quinn_2012,Ruyer_2020}. Filament-like magnetic-field structures were observed to persist for an extended time period, a finding that could explain sustained, spatially elongated structures observed in various astrophysical environments.

\subsection{Shocks}
\label{Sec:apps:Shocks}

Proton imaging has been instrumental in studying the field structures of shocks in laboratory astrophysics experiments.  These shocks act to dissipate kinetic ram pressure in systems with supersonic flows, and are commonly found in heliospheric and astrophysical systems, including planetary bow shocks, jets, supernova remnants, and galaxy clusters, and are often associated with extremely energetic particles.  A key component of shocks are their strong electromagnetic fields.  In magnetized shocks, which propagate through a pre-existing magnetic field, the global structure of the shock is defined by a jump in the magnetic field on ion kinetic scales, while strong electric fields in the shock layer can help mediate dissipation by reflecting incoming ions.  Similarly, in electrostatic shocks, the shock layer is defined by electric fields on electron kinetic scales.  Meanwhile, in electromagnetic shocks, initially unmagnetized counter-propagating plasmas can spontaneously generate magnetic fields through streaming instabilities (e.g., Weibel), leading to shock formation.

\begin{figure*}
\includegraphics[width=17 cm]{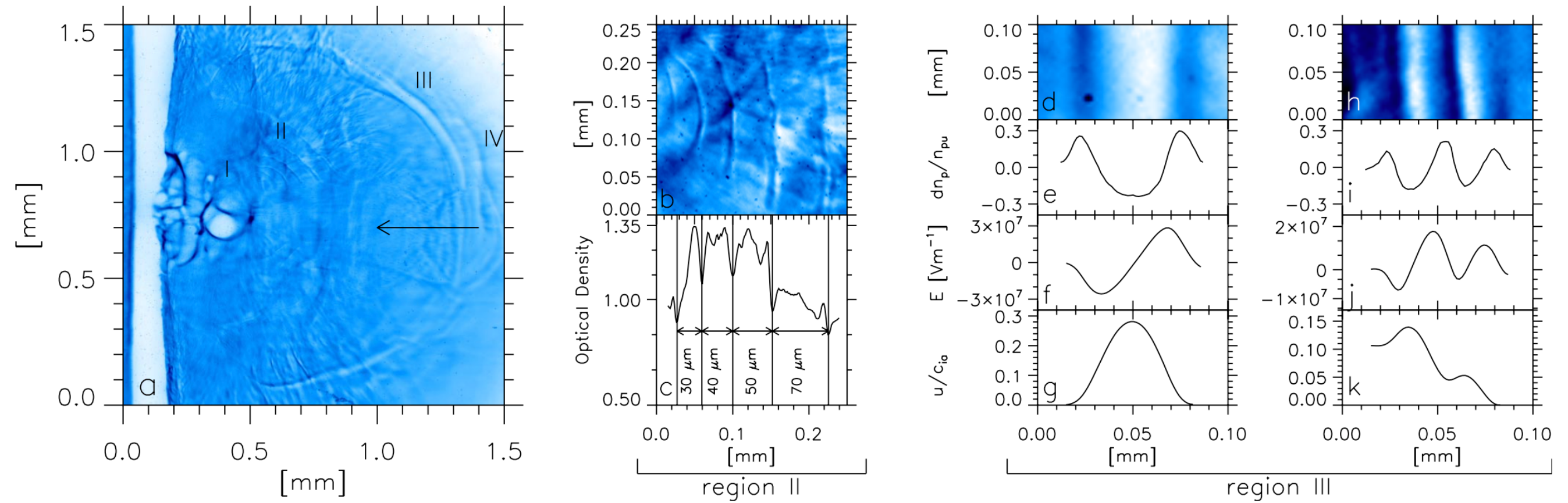}
\caption{Example data from electrostatic shock experiments. (a) Typical proton imaging data taken at the peak of the interaction pulse with 7~MeV TNSA protons.  Note the strong modulation associated with the ablating plasma in region I and the modulated pattern ahead of the shock front possibly associated with a reflected ion bunch in region IV. The arrow indicates the laser beam direction. (b)–(c) Detail and RCF optical density lineout corresponding to region II, showing modulations associated with a train of solitons. (d)–(k) Details of region III and corresponding lineouts of the probe proton density $\delta n_p/n_{pu}$, reconstructed electric field $E$, and reconstructed normalized ion velocity $u_/c_{ia}$ in the case of an ion acoustic soliton (d)–(g) and of a collisionless shock wave (h)–(k) (the collisionless shock detail corresponds to a different shot not shown here). Adapted from \onlinecite{Romagnani_2008}.}
\label{fig:romagnani_2008}
\end{figure*}

Romagnani \textit{et al.} performed the first experiments with proton imaging to study shocks. They used a high-intensity laser to create a supersonic plasma plume that expanded into an unmagnetized low-density ambient plasma, driving a collisionless electrostatic shock \cite{Romagnani_2008}. TNSA protons were then used to probe the interaction.  Proton images, and electric fields reconstructed from the images, showed modulations of the shock front consistent with shock theory and electron kinetic scales (see Fig.~\ref{fig:romagnani_2008}). The shock speed was estimated by comparing features between different proton images within an RCF stack. A follow-up experiment by Ahmed \textit{et al.} used TNSA proton imaging to provide further details about how the electrostatic potential in the shock layer evolves during electrostatic shock formation \cite{Ahmed_2013}.

\begin{figure}
\includegraphics[width=7 cm]{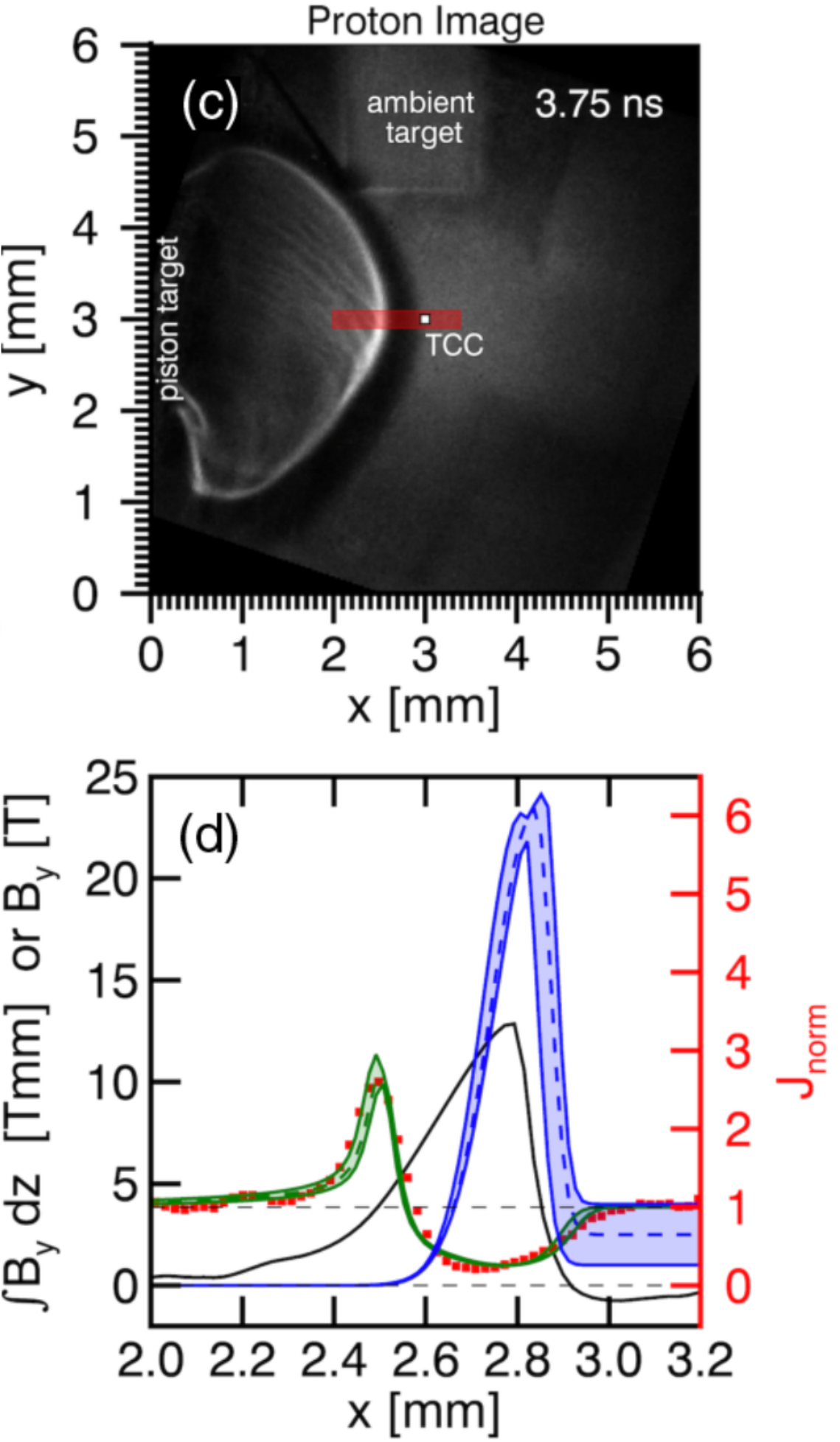}
\caption{Example data from magnetized shock experiments. (c) Proton image taken at time $3.75$~ns using 14.7~MeV \DHe{} protons. (d) Proton intensity (red squares) taken from the red region in (c), normalized to the mean intensity, and the associated reconstructed path-integrated magnetic field $\int B_y dz$ (black). Also shown is the normalized proton intensity (green dashed) forward-modeled from a 2D synthetic magnetic field $B_y(x,z)$, which has the dashed blue profile at $z=0$. The model uncertainties are shown as shaded regions. Adapted from \onlinecite{Schaeffer_2019}.}
\label{fig:schaeffer_2019}
\end{figure}

Schaeffer \textit{et al.} first probed magnetized collisionless shocks shortly thereafter, using a high-energy laser to drive a super-magnetosonic piston plasma through a magnetized ambient plasma, generating a collisionless shock in the ambient plasma \cite{Schaeffer_2017}. The shock was probed with TNSA protons, and the resulting proton images showed large proton fluence variations followed by uniform fluence.  Using a 1D reconstruction technique, the fluence variations were shown to correspond to strong magnetic field compressions at the shock front and a diamagnetic cavity created by the piston behind the shock, consistent with features observed in PIC simulations.  Further experiments \cite{Schaeffer_2019} used \DHe~protons to image the fields in a magnetized shock precursor (see Fig.~\ref{fig:schaeffer_2019}).  These were compared to Thomson scattering data to show how the fields coupled energy from the supersonic piston to an ambient plasma.  Piston-driven magnetized shocks were also studied with TNSA protons by Yao \textit{et al.}, who inferred the electric field structure at the shock front by comparing proton data with a particle-tracing algorithm \cite{Yao_2022}.

Experiments by Li \textit{et al.} investigated electromagnetic shocks by using a high-energy laser to ablate a target, which generated a jet plasma that expanded into a gasbag \cite{Li_2019}.  The collision created a counter-propagating plasma that was imaged with \DHe~protons.  The proton images showed the formation of a shock and Weibel filaments on timescales significantly faster than expected from theory, which was attributed to seed Biermann battery fields embedded in the ablated jet plasma.  Other experiments by Hua \textit{et al.} studied self-generated electromagnetic fields in shocks using a shock tube \cite{Hua_2019}.  High-energy lasers incident on one end of the tube created a strong collisional shock, which was probed from multiple angles with TNSA protons.  By comparing the proton images from different directions, they showed that magnetic fields, self-generated through the Biermann battery effect, dominated the shock structure, and that electric fields were relatively insignificant.  

Levesque \textit{et al.} utilized proton imaging to study laser-driven bow shocks \cite{Levesque_2021}, which led to the development of a new technique for analyzing the proton data \cite{Levesque_2022}.  The technique utilizes caustic features to help reconstruct the path-integrated magnetic fields, as well as two proton energies (times) to to break the degeneracy of the solutions (see Sec.~\ref{sec:frontiers:schemes} for further discussion).

\subsection{Jets}
\label{Sec:apps:Jets}

An important use for proton imaging has been in laboratory astrophysics experiments that have investigated the dynamical effect of magnetic fields on supersonic plasma jets. Various astrophysical systems -- including active galactic nuclei (AGN), pulsar wind nebulae (PWN), and young stellar objects (YSOs) -- are associated with magnetized jets and outflows. The magnetic fields are thought to explain a number of observed phenomena in these jets, including, for example, collimation, clumping, and kinking. In certain conditions, laser-produced plasma jets can be treated as re-scaled analogues for astrophysical jets, meaning that tailored laboratory experiments can shed light on these astrophysical phenomena~\cite{Blackman_2022}. 

Loupias \textit{et al.} carried out the first such experiment using proton imaging to compare the expansion of a front-side blow-off plasma jet into vacuum with that of a similar jet into an ambient gas, and found tentative evidence for electromagnetic fields at the gas-jet boundary from their ${\sim}$3-5~MeV proton imaging data~\cite{Loupias_2009}. More recently, the evolution of the MHD interchange and kink instabilities in a jet created by the irradiation of a cone-shaped target were observed by \onlinecite{Li_2016}. The perturbed magnetic fields associated with both MHD instabilities manifested as quasi-periodic proton-fluence structures in the proton imaging data. It was then demonstrated that this laboratory jet was a reasonable analogue to the Crab Nebula jet under appropriate re-scaling, supporting the idea that MHD instabilities provide a plausible explanation for the periodic oscillations in the Crab Nebula jet's direction that were previously detected by the Chandra X-ray Observatory. 

\begin{figure}
\includegraphics[width=\linewidth]{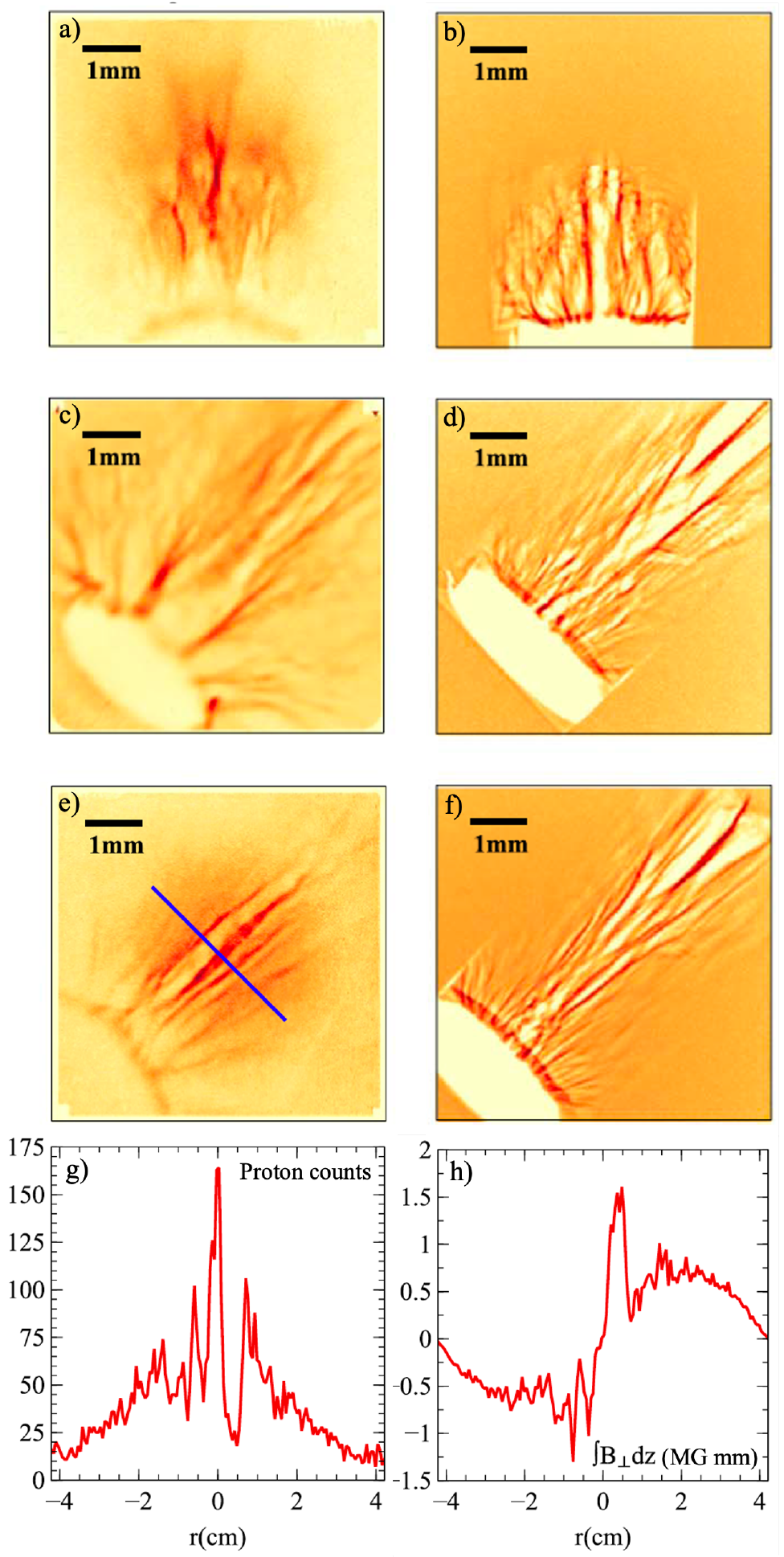}
\caption{Proton images of laser-driven magnetized, supersonic jets. For this experiment, 10~kJ of laser energy (20 beams, each with 500~J of energy) was focused over 1~ns into a hollow ring (radius $800\, \mu$m) to produce the jet. 14.7~MeV proton-imaging data was then collected at different times: a) 1.6~ns, c) 2.8~ns, and e) 3.6~ns after the initiation of the drive beams. 3D FLASH simulations of the experiment were also performed, and synthetic proton images [b), d), f)] then generated at these times. In addition, 1D direct inversion analysis was performed on a lineout of the experimental data; the position of the lineout is indicated by the blue line on e), the lineout itself shown in g), and the path-integrated magnetic field recovered by the inversion shown in h). Adapted from \onlinecite{Gao_2019}.} 
\label{jets_examplePRAD}
\end{figure}

By contrast, another experiment by Gao \textit{et al.} successfully realized magnetically collimated, stable supersonic jets using laser beams focused into in a hollow ring configuration \cite{Gao_2019}. The combined use of an electromagnetic field-reconstruction algorithm applied directly to the proton imaging data and particle-tracing with FLASH simulations showed that the experiment realized ${\sim}$MG magnetic fields (see Fig.~\ref{jets_examplePRAD}); given other plasma jet parameters, fields of this strength were sufficient to efficiently collimate the jet, as well as realize magnetization parameters (such as the plasma beta and the Hall parameter) of direct relevance to astrophysical systems. Previously, \onlinecite{Manuel_2015} first used proton imaging to look at a magnetized jet with inconclusive results.

In addition to understanding the dynamics of individual jets, Li \textit{et al.} studied the evolution of magnetic fields in colliding plasma jets at both collinear and noncollinear angles with the aid of proton imaging~\cite{Li_2013}. These measurements have been used to show that the underlying physics of collisions between sufficiently supersonic jets cannot be adequately described by single-fluid hydrodynamics due to the low inter-jet collisionality between constituent particles, instead requiring two-fluid or kinetic models.

\subsection{Turbulence and Dynamos}
\label{Sec:apps:Dynamos}

In recent years, proton imaging has come to play an important role in diagnosing magnetic fields in experiments investigating the evolution of turbulent laser-plasmas. Of particular note are a series of experiments that have investigated magnetic-field amplification by turbulent plasma motions on various high-energy laser facilities. It is a long-standing theoretical prediction that turbulent, (weakly) magnetized plasmas with sufficiently large magnetic Reynolds numbers $\mathrm{Rm} \equiv u_{\rm rms} L/\eta$ (where $u_{\rm rms}$ is the root-mean-square (rms) turbulent velocity, $L$ the driving scale of the turbulence, and $\eta$ the plasma's resistivity) can support sustained magnetic-field amplification until dynamical magnetic-field strengths are attained, a mechanism known as the \emph{fluctuation} or \emph{small-scale turbulent dynamo}. This mechanism, which provides a plausible explanation for the magnetic fields ubiquitously observed in various different astrophysical environments, has been seen in numerous MHD and more recently kinetic simulations (see \onlinecite{Rincon_2019} for a recent review), but had not been observed in any laboratory experiments. 

Tzeferacos \textit{et al.} first realized a small-scale turbulent laser-plasma dynamo on the OMEGA laser~\cite{Tzeferacos_2018}. Proton imaging with a \DHe{} source helped confirm the formation of a dynamo via measurements of magnetic fields both at the formation of the turbulent plasma and several ns later. More specifically, the application of magnetic-field reconstruction algorithms to 15~MeV proton radiographs yielded two-dimensional maps of path-integrated stochastic magnetic fields at both times. Further analysis of these maps under various assumed statistical symmetries allowed for values of the rms strength of the magnetic field, the magnetic-energy spectrum, and a bound on the maximum magnetic field strength to be inferred. This analysis showed that magnetic energy was amplified ${\sim}$600 times during the course of the experiment, with the characteristic magnetic energies post-amplification being a finite fraction of the turbulent kinetic energy. Particle-tracing simulations applied to the magnetic field outputted by MHD simulations of the experiment using the code FLASH corroborated these findings \cite{Tzeferacos_2017}. 

\begin{figure*}
\includegraphics[width=\linewidth]{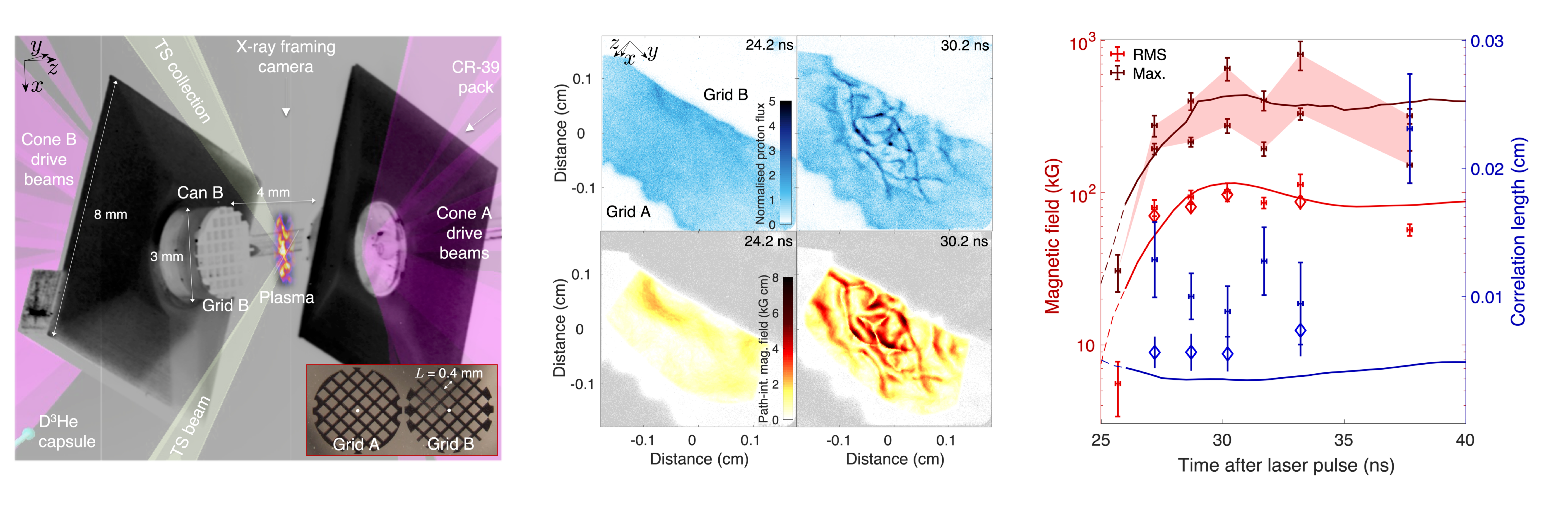}
\caption{Proton images of stochastic magnetic fields amplified by a turbulent laser-plasma dynamo. A annotated photograph of the experimental platform used to create the dynamo is shown in the left panel; the 14.7~MeV proton images collected during the experiment are shown on the top middle, with corresponding path-integrated magnetic field maps recovered from these images shown on the bottom middle. Estimates of the rms and maximum magnetic field strengths, as well as correlation lengths, were then recovered from these maps using statistical methods (see right panel, solid errorbars); the results were compared with similar quantities inferred from 3D FLASH simulations of the experiment (see right panel, triangle markers), as well as the same quantities computed directly (solid lines). Adapted from \onlinecite{Bott_2021}. } 
\label{TDYNO_examplePRAD}
\end{figure*}

Bott \textit{et al.} also used proton imaging in a related manner for several subsequent experiments on this topic \cite{Bott_2021}. Time-resolved measurements in a turbulent plasma with order-unity magnetic Prandtl number showed the evolution of stochastic magnetic fields being amplified by the fluctuation dynamo.  The proton data was characterized by applying direct inversion analysis to a time sequence of proton images, obtaining path-integrated magnetic field maps (see Fig.~\ref{TDYNO_examplePRAD}).

Other related experiments include observations of (inefficient) magnetic-field amplification by supersonic plasma turbulence~\cite{Bott_2021b}, measurements of the transport of high-energy charged-particle through intermittent magnetic fields~\cite{Chen_2020}, and a demonstration that the key properties of a particular laser-plasma dynamo were insensitive to the plasma's initial conditions~\cite{Bott_2022}. Proton imaging was also fielded as part of a recent experiment on the NIF by Meinecke \textit{et al.} that investigated the suppression of heat conduction in magnetized turbulent plasmas~\cite{Meinecke_2022}; however, the ${\sim}$MG fields realized in that experiment were sufficiently strong, and the characteristic deflection angle of 14.7~MeV protons sufficiently large, that electromagnetic field-reconstruction algorithms used in previous experiments could not reasonably be applied. To overcome this, alternative diagnostic approaches including proton-beam truncation using slits and pinholes were employed to recover the rms and maximum magnetic-field strengths realized in the experiment.

Liao \textit{et al.} proposed a different dynamo experiment using turbulent ablated blow-off plasmas \cite{Liao_2019} and conducted experiments showing a magnetic dynamo is created \cite{Liao_2022}; this, and other new experiments, might open up more potential platforms to study HED dynamos with astrophysical relevance in the laboratory.

\subsection{Ultrafast Dynamics}
\label{Sec:apps:Dynamics}
\label{PRADexp_ultrafast}

\begin{figure}
\includegraphics[width=7 cm]{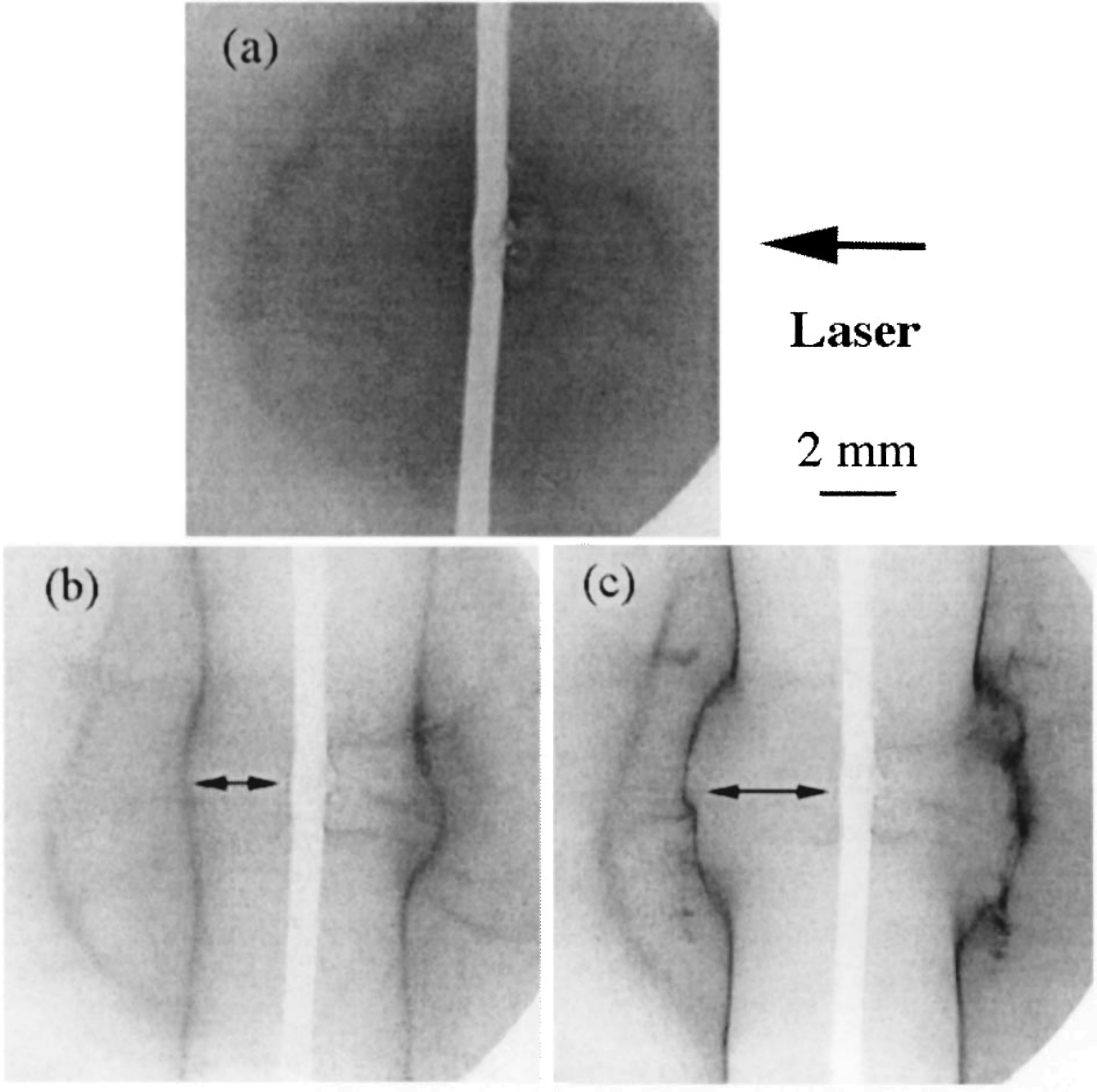}
\caption{Proton images taken during laser irradiation of a 50-mm Ta wire. The frames are three active layers from the same shot and refer to different probing times ahead of the peak of the interaction pulse: (a) $E_p\sim8$~MeV,
$t\sim-8$~ps; (b) $E_p\sim7$~MeV, $t\sim-12$~ps; (c) $E_p\sim6$~MeV, $t\sim-3$~ps. Adapted from \onlinecite{Borghesi_2003}.}
\label{fig:wire_chargeup}
\end{figure}

The ps-scale temporal resolution obtainable when using a TNSA probe has been exploited in several experiments to investigate the ultrafast dynamics following  high intensity, short pulse  laser interaction with a target or a plasma. Very large and transient electromagnetic fields are generated in these interactions, in connection with the large flows of relativistic electrons generated in the irradiated portion of the target. The most energetic electrons typically escape from the target, charging it positively on ps time scales. The process of target charge-up and subsequent discharge was observed in some of the earliest proton imaging experiments investigating 50~TW interactions with wire targets \cite{Borghesi_2003} (see Fig.~\ref{fig:wire_chargeup}).

Quasi-instantaneous target charge-up was observed via strong deflection of protons away from the target surface, causing the appearance of caustics, which were associated with a transverse electric field with amplitude at the surface of order $10^{10}$~V/m. The charge-discharge cycle was characterized more extensively in follow up experiments by Quinn \textit{et al.}, which employed a high energy proton probe (up to 40~MeV) generated by the VULCAN Petawatt laser \cite{Quinn_2009}. This probed a portion of the wire away from the interaction region, which allowed reconstructing the characteristic time (${\sim}20 \, \mathrm{ps}$) over which target neutralization occurs. The target charging measured in these studies was found to be consistent with the number of escaping electrons evaluated from self-consistent models. A modified proton imaging setup \cite{Quinn_2009b}, in which the wire was tilted away from the vertical position, within the plane containing the proton beam's axis, allowed resolving the very early phases of these dynamics, in which an electric field is seen to spread from the interaction region along the target at a velocity close to the speed of light \cite{Quinn_2009c}. The field was interpreted, with the help of PIC simulations, as a surface electromagnetic mode generated by the ultra fast motion of the electrons escaping from the target (similarly to the emission from a transient antenna).

Kar \textit{et al.} further investigated the dynamics of this surface mode by characterizing the propagation along mm-length wires connected to the laser-irradiated target \cite{Kar_2016}. These studies, mostly carried out employing a self-imaging scheme, where the proton probe is provided by the same laser-irradiated target which produces the electromagnetic mode \cite{Ahmed_2016}, have highlighted its nature as an unipolar electromagnetic pulse of temporal duration comparable to the target discharge time characterized in the earlier experiments \cite{Quinn_2009}. This characterization, carried out with bespoke targets where the length of wire within the probe field of view is maximized, has been at the basis of methods for TNSA proton beam conditioning in suitably designed helical coil targets, also presented in \onlinecite{Kar_2016}.

Electron energization at the irradiated target surface is also at the basis of the TNSA mechanism for proton acceleration, which has been discussed in detail in Sec.~\ref{sec:exp:sources:tnsa}. Experiments by Romagnani \textit{et al.} employed proton backlighting (using a TNSA probe from a separate foil) to detect the electric fields associated with TNSA acceleration from the rear surface of a laser-irradiated target at $I\sim10^{19}$~W~cm$^{-2}$ \cite{Romagnani_2005}. Careful temporal synchronization, as well as exploitation of the multiframe capability of RCF stack detectors, led to the detection of the highly transient, Gaussian-shaped TNSA sheath field. The data were used to benchmark TNSA expansion models, which were in substantial agreement with the experimental results.

Romagnani \textit{et al.} also reported the characterization of the field associated with the propagation of relativistic electrons in the interior of a target \cite{Romagnani_2019}. This experiment, employing a PW-driven high-energy proton probe, exploited the capability of high energy protons to penetrate through dense matter, while at the same time using a target design aimed to minimize collisional scattering and the corresponding spatial resolution degradation. The data provided evidence of magnetized filamentation of the electron current within the target, which, via comparison with hybrid simulations, was attributed to resistive processes. The angular opening of the electron beam injected in the target is another quantity that could be directly inferred from the data. 

Complex plasma dynamics are also initiated when an intense laser pulse propagates through an underdense plasma. Several experiments were carried out to investigate this scenario, with the aim of measuring the electric and magnetic field generated as a plasma channel is formed, and, additionally, to obtain information on the propagation and channel features in near-critical plasma where optical probing may be difficult. Experiments by Kar \textit{et al.} investigated the formation of a charge-displacement channel in near-critical plasma following the propagation of an intense 100~TW, ps pulse through a gas jet \cite{Kar_2007}.  A moving evacuated region was observed in the proton images, in coincidence with the position of the laser pulse, which was consistent with a radial space-charge field within the channel set-up by electron displacement by the pulse's ponderomotive force. The walls of a channel expanding from the interaction regions are seen to develop after the pulse has passed, together with the appearance of a region of proton accumulation along the propagation axis, which was later interpreted \cite{Romagnani_2010} as the signature of a long-lived azimuthal magnetic field within the channel. 

More complex channel structures were observed in experiments by Willingale \textit{et al.} performed on the OMEGA-EP laser, investigating the propagation of 1--8~ps, kJ class laser through an underdense, mm-scale preformed plasma plume \cite{Willingale_2011b,Willingale_2013}. The time-resolved formation of an evacuated channel was also observed in this experiment,  which highlighted a number of additional features, such as  filamentation at the channel's end, channel wall modulations (tentatively associated to the formation of surface waves), as well as the copious appearance of bubble-like structures within the interaction region. This is a recurring occurrence in these types of experiments \cite{Borghesi_2002b,Romagnani_2010,Sarri_2010}. Through comparison with PIC simulations, these structures have been identified as late-time remnants (electromagnetic \textit{post-solitons}) of solitary structures (\textit{solitons}) originating from local trapping of frequency down-shifted laser-radiation in cavitated plasma regions \cite{Bulanov_1992,Naumova_2001}.


\subsection{HED Hydrodynamic Instabilities}
\label{Sec:apps:Hydro}

\begin{figure*}
\includegraphics[width=17cm]{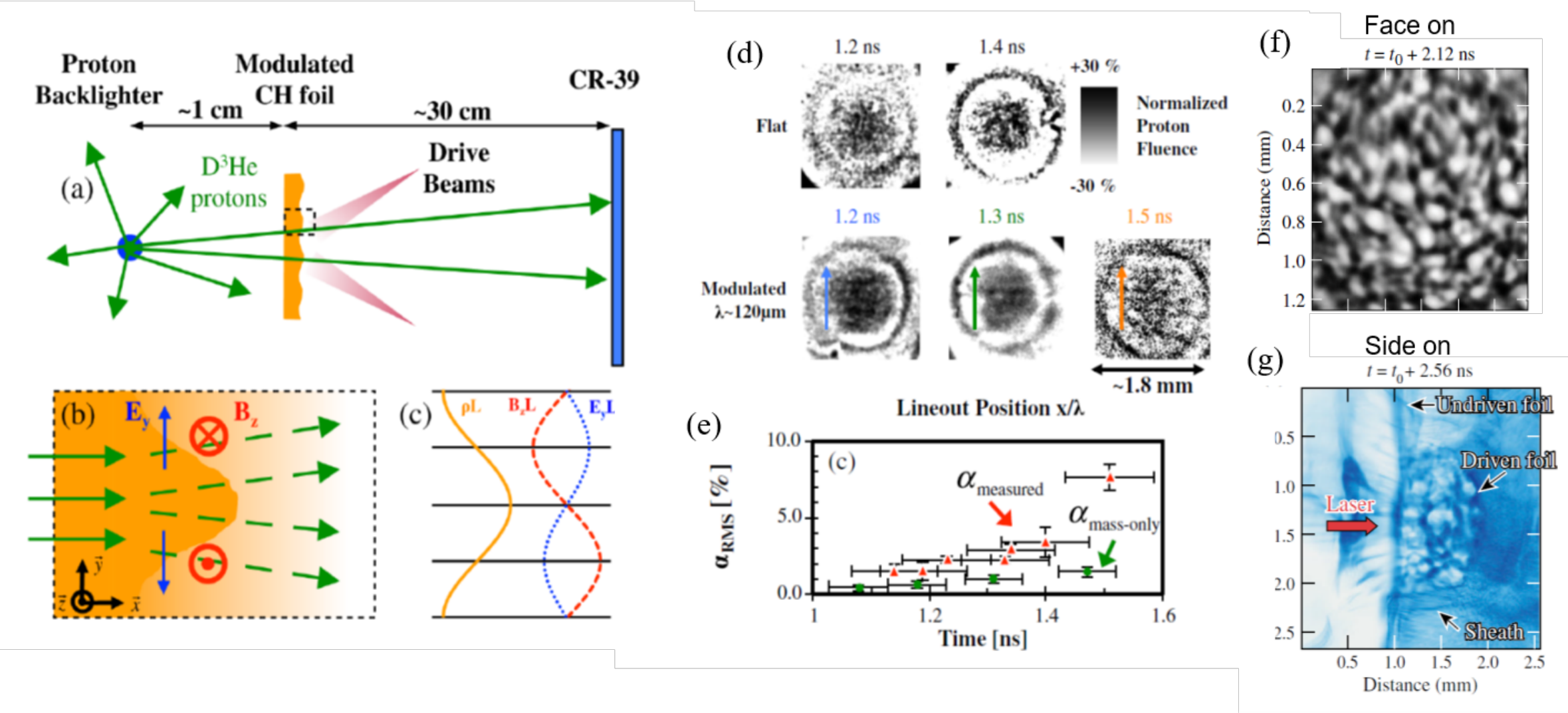}
\caption{(a) A schematic drawing of the experimental setup used to image CH foils with 2D seed perturbations using \DHe{} proton sources. (b) An expanded view of proton deflections (green arrows) due to RT-induced density, as well as electric field (blue arrows) and magnetic field (red symbols) modulations in the target. (c) Path-integrated quantities (arbitrary units) are shown during the linear growth phase. (d) Sample proton fluence images for flat and modulated foils; scale size is given in the target plane and lineout direction is indicated. Proton fluence is normalized for comparison across different shots. (e) Measured rms fluence variations (red triangles) in proton images. Expected rms variation due to the mass only (green circles) was calculated using density distributions from x-ray data. Adapted from \onlinecite{Manuel_2012b}. (f) A face-on proton image of a 15~$\mu$~m-thick CH foil taken with 25-MeV TNSA protons at 2.12~ns after irradiation. Adapted from \onlinecite{Gao_2013}]. (g) A side-on image using 13~MeV TNSA protons at 2.56~ns after irradiation. Adapted from \onlinecite{Gao_2012}.} 
\label{fig:RTI_2012b}
\end{figure*}

Creating and studying hydrodynamic instabilities in HED environments -- including Rayleigh-Taylor (RT) \cite{Rayleigh_1883,Taylor_1931}, Richtmyer-Meshkov (RM)~\cite{Richtmyer_1960, Meshkov_1969}, and Kelvin-Helmhotz (KH)~\cite{Kelvin_1880} -- is of particular interest for the study of HED phenomena such as core-collapse supernova~\cite{Swisher_2015}, accretion disks~\cite{Balbus_1991}, ICF implosions~\cite{Lindl_2004, Sadler_2020a}, or pulsed power pinches~\cite{Harris_1962}. These instabilities are present in many strongly-driven plasma systems and can lead to turbulence (e.g., see \onlinecite{Flippo_2016}) and mixing of materials that can significantly change the behavior and understanding of experiments and phenomena. The addition of self-generated electromagnetic fields makes these systems especially complicated and not well studied experimentally, as access is often limited to highly penetrating x-rays.

Proton imaging, as opposed to x-ray imaging, is uniquely suited to observe these electromagnetic fields in HED experiments and has been employed with some success to date. These self-generated fields, as well as applied fields, can change the plasma properties and can be crucial to understanding the evolution of instabilities like RT \cite{Song_2020,Srinivasan_2012b,Modica_2020}, ablative RT \cite{Garcia-Rubio_2021}, RM \cite{Samtaney_2003,Shen_2019,Shen_2020}, and KH \cite{Ryu_2000,Modestov_2014,Sadler_2022}, along with a newly discovered composition instability\cite{Sadler_2020c}. This includes the possibility of stabilizing these instabilities or curtailing their growth with external fields \cite{Rosensweig_1979,Srinivasan_2013,Sano_2013,Praturi_2019}. Some of the first HED experiments to use proton imaging to study hydrodynamic instabilities were done using a laser-driven ablative RT platform and a \DHe{} proton source by \onlinecite{Manuel_2012b}.  The results showed that the RT instability can lead to self-generated fields as predicted. A summary is shown in Fig.~\ref{fig:RTI_2012b}(a-e), where a CH target with sinusoidal perturbations was driven by lasers (a), and the observed RT growth caused Biermann generated magnetic fields (b) to grow from 3 to 10~T (d-e); however, these fields were $10^{4}$ times too small to affect the hydrodynamics directly.  Other experiments by Gao \textit{et al.} showed RT bubble growth using a laser-driven foil (see Fig.~\ref{fig:RTI_2012b}(f-g)) with either transverse \cite{Gao_2012} or longitudinal \cite{Gao_2013} proton imaging of CH targets.
More recent experiments have attempted to look at the self-generated magnetic fields inside denser HED shock-tube targets, where Coulomb scattering is an issue~\cite{Lu_2020,Lu_2020a}, and where the fields can change the heat-flow in these targets, changing the instability growth \cite{Sadler_2022}.
\subsection{Inertial Confinement Fusion}
\label{Sec:apps:ICF}

The ultimate goal of inertial-confinement fusion (ICF) is ignition and high gain, which requires that a cryogenic deuterium-tritium (DT) spherical capsule be symmetrically imploded to reach sufficiently high temperature and density. Such an implosion results in a small mass of low density, hot fuel at the center, surrounded by a larger mass of high density, low temperature fuel \cite{Nuckolls_1972,McCrory_1988,Lindl_1995,Atzeni_2004,Hurricane_2014,Betti_2016}. Shock coalescence ignites the hot spot, and a self-sustaining burn wave subsequently propagates into the main fuel region. The symmetry requirements impose strict constraints for achieving fusion ignition \cite{Nuckolls_1972,McCrory_1988,Lindl_1995,Atzeni_2004,Hurricane_2014,Betti_2016,Glenzer_2010,Li_2010}. The tolerable drive asymmetry of an implosion, in a time-integrated sense, is less than 1--2\%, depending on the ignition margin \cite{Lindl_1995,Atzeni_2004,Hurricane_2014,Betti_2016,Glenzer_2010,Li_2010}. Consequently, understanding and controlling implosion dynamics is essential for ensuring success. Proton radiography has been developed as an important method for diagnosing ICF implosions because it is sensitive both to plasma density and to electromagnetic fields. Additionally, there are promising indications that externally applied magnetic fields can improve hydrodynamic conditions in ICF implosions and thereby increase capsule performance~\cite{Srinivasan_2013, Mostert_2014,Perkins_2013,Strozzi_2015,Walsh_2019,Walsh_2020,Moody_2022}, and proton imaging provides a vital tool for assessing how such imposed fields evolve as the implosion proceeds \cite{Heuer_2022}. 

\textbf{Direct-Drive Implosions}: In direct-drive ICF, a fuel capsule needs to be compressed through illumination by laser light in order to bring the fuel to high temperature and density conducive to fusion and ignition. Earlier work by Mackinnon \textit{et al.} successfully demonstrated the feasibility of imaging implosions with TNSA protons, backlighting plastic (CH) capsules that were imploded by six 1-$\mu$m-wavelength laser beams \cite{Mackinnon_2006}. These were followed by nuclear observations of implosion dynamics for direct-drive spherical capsules on the OMEGA laser using monoenergetic proton imaging \cite{Li_2006a,Li_2006b}.  These experiments aimed to probe distributions of self-generated electric and magnetic fields \cite{Igumenshchev_2014}, determine areal density $\rho R$ by measuring the energy loss of backlighting protons, and sample all the implosion phases from acceleration, through coasting and deceleration, to final stagnation, in order to provide a more comprehensive picture of ICF spherical implosions.


\begin{figure}
\includegraphics[width=7cm]{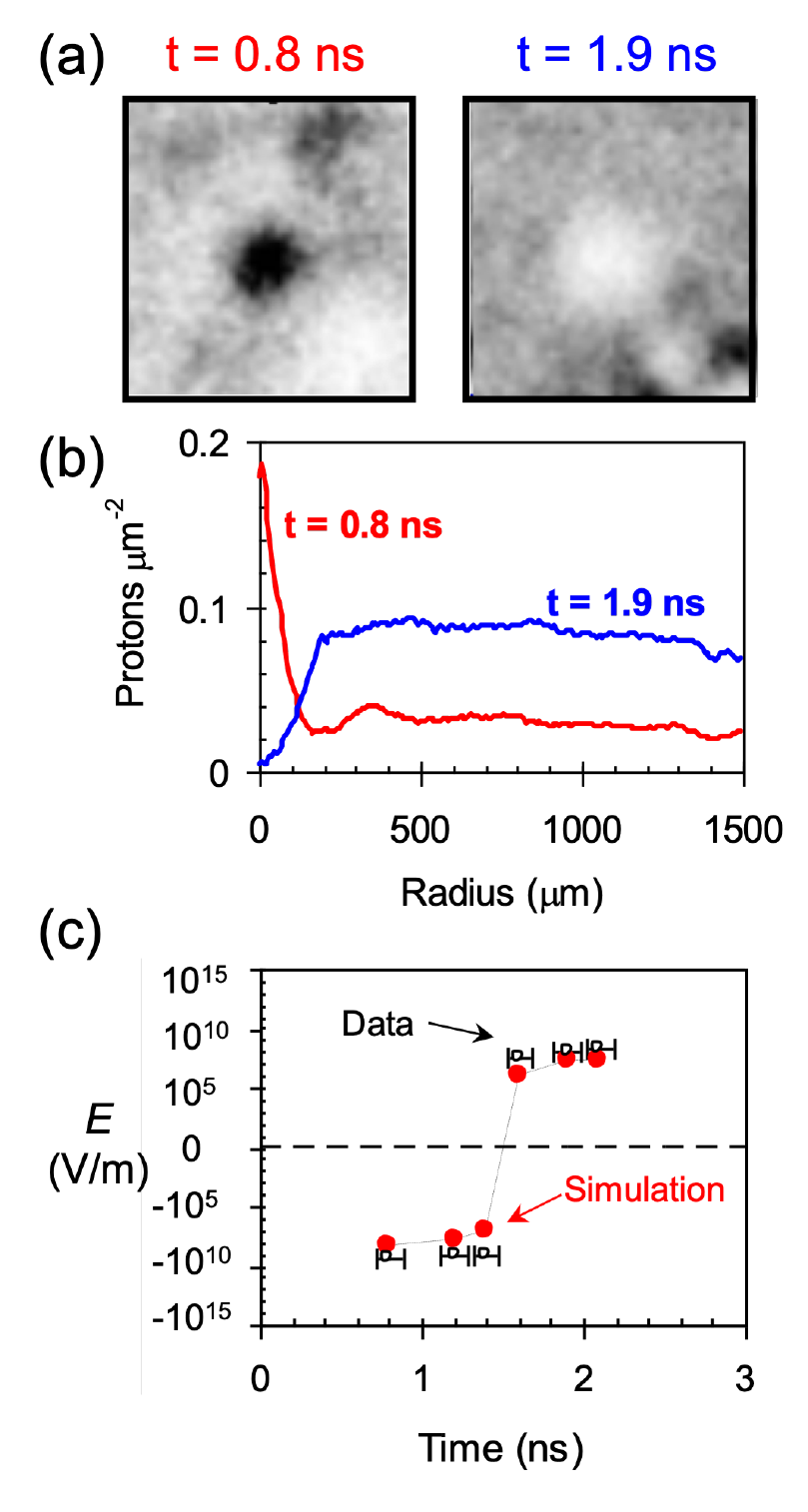}
\caption{(a) 15.1-MeV proton images of imploding capsules at two times, $t = 0.8$ ns and 1.9 ns. In the fluence images, darker means higher fluence. Comparatively, a fluence peak occurs in the image centers during the early stages of implosion, indicating a ``focusing'' of imaging protons there, while at later times, the fluence is extremely low, or defocused, at the image centers. (b) Radial profiles of the proton fluence images from (a). (c) Radial electric fields estimated from experimental measurements (open circles) and from LILAC simulations (solid circles) vs. implosions times. Horizontal error bars represent uncertainties in backlighter burn time. The differences between simulation and data can be largely related from effects of proton scattering. Adapted from \onlinecite{Li_2008}.}
\label{fig:ICF_direct_imploding}
\end{figure}

Further proton imaging experiments by Li \textit{et al.} revealed the existence of a radial electric field inside the imploding capsules \cite{Li_2008}. As shown in Fig.~\ref{fig:ICF_direct_imploding}, proton images showed both an inward and outward directed radial electric field, suggesting the radial electric field has reversed direction during an ICF implosion. The magnitude of these electric fields compared well with the field calculated from the pressure gradients predicted by the 1D hydrodynamic code LILAC \cite{Delettrez_1987}, indicating that the fields are a consequence of the evolution of the electron pressure gradient.  The proton images were also utilized to extract quantitative information about capsule sizes and $\rho R$'s at different times, which indicated that the implosions had approximately 1D performance, with little impact from hydrodynamic instabilities before deceleration.

Additional experiments by Li \textit{et al.} indicated that the apparent degradation of capsule performance at later times relative to the 1D simulation could be largely a consequence of fuel-shell mixing \cite{Li_2002} and implosion asymmetry \cite{Li_2004}. Proton images from experiments by S{\'e}guin \textit{et al.} also revealed that electromagnetic fields induced filaments inside the capsule shell \cite{Seguin_2012a}. These field structures were discussed further by \onlinecite{Manuel_2013}, who interpreted them as heat-flux-type instabilities driven by the heat conduction from the corona down to the solid ablation layer.

Proton imaging was used by \onlinecite{Chang_2011} on the first ICF direct-drive experiment to employ an external applied magnetic field. The experiments aimed to increase the peak ion temperature of an ICF capsule by strongly magnetizing the plasma's constituent electrons, reducing thermal conduction across magnetic-field lines and thereby reducing heat losses from the imploded capsule's hotspot. A ${\sim}$30\% increase to the neutron yield was indeed observed when the external magnetic field was applied. While the proton-imaging data collected were not of sufficient quality for an unambiguous magnetic-field measurement to be made, the characteristic proton-flux inhomogeneity that was observed was consistent with fields that were both trapped and amplified in the imploding capsule.

\textbf{Laser-Driven Hohlraums}: In the indirect-drive approach to ICF, the capsule implodes in response to a quasi-uniform, nearly Planckian x-ray radiation field (hundreds of electron-volts), which is generated by multiple high-power laser pulses irradiating a high-Z enclosure called a hohlraum \cite{Lindl_1995}. The x-ray radiation drives the implosion of a cryogenic DT capsule contained within a low-Z ablator, leading to the achievement of high temperature, high density, and tremendous plasma pressure in the compressed core, and potentially resulting in hot-spot ignition and a self-sustaining fusion burn wave that subsequently propagates into the main fuel region for high-energy gain. In addition, hohlraum-generated x-ray drive can create extreme plasma conditions and has served as an important platform for studying a wide range of basic and applied high-energy-density physics \cite{Lindl_1995,Atzeni_2004}, including laboratory astrophysics, space physics, nuclear physics, and material sciences.

In diagnosing plasma conditions and field structures generated in a laser-irradiated hohlraum, proton imaging plays an important role in providing physics insight into the hohlraum dynamics and x-ray-driven implosions, impacting the ongoing ignition experiments at the National Ignition Facility. For example, experiments by \onlinecite{Li_2010,Li_2009,Li_2012} utilizing proton imaging to measure the spatial structure and temporal evolution of plasma blowing off from hohlraum wall revealed how the fill gas compresses the wall blowoff, inhibits plasma jet formation, and impedes plasma stagnation in the hohlraum interior. These experiments also showed that the magnetic field is rapidly convected by the heat via the Nernst effect, which was $\sim$10 times faster than convection by the plasma fluid from expanded wall blowoff.  This results in the inhibition of heat transfer from the gas region in the laser beam paths to the surrounding cold gas and in a subsequent increase in local plasma temperature. The experiments further showed that interpenetration of the two materials (gas and wall) occurs due to the classical Rayleigh-Taylor instability as the lighter, decelerating ionized fill gas pushes against the heavier, expanding gold wall blowoff.  

\begin{figure}
\includegraphics[width=7 cm]{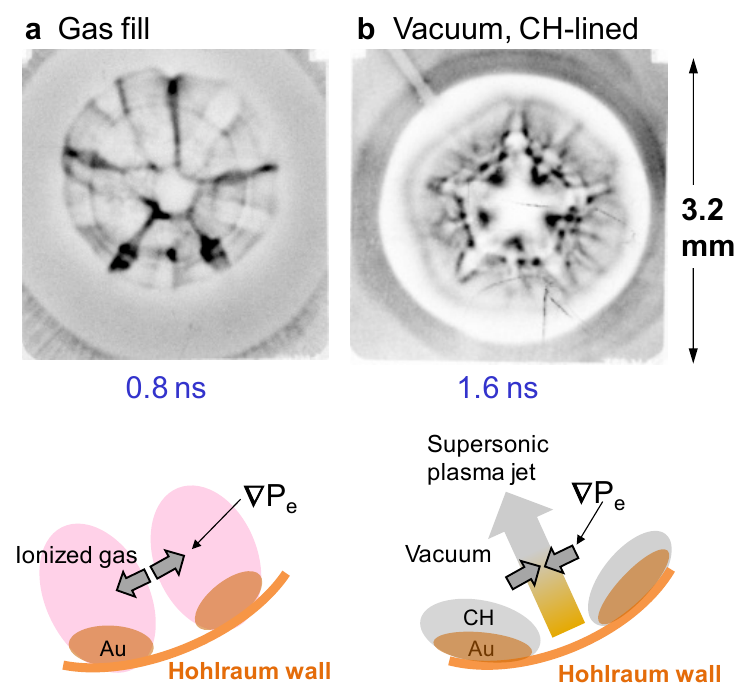}
\caption{Proton images show surpluses in the regions between the pairs of expanding plasma plumes in a gas-filled, Au hohlraum (a) but show deficits in a CH-lined, vacuum Au hohlraum (b). They indicate opposing directions of the self-generated electric fields, as illustrated schematically by the corresponding cartoons.  Adapted from \onlinecite{Li_2012}.}
\label{fig:ICF_hohlraum_proton_fluence}
\end{figure}

Further experiments by Li \textit{et al.} deployed proton imaging to address plasma flow dynamics in hohlraums by providing the first physics picture of the process of hohlraum plasma stagnation \cite{Li_2012}. Using a Au hohlraum filled with neopentane gas (C$_5$H$_12$), the resulting proton fluence patterns indicated that no high-density plasma jets were formed, that the fill gas along the laser beam path is fully ionized, and that the interfaces between the gas plasma and the Au wall blowoff are constrained near the wall surface. The proton images also revealed a unique 5-prong, asterisk-like pattern that is a consequence of the OMEGA laser beam distribution. Spherical CH targets driven in both gas-filled Au hohlraums and CH-lined vacuum Au hohlraums were used to explore this mechanism further, as shown in Fig.~\ref{fig:ICF_hohlraum_proton_fluence}. The results show that protons were focused into the gaps (high-fluence spokes) for the gas-filled hohlraum (Fig.~\ref{fig:ICF_hohlraum_proton_fluence}a) but were deflected away from the spokes in the CH-lined vacuum hohlraum (Fig.~\ref{fig:ICF_hohlraum_proton_fluence}b).  By symmetry these deflections were not due to spontaneously-generated magnetic fields; instead, lateral electric fields associated with azimuthally oriented electron pressure gradients ($\nabla P_e$) in the plasma plumes and in the radial plasma jets, $\boldsymbol{E}=-\nabla P_e/en_e$, may have been the source of these deflections.  Alternatively, the electric field associated with a supersonic heat front generated by the laser-heated gas channels that are in close proximity to the capsule might have caused the deflections. 

\section{Frontiers} \label{sec:frontiers}

While the past two decades have seen the invention and wide-spread adoption of proton imaging as a diagnostic tool for HED plasma experiments, there remain several key challenges that need to be overcome to extend proton imaging into new experimental frontiers.  One challenge is how to probe and analyze ever more complicated electromagnetic fields, including how to measure fields that strongly deflect protons or create caustics or how to disentangle electric and magnetic fields.  A related challenge is how to combine proton images to extract more information, including in 3D.  Another challenge is how to probe higher-density plasmas, where scattering is a significant issue.  A final grand challenge is how to operate a proton imaging diagnostic in a high-repetition-rate regime.

These challenges can be addressed by advancing the field in four key areas: Sources (Sec.~\ref{sec:frontiers:sources}), Detectors (Sec.~\ref{sec:frontiers:detectors}), Algorithms (Sec.~\ref{sec:frontiers:algorithms}), and Schemes (Sec.~\ref{sec:frontiers:schemes}).

\subsection{Advanced Sources} \label{sec:frontiers:sources}

Extending the proton image capabilities demonstrated so far to plasmas with stronger fields or higher densities will require an enhancement of the energy of the protons beyond what is currently available.

As discussed earlier, existing TNSA proton sources extend up to 85~MeV with $\sim$ps pulses \cite{Wagner_2016} and to about 60~MeV with 10's~fs pulses \cite{Ziegler_2021}, although typically the energies that can be used as an efficient probe for imaging will be lower than these cutoffs (as the beam component at the higher energy will have a very small divergence and low number of particles). An obvious route to increasing the proton energies is to use higher power/energy laser drivers. There are a number of multi-PW systems currently being developed or commissioned, with some aiming to deliver up to 10 PW power \cite{Danson_2019}. Most of these systems will be based on ultrashort pulse technology (10's fs pulses), but there are also developments with 100's~fs duration (such as the 10~PW ATON laser at ELI Beamlines \cite{Jourdain_2021}).

The progress achievable in terms of proton energies will depend on the applicability of scaling laws, as well as secondary factors such as how well the beams can be focused and the extent of pulse contrast that can be obtained on these systems.  Various attempts have been made to develop reliable scaling laws for proton energies (see, for example, \onlinecite{Fuchs_2006,Passoni_2010}), which typically indicate dependencies on laser intensity or laser energy (see Fig.~\ref{fig:TNSA_scaling}). For example, faster scalings than the ponderomotive scaling predicted by earlier TNSA theories \cite{Wilks_2001} have been reported, within given intensity ranges, with ultrashort (10's~fs) \cite{Zeil_2010}, multi-ps \cite{Simpson_2021}, or multi-kJ \cite{Flippo_2007, Mariscal_2019} laser pulses.  There is an expectation that experimental results in the multi-PW regime, providing validation to  these scaling predictions, will become available soon as some of the new facilities ramp up their operations.

Additionally, there are a variety of approaches that aim to increase TNSA proton energies by enhancing the energy coupled into relativistic electrons, and by increasing their number density and/or energy (see \onlinecite{Macchi_2013} for a review). These approaches are mostly based on target engineering and can involve, for example, reducing the mass \cite{Buffechoux_2010} or density of the target, structuring  the target surfaces \cite{Margarone_2012}, or adding controlled pre-plasmas \cite{McKenna_2008}.  The electrons can also be enhanced through additional mechanisms such as direct laser light pressure acceleration (DLLPA) \cite{Kluge_2010} or acceleration by surface waves \cite{Ceccotti_2013,Shen_2021}.  Many of these approaches have provided evidence of some proton energy increase from flat foil comparators on a proof of principle basis and under specific experimental conditions. Although some of these schemes may have a role to play in the development/optimization of future proton imaging sources, complications and constraints associated with their implementation may limit their applicability and usefulness. 

Beyond TNSA, there are a number of alternate mechanisms that are being investigated, which aim at increasing the acceleration efficiency and the accelerated ion energy, or at accelerating ion species different from protons. These include, amongst others, Radiation Pressure Acceleration (RPA) \cite{Esirkepov_2004, Robinson_2008} (in the Hole Boring \cite{Robinson_2012} and Light Sail \cite{Macchi_2009} implementations), shock acceleration \cite{Fiuza_2012}, as well as schemes taking place in Relativistic Induced Transparency (RIT) regimes \cite{Henig_2009, Poole_2018}, such as the Break-Out Afterburner approach \cite{Yin_2007}. Hybrid regimes involving a combination of these processes have been highlighted in experiments, and have, for example, led recently to record proton energies approaching 100~MeV, through a combination of RPA, TNSA, and RIT acceleration \cite{Higginson_2021}.

Although these processes are promising in terms of energy enhancement, e.g., in view of potential medical use, they typically generate beams which do not possess the laminarity and homogeneity of TNSA beams, and their potential usefulness for proton imaging is therefore unclear at present, at least in the backlighting implementation discussed in this review. Nevertheless, if very-high-energy beams will be produced through any of these processes, there may be prospects for different radiography/deflectometry approaches in which a small portion of a beam is spatially/ angularly selected (e.g., with an aperture) and used to sample a finite region within a target or plasma (see also Sec.~\ref{sec:exp:diag:mesh}).

Spatial structuring of the beam may also have a role to play in future proton imaging sources. For example, methods have been put forward for generating multiple separate sources from a single TNSA beam \cite{Zhai_2019} , which may lead to additional backlighting capabilities. Similarly, techniques for producing collimated, quasi-mono-energetic beamlets \cite{Kar_2016} may lead to opportunities for "spot-scanning" of an extended field distribution, \textit{i.e.} sampling portions of the field region in separate, consecutive shots, and tracking for each shot the beamlet's deflection.

Increasingly, state-of-the-art ultrashort high-power laser systems provide the opportunity to operate at high repetition rates (1-10 Hz) \cite{Danson_2019}. This could enable the generation of secondary particle sources at a commensurately high rep-rate, provided suitable targets can be used, which offer a refreshed surface for irradiation by consecutive pulses. This may be exploited in future proton imaging experiments, where a repeated proton pulse is coupled to a high repetition interaction pulse, leading to increased data throughput, higher statistics, and/or shorter experiments. 
A number of targetry solutions suitable for high-rep operations are currently being developed and tested. These include tape targets, where a continuous moving foil tape allows mechanical refreshment of the laser-impacted surface between shots \cite{Dover_2020,Noaman-ul-Haq_2017}, free-flowing water sheets \cite{Valdes_2022}, and cryogenic hydrogen targets \cite{Obst_2017, Chagovets_2022}. The above approaches can provide planar targets with thicknesses in the $\mu$m to 10's~$\mu$m range, leading to beams with the expected TNSA properties, with  beam production demonstrated so far at repetition rates of up to 1 Hz employing PW-class laser pulses. Water targets have also been shown to be capable of sustained kHz operation at much lower laser energies \cite{Morrison_2018}.
Another approach explored is the in-situ formation of liquid crystal foils \cite{Poole_2016, Poole_2018}, which can operate at a more moderate repetition of 0.1~Hz, and which provides the capability of varying thickness on demand (from 10~nm to 50~$\mu$m).
An issue with any high rep-rate target is the production of debris in the interaction chamber, potentially leading to optics degradation, which may make solutions based on low-density or thinner targets more attractive. 

In addition to protons and other heavier ions, electrons are another possible source for imaging, using various electron acceleration schemes, which are too numerous to list and outside of the scope of this review. Electron sources are attractive because it is generally easier to create very high ($\sim$GeV) energy electrons that can subsequently probe large field strengths.  Electron sources can also have higher temporal resolution, are easier to operate at high rep-rate, and are easier to incorporate into existing detector designs. This is demonstrated in experiments by Zhang \textit{et al.}, who generated the electron-Weibel instability in a low-density gas jet plasma using a circularly polarized laser \cite{Zhang_2020}, and utilized electron imaging with bunches of electrons from a linear accelerator to image the Weibel fields. It is also straightforward to apply the theory of proton imaging to electrons, including particle-tracing algorithms, although the theory would need to account for relativistic effects given the larger electron speeds involved.  However, electron sources suffer several disadvantages as well.  It will typically be more difficult to discriminate between electric and magnetic fields with electrons due to their smaller mass.  Electrons also scatter more easily, requiring material (e.g., filters) between the source and detector to be thinner.

\subsection{Advanced Detectors} \label{sec:frontiers:detectors}

Currently in HED proton imaging, CR-39 (see Sec.~\ref{sec:exp:detectors:cr39}) and RCF (see Sec.~\ref{sec:exp:detectors:film}) detectors are the most commonly used, and are highly efficient; however, they require extensive chemical processing (in the case of CR-39) or a high proton fluence (in the case of RCF), and both require manually labor-intensive digitization efforts. In contrast, outside of HED, various other proton detectors have been utilized (e.g., see \onlinecite{Bolton_2014,Poludniowski_2015}). Consequently, more advanced detector systems are currently being developed, especially those that can be used with high-rep-rate laser systems (of order 10 shots per hour or higher). Here we consider a few systems for electronic imaging of protons.  

Many position sensitive electronic and solid state detectors have been developed over the years, mainly to support hadron therapy (see, for example, \onlinecite{Poludniowski_2015,Johnson_2017}). These include diodes \cite{Briz_2021,Wang_2016}, multi-wire gas-proportional counters \cite{Sauli_2014}, and Cadmium-Zinc Telluride (CZT) detectors \cite{Simos_2009}. However, these detectors suffer from lower spatial resolution and lower sensitivity compared to CR-39 and RCF, and conversely more sensitivity to noise from EMP due to short-pulse lasers.  Consequently, they have not yet been used for proton imaging.

However, one of the simplest methods for a reusable, rep-rated electronic proton detector is to convert the proton flux into UV or optical light with a luminescent material, i.e., a scintillator. Scintillators include inorganic or organic phosphorescent compounds, gases, solids or liquids, the choice of which depends on the needed attributes (see http://scintillator.lbl.gov/ for an extensive list). Common solid inorganic examples include CsI(Na), bithsmuth germanate (BGO), lutetium Oxyorthosciilicate(Ce) (LSO),lutetium-yttriumoxyorthosilicate (LYSO),
or phosphor screens like rare-earth doped gadolinium oxysulphide (Gadox) with brand-names like Lanex, Luminex or Rapidex. Also, YAG(Ce), GAGG(Ce) and LuAG(Ce) thin screen crystals are made specifically for proton beam applications (\href{https://www.shalomeo.com/Scintillators/Scintillator-Screens}{Shalom EO}). Examples of organic compounds are doped plastics such as BC plastics (\href{https://www.crystals.saint-gobain.com/radiation-detector-assemblies/standard-detectors}{Bicron}) and EJ plastics (\href{https://eljentechnology.com/products/plastic-scintillators}{Eljen}). Scintillators can have decay times from 1500 $\mu$s down to 10’s~ns and as short as 2~ns for ZnO:Ga (\href{https://www.phosphor-technology.com/scintillation-phosphors/}{PhosTech}).

\begin{figure}
    \centering
    \includegraphics[width=7.5cm]{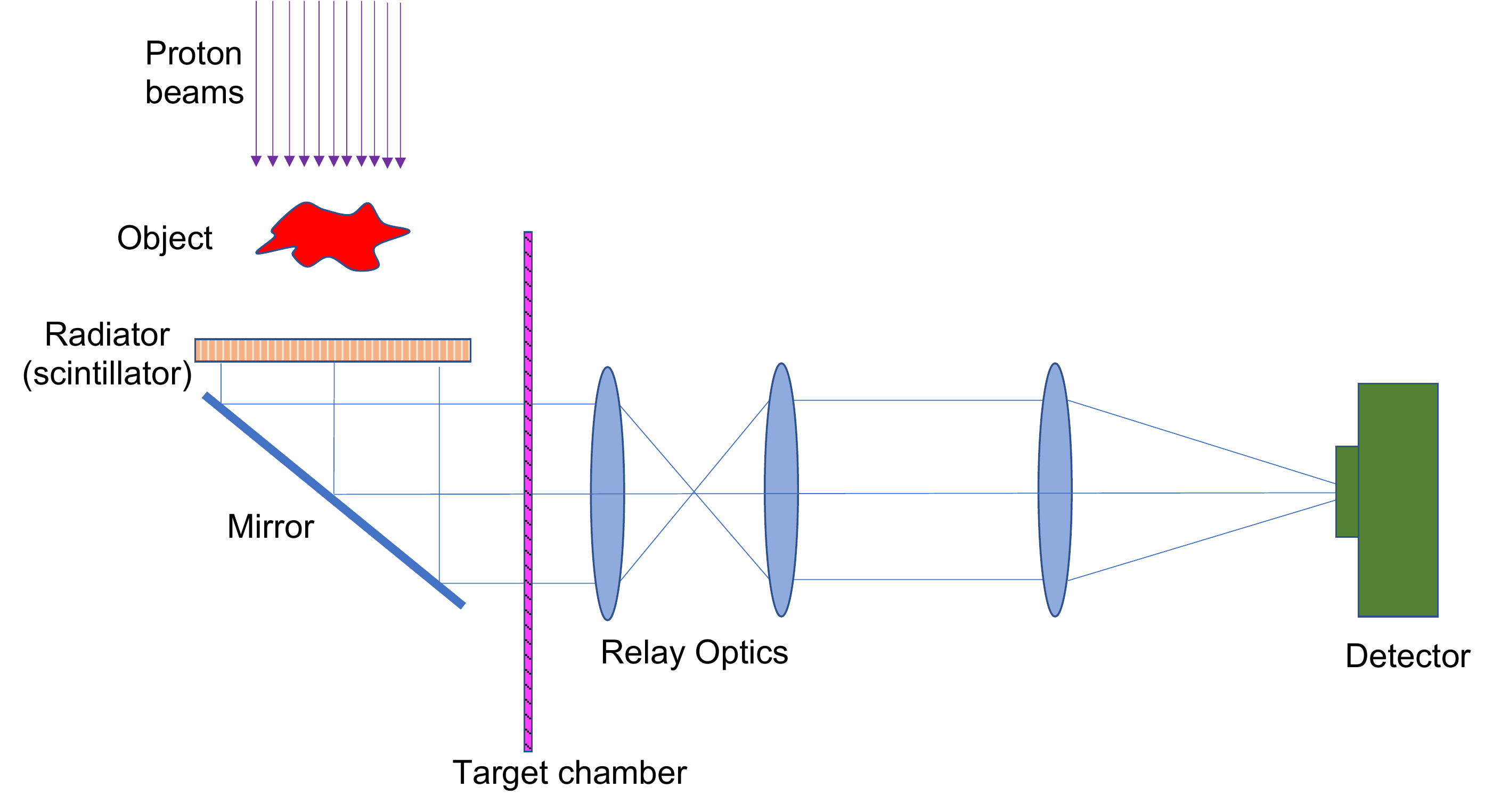}
    \caption{Sketch of an advanced proton imaging detector.}
    \label{fig:AdvDetector}
\end{figure}

The concept of an electronic scintillator-based proton detector system is sketched in Fig.~\ref{fig:AdvDetector}. The proton beam goes through the plasma being imaged and is recorded by a radiator. The role of the radiator is to convert the proton flux into optical signal. The optical image is then transferred outside the target chamber via relay optics and recorded on an optical imaging system. A careful shield or collimator design is needed so that the radiator does not produce unwanted signal from the spurious interactions by the background particles.  Converted light can be relayed via optics or fiber bundle outside the vacuum chamber, where advanced electronics are often situated.  The light is then coupled to a CCD camera or CMOS detector, and multi-channel plates (MCPs) can be added to provide gating functionality to further eliminate background signal.

To date, scintillators have been utilized mainly to understand proton beam profiles in terms of their spatial and energy characteristics. Numerous scintillators have been tested for their optical emission spectrum using monoenergetic proton beams, as well as their response to proton energies. Experimental measurements indicate that the scintillators have a non-linear scaling with proton energy but a linear response to incident flux \cite{Green_2011}. These scintillators were also utilized to characterize the TNSA proton beam profile of different energy bands generated by different stopping material thicknesses \cite{Green_2011, Bolton_2014, Metzkes_2016, Dover_2017, Huault_2019}.

Spatial resolution of the scintillators has been characterized for TNSA proton beams using Eljen Technology organic scintillators \cite{Tang_2020, Manuel_2020}. A few different thicknesses of the scintillators were tested with an effective proton source size of 10 $\mu$m. The spatial resolution measurement was performed by measuring the point spread function and the contrast of a grid pattern that was placed in between the proton source and the detector. The scintillator performance was simultaneously compared with imaging using a RCF detector. The effective resolution limit for the scintillator was measured to be $\sim$22~$\mu$m compared to $\sim$12~$\mu$m for the RCF \cite{Tang_2020}.  The spatial resolution was only weakly dependent on scintillator thicknesses between 50~$\mu$m to 500~$\mu$m for $>2$~MeV protons, while thicker scintillators showed improved imaging contrast.  To further improve the spatial resolution, a pixelated scintillator can be used. The pixelation is achieved by laser cutting a grid pattern into the scintillator to optically isolate regions.  Experiments with TNSA protons measured the performance of different grid patterns and demonstrated a 20\% improvement of the spatial resolution \cite{Manuel_2020}. The current spatial resolution of the scintillators may be sufficient for imaging, but further improvement is required for the spectrum measurements.

Recently, a volumetric prototype imaging system using a liquid scintillator was developed for hadron therapy dose applications \cite{Darne_2019}, but could be adapted and modified for proton imaging. Imaging from several directions would give a tomographic view of the volume, and would have information to reconstruct a detailed and possibly 3D proton image with more information than available with current 2D detectors. Such a system has the possible advantage of capturing all the beam energies efficiently and reconstructing the object in a single shot.  It could also be time resolved with high frame-rate CMOS chips, framing cameras, or steak cameras (depending on the length of emission), and the liquid scintillator has the advantage of being able to be continually replaced in a high-rep-rate system.

\subsection{Advanced Algorithms/Analysis} \label{sec:frontiers:algorithms}

While there have been various developments in the approaches used to analyze proton imaging data in recent years -- for example, using analytically derived field reconstruction algorithms to perform inverse analysis (see Sec.~\ref{sec:theory:inverse:alg}) -- there remains  scope for further advances in several different areas. We do not provide an exhaustive list of these here, but instead highlight a few particularly notable areas. 

One of these areas is successful differentiation between proton imaging measurements of electric and magnetic fields. It is not possible to identify whether electric or magnetic fields are responsible for proton fluence inhomogeneities seen in a single proton image without further assumptions based on either considerations of geometry or the physics of the plasmas being imaged. These assumptions are often well founded~\cite{Li_2010,Kugland_2012,Huntington_2015,Schaeffer_2019}, but in situations where they are not appropriate, establishing alternative approaches would be of great value. One promising possibility involves the simultaneous analysis of two (or more) images using protons with different initial energies. In situations where the electromagnetic fields being imaged evolve over much longer timescales that the transit delays between protons of all energies, it is reasonable to consider that all proton images collected are (approximately) of the same electromagnetic field. The distinct energy scalings of deflection angles due to magnetic and electric fields then allows for the degeneracy between the two fields that is characteristic of a single proton image to be overcome. \onlinecite{Du_2021} recently presented an algorithm that implements this schema, and also proposed a method for minimizing inaccuracies introduced by the temporal evolution of electromagnetic fields. Because both TNSA and \DHe{} proton sources characteristically produce protons with differentiated energies (see Sec.~\ref{sec:exp:sources}), which can be detected independently (see Sec.~\ref{sec:exp:detectors}), this approach should be straightforward to include in future analysis (although more work needs to be done to quantify inherent uncertainties). 

Simultaneous analysis of multiple proton images with different initial energies also provides a route toward making progress on another outstanding issue: analyzing images in which there are caustics. As discussed in Sec.~\ref{sec:theory:inverse:alg}, it is not possible to uniquely reconstruct path-integrated electromagnetic fields from a single proton image if there are caustics present. However, for slowly evolving electromagnetic fields (in the sense just described in the previous paragraph), multiple-energy proton images can be used to provide more restrictive constraints on the possible solution space of path-integrated fields. Levesque and Beesley~\cite{Levesque_2021} recently proposed a differential evolution algorithm that realized this idea for proton images (two different energies) of simple magnetic fields that were either quasi-one-dimensional, or possessed spherical symmetry; the algorithm was then successfully used to reconstruct the magnetic fields associated with a bow shock in a laser-produced plasma on its collision with a magnetized obstacle~\cite{Levesque_2022}. At present, it is too computationally expensive to apply this particular differential evolution algorithm to more general (electro)magnetic fields, although it seems likely that other algorithms may be able to address this limitation. That being said, the number of proton energies required to reduce the possible solution space to a `unique' solution for more complicated path-integrated electromagnetic fields remains uncertain.  

The successful use of a differential evolution algorithm to overcome the issues posed by caustics is emblematic of the promise of the application of machine-learning algorithms for enhanced analysis of proton images. \onlinecite{Chen_2017} provided a proof of concept in this regard by training an artificial neural network to reconstruct successfully the three-dimensional structure of a simple magnetic field from proton images. It is unclear how well this particular approach would generalize to more complicated electromagnetic fields, but convincing arguments are made that, once they have been trained, similar neural networks are more computationally efficient than classical analysis algorithms. Similar neural networks could be used to provide improved path-integrated field reconstruction algorithms by more easily accounting for known limitations in current analysis procedures (for example, uncertainties in the initial beam profiles), as well as quantifying uncertainties. They could also reduce significantly the time taken to perform field reconstructions; the current generation of algorithms typically run for a few hours on a standard laptop~\cite{Bott_2017}, which is long enough to make repeated reconstruction analyzes impractical. Finally, image-recognition-focused machine-learning algorithms could enable more systematic comparisons between synthetic images and measured ones, in turn driving improved standards in the accuracy of field reconstructions.  

One final area in which the analysis of proton images could be developed further is the systematic inclusion of Coulomb scattering models into electromagnetic-field reconstruction algorithms. Such scattering cannot be neglected in many HED experiments when current-generation proton sources are used; and while the effect of Coulomb scattering could be minimized by using next-generation beams with higher energies (see Sec.~\ref{sec:frontiers:sources}), such beams also experience smaller deflections due to electromagnetic fields, which could reduce the feasibility of a successful measurement of the latter. High-quality models of Coulomb scattering are quite commonly incorporated into particle-tracing codes, and have been used previously to extract information about plasma densities in ICF experiments~\cite{Mackinnon_2004}, but simultaneous analysis of electromagnetic fields has not typically been done. Some recent research suggests that this could have been an oversight. \onlinecite{Lu_2020} provided a proof of concept for measuring magnetic fields in high-density ($>$1 g/cc) plasmas by utilizing very small aperture proton beams in higher density objects like a laser-driven shock-tube, while \onlinecite{Bott_2021b} used the broadening of caustic features by scattering in multiple proton images to simultaneously measure magnetic fields and areal densities. If such approaches could be refined and extended, proton imaging could become a vital diagnostic on a wider set of experiments than is the case currently. 

\subsection{Advanced Schemes} \label{sec:frontiers:schemes}

In addition to advanced sources and detectors, we briefly discuss here possible new schemes for setting up proton imaging experiments.  These schemes may allow for some of the most significant current limitations of proton imaging setups to be overcome: specifically, providing characterization of the undisturbed proton fluence, fiducial images for proton deflectometry registration, novel imaging setups utilizing proton optics, and tomographic imaging to obtain 3D measurements.

Characterizing the undistributed proton fluence is critical for implementing numerical reconstruction techniques, but this is often hampered by shot-to-shot variations and non-uniform initial proton distributions.  A straightforward way to address this by extending existing setups is to place one detector (e.g., a piece of RCF or a thin scintillator) directly in front of an imaged object and another detector behind.  This would allow one to record an image of both the undisturbed proton fluence as well as the proton deflections on each shot.  Similar work has been pursued in hadron therapy (see, for example, \onlinecite{Johnson_2017}).

Another key opportunity is to combine information from both x-ray and proton images (e.g., \onlinecite{Johnson_2022,Orimo_2007}).  In this technique, the last piece in an RCF or CR-39 detector stack is an image plate (see Fig.~\ref{fig:PRAD_diagram}), which can record high-energy x-rays from the \DHe{} implosion or TNSA target.  This can be used as an alignment or registration fiducial for proton deflectometry \cite{Johnson_2022}. Using lower intensity lasers \cite{Orimo_2007}, one can image smaller or less dense objects, and a variant of this scheme uses a thin needle to produce protons and x-rays \cite{Ostermayr_2020} along the same line-of-sight. This dual imaging has been shown with electrons as well \cite{Nishiuchi_2014,Faenov_2016}, which could be used to break the degeneracy between electric and magnetic fields. In this vein, the use of an x-ray free electron laser (XFEL) or coherent source co-located at laser facilities could provide significant advantages when combining proton and high resolution x-ray images along the same line-of-sight.

The development of alternative imaging schemes based on scattering and/or diffraction could be useful for enhanced data acquisition. Possibilities include x-ray and electron analogs for Fourier plane imaging \cite{Smalyuk_2001}, coded apertures \cite{Ignatyev_2011}, or scattering using a proton microscope for dark field imaging (as done with electrons \cite{Martin_2012, Klein_2015}). Many of these applications would require the development of compact permanent magnet proton optics \cite{Schollmeier_2014} or dynamic laser driven optics \cite{Toncian_2006}. These would also be useful for generally improving imaging capabilities by using charged particle optics (i.e., a proton microscope, much smaller than, but similar to those at FAIR \cite{Mottershead_2003}, LANL \cite{Prall_2016,Merrill_2009,Zellner_2021}, or PRIOR \cite{Varentsov_2016}) to image the object, in contrast to the simple point-projection imaging currently employed. Small permanent optics have already been used for energy selection \cite{Schollmeier_2014}, as well as pulse solenoid optics \cite{Brack_2020} where a particular energy can be selected, but we are not aware of their use for the implementation of a proton microscope. 

Another advanced imaging scheme that, if realized successfully, would lead to much more detailed measurements of electromagnetic fields is tomographic imaging. At present, the main limiting factor on tomography is the simultaneous production (and then detection) of multiple high-energy proton beams; the number of high-energy laser facilities equipped with multiple high-intensity laser beams suitable for proton acceleration is currently small~\cite{Danson_2019}, while fielding more than two \DHe{} capsules at once is not feasible even at the the largest facilities due to the number of beams required per capsule. Novel targets have been proposed for overcoming this issue \cite{Spiers_2021}, which in principle would allow a single short-pulse beam to produce multiple proton beams. Future experiments are needed to confirm whether such a scheme works in practice. Even if a few images of the same electromagnetic structure were obtained successfully, such a sparse number of lines of sight falls well short of a standard tomographic imaging setup; this suggests that specific work on sparse-angle tomography algorithms would be warranted. Two possible approaches have been demonstrated to address this problem that improved the performance of more conventional filtered back-projection schemes \cite{Spiers_2021}. Further improvements could be derived by considering studies in related areas, such as holographic reconstruction via scattering as done with electrons~\cite{Mankos_1996}. 


\section{Summary and Outlook} \label{sec:sum}

In this paper, we have reviewed the use of proton imaging as a diagnostic for electric and magnetic fields in high-energy-density (HED) laser plasmas. In Sec.~\ref{sec:exp}, we described the experimental techniques that underpin proton imaging, including the two primary sources of multi-MeV protons (high-intensity-laser-driven sources, and \DHe{}-fusion capsule sources) and the two primary approaches for detecting them (radiochromic film and CR-39 nuclear-track detectors). The characteristic geometry of proton-imaging setups and other important considerations for successful imaging in laser-plasma experiments were also outlined. The theory of how proton images are analyzed in order to extract information about the electric or magnetic fields present in such experiments was reviewed in Sec.~\ref{sec:theory}; we explained how a basic physical description of the interaction between charged particles and arbitrary electromagnetic fields allows for both numerical simulation of synthetic proton images of pre-specified fields, as well as inverse-analysis techniques (using numerical and/or analytical modelling) that allows for the unique characterization of electromagnetic fields in some -- though not all -- situations. Section~\ref{sec:apps} presented a broad overview of experiments that have successfully used proton imaging to elucidate many different physical processes of interest in HED plasmas.
 
While the efficacy of proton imaging as a diagnostic of electromagnetic fields in some HED plasmas is already beyond doubt, there exists various different avenues for extending the capabilities of the diagnostic further, which we outlined in Sec.~\ref{sec:frontiers}. One of the primary drivers of these improvements is ongoing technological progression in high-power laser technology, as well as improved understanding of the interaction of these lasers with matter. Taken together, these advances have led to a number of promising new proton sources, including some with higher characteristic energies (which can be used to characterize stronger electromagnetic fields than conventional sources), others with better controlled beam qualities, and the paradigm-shifting prospect of high-repetition-rate sources. The latter prospect in particular has generated much interest in researching new detector technologies that (unlike existing ones) can reliably output sufficiently resolved proton images at equivalent rates to the repetition rate of the laser driving the source. Concurrently, there has been renewed effort in the last five years towards developing new techniques for extracting information from proton-imaging data systematically and automatically. Given the recent rate of progress in this area, and broader scientific advances in data analysis derived from machine learning, it is not unreasonable to anticipate that there will exist within ten years a plethora of new, sophisticated algorithms beyond anything we have described here. Finally, although moving to imaging schemes that are more advanced than the current standard -- such as tomographic schemes, or schemes attempting ``proton optics'' -- presents several serious practical challenges, the latest research suggests that progress towards realizing such schemes is not an impossible dream. 

In short, during the (just over) two decades since it became practically realizable, proton imaging has proven to be a powerful approach for measuring two of the key physical fields that characterize HED plasmas. Amongst its many successful applications, it has been used to show magnetic-field generation in both direct-drive and indirect-drive ICF experiments, with significant ramifications for heat transport; it has helped probe the mechanism for kinetic processes in collisionless or weakly collisional plasmas; and it has played a key role in numerous laboratory astrophysics experiments. Looking forward, the ongoing development of high-intensity-laser and fusion-capsule-backlighter proton sources on the highest-energy lasers in the world suggest that proton imaging will continue to be used to probe electric and magnetic fields in the most exciting new HED plasma experiments for the next decade and beyond.

\begin{acknowledgments}

DBS acknowledges support from the U.S. Department of Energy (DOE) National Nuclear Security Administration (NNSA) under Award No. DE-NA0004033. AFAB acknowledges support from the UKRI (grant number MR/W006723/1). MB acknowledges support from the Engineering and Physical Sciences Research Council (grant number EP/P010059/1). KF acknowledges support from the Los Alamos National Laboratory (LANL) Directed Research and Development program under project 20180040DR.  LANL is operated by Triad National Security, LLC for the NNSA under Contract No. 89233218CNA000001. WF acknowledges support from the NNSA under Award No. DE-NA0004034. JF acknowledges the support of the European Research Council (ERC) under the European Union’s Horizon 2020 research and innovation program (Grant Agreement No. 787539). CKL and FHS acknowledge support from the U.S. DOE under Contract No. DE-NA0003868 to the NNSA Center of Excellence at the Massachusetts Institute of Technology. HSP's work was performed under the auspices of the U.S. DOE by Lawrence Livermore National Laboratory (LLNL) under Contact No. DE-AC52-07NA27344. PT acknowledges support for the FLASH Center by the NNSA under Awards DE-NA0002724, DE-NA0003605, DE-NA0003842, DE-NA0003934, DE-NA0003856, and Subcontracts 536203 and 630138 with LANL and B632670 with LLNL; the NSF under Award PHY-2033925; and the U.S. DOE Office of Science Fusion Energy Sciences under Award DE-SC0021990. LW acknowledges support from the U.S. National Science Foundation under Award No. 1751462. 

\end{acknowledgments}

\bibliography{references.bib}
\end{document}